\documentclass[
pre,
reprint,
twocolumn,
10pt,
tightenlines,
superscriptaddress,
amsmath,
amssymb,
noshowkeys,
nofootinbib,
floatfix,
]{revtex4}
\usepackage{graphicx}          
\usepackage{subfigure}         
\usepackage{dcolumn}           
\usepackage{capt-of}           
\usepackage{makecell}          
\usepackage{bbm}               
\usepackage{bm}                
\usepackage{time}              
\usepackage[mathlines]{lineno} 
\usepackage{stackrel}          
\usepackage{physics}           
\usepackage[mathscr]{eucal}    
\usepackage{hyperref}          
\hypersetup{
    pdfnewwindow=true,         
    colorlinks=true,           
    linkcolor=red,             
    citecolor=blue,            
    filecolor=green,           
    urlcolor=cyan              
}
\newcommand{\Schrodinger}{Schr{\"o}dinger}     
\newcommand{\ansatz}{ans{\"a}tz}               
\newcommand{\rmi}{{\rm i}}                     
\newcommand{\rme}{{\rm e}}                     
\newcommand{\Partial}[4]
   {\Bigl ( \frac{\partial #1 }{\partial #2 } \Bigr )_{\! #3, #4 }}
\newcommand{\tint}{\!\int\!}                   
\DeclareMathOperator{\Li}{Li}                  
\newcommand{\Tproduct}[1]%
   {\ensuremath{\mathcal{T} \{ \, #1 \, \} } }
\usepackage{overpic}
\newcommand{\bq}{\begin{equation}}
\newcommand{\be}{\begin{equation}}
\newcommand{\ba}{\begin{eqnarray}}
\newcommand{\eq}{\end{equation}}
\newcommand{\ee}{\end{equation}}
\newcommand{\ea}{\end{eqnarray}}
\begin{document}
%
%
\preprint{LA-UR-23-21703} 
\title{Uniform Bose-Einstein Condensates as Kovaton solutions of the Gross-Pitaevskii 
Equation through a Reverse-Engineered Potential} 
\author{Fred~Cooper}
\email{cooper@santafe.edu}
\affiliation{Santa Fe Institute, Santa Fe, NM 87501, USA}
\affiliation{Theoretical Division and Center for Nonlinear Studies, %
Los Alamos National Laboratory, Los Alamos, NM 87545, USA}
\author{Avinash Khare}
\email{avinashkhare45@gmail.com} 
\affiliation{Physics Department, Savitribai Phule Pune University, Pune 411007, India}
\author{John F. Dawson}
\email{john.dawson@unh.edu}
\affiliation{Department of Physics, University of New Hampshire, Durham, NH 03824, USA}   
\author{Efstathios G. Charalampidis}
\email{echarala@calpoly.edu}
\affiliation{Mathematics Department, California Polytechnic State University, %
San Luis Obispo, CA 93407-0403, USA}
\author{Avadh Saxena} 
\email{avadh@lanl.gov}
\affiliation{Theoretical Division and Center for Nonlinear Studies, %
Los Alamos National Laboratory, Los Alamos, NM 87545, USA}
\date{\today, \now \ EST}
\begin{abstract}
In this work, we consider a ``reverse-engineering'' approach to construct confining
potentials that support exact, constant density kovaton solutions to the classical 
Gross-Pitaevskii equation (GPE) also known as the nonlinear \Schrodinger\ equation (NLSE).
In the one-dimensional case, the exact solution is the sum of stationary kink and anti-kink solutions, i.e. a kovaton, and in the overlapping region, the density is constant. In higher dimensions, the exact solutions are generalizations of this wave function. In the absence of self-interactions, the confining potential is similar to a smoothed out finite square well with minima also at the edges. When self-interactions are added, a term proportional to $\pm g \psi^{\ast}\psi$ gets added to the confining potential and $ \pm g M$, where $M$ is the norm, gets added to the total energy. In the 
realm of stability analysis, we find (linearly) stable solutions in the case with repulsive 
self-interactions which also are stable to self-similar deformations. For attractive interactions, 
however, the minima at the edges of the potential get deeper and a barrier in the center 
forms as we increase the norm. This leads to instabilities at a critical value of $M$ (related to the 
number of particles in the BEC). Comparing the stability criteria from Derrick's theorem  
and  Bogoliubov-de Gennes analysis stability  results, we find that both predict stability for 
repulsive self-interactions and instability at a critical mass $M$ for attractive interactions. However, the numerical analysis gives a much lower critical mass. The numerical analysis shows further that the initial instabilities violate the symmetry $x\rightarrow-x$ assumed by Derrick's theorem.
\end{abstract}

\maketitle
%
\section{\label{s:Intro}Introduction}

The study of Bose-Einstein condensates (BECs)~\cite{stringari,pethick} plays 
a fundamental role in many investigations related to addressing timely questions
in Physics. Indeed, it has recently been suggested that some fundamental questions 
concerning the unification of  the theory of General Relativity (GR) and Quantum Mechanics 
(QM) can be explored by considering the gravitational interaction between two 
BECs~\cite{Howl-2019}. One problem that has not been addressed thoroughly 
in the BECs' literature, and which has been an experimental challenge is how one can confine 
BECs in configurations which have constant density. Recent efforts in this direction
through the use of an optical box trap have been reported by Gaunt, et.al.~\cite{Gaunt-2013}, and by Lin, et.al.~\cite{Lin-2009} for a BEC in a uniform light-induced vector potential.

Our approach in the present work is to consider the Gross-Pitaevskii equation
(GPE)~\cite{Gross1961,Pitaevskii-1961} (i.e., the nonlinear \Schrodinger\ equation (NLSE) with an external potential), and first construct a wave function that has a constant density in one, two and three spatial dimensions (respectively denoted as 1D, 2D, and 3D, hereafter). We then determine the confining potential which makes this wave function an exact solution by ``reverse engineering''. This approach for obtaining exact solutions has been used previously by the present authors~\cite{Cooper-2022} to study blowup phenomena in the NLSE  with Gaussian initial data. The method for finding exact solutions by this ``reverse engineering'' approach is implicit in the result of homotopy perturbation theory~\cite{He-1999,Antar-2013}.  After finding the relevant potentials which make these wave functions exact solutions to the NLSE, we numerically study their stability by using spectral stability (or Bogoliubov-de Gennes) analysis~\cite{Bogolyubov-1947}. We also study their stability with respect to self-similar deformations of the wave functions (Derrick's Theorem)~\cite{Derrick-1964}. Both approaches lead to the conclusion  that when the self-interactions are repulsive the solutions are stable. (These are the dark solitons commonly found in most BECs).  For the case of attractive self-interactions, for which the NLSE supports bright solitons found in $^7\mathrm{Li}$ BECs~\cite{Wadati-1998}, our analysis shows there is a critical mass $M$ related to the number of atoms $N$ in the BEC above which the solution becomes unstable. The numerical analysis shows that the most unstable modes break parity symmetry and that the soliton then travels toward the boundary of the confining potential. This occurs at a mass $M$ much lower than the mass found by Derrick's theorem which preserves parity. A variant of Derrick's Theorem which studies how the energy landscape changes when we vary the position of one of the kinks  gives results more in accord with the numerics.  

The constant density solutions we study in this work are trapped versions of \textit{kovaton} 
solutions that have been found previously in certain nonlinear partial differential equations 
(PDEs). Kovatons in 1D are kink-antikink solution pairs having a plateau of arbitrary width. 
They were first discovered numerically by Pikovsky and Rosenau~\cite{Rosenau-2005,Pikovsky-2006} 
in the so-called $K(\cos)$ equation:
\begin{equation}\label{e:Kcos}
   \partial_t u + \partial_x \cos{u} + \partial_{xxx} \cos{u} = 0 \>.
\end{equation}
More recently, Eq.~\eqref{e:Kcos} has been generalized by Popov~\cite{Popov-2017} to the extended 
$K(n,m)$ equations for compactons~\cite{PhysRevLett.70.564}:
\begin{equation}\label{e:Kcosn}
   \partial_t u + \partial_x \cos^m{u} + \partial_{xxx} \cos^n{u} = 0 \>,
\end{equation}
and these traveling wave solutions (i.e., compactons and kovatons) have been found in numerical
simulations in e.g.,~\cite{Garralon-2012,Garralon-2013}. Here we are interested in studying 
\textit{stationary} yet trapped kovaton solutions of the NLSE having constant density in a specified 
1D, 2D, and 3D domain. These stationary kovatons are trapped in particular external potentials that 
we find by ``reverse engineering", and consist of the sum of two terms. The first term is present 
in the {\em linear} \Schrodinger\ equation, and is similar to a finite ``square well" and its generalizations, except the hard edges of the potential are smoothed out. There are also shallow minima near the edges of the potential. The second term therein is proportional to $\pm |\psi(x,t)|^2$, and so depends on the norm $M$ or number of particles $N$. In the repulsive case the second term makes the well progressively deeper, and the kovaton solutions are always stable.  In the attractive case the second term adds a positive term proportional to the density which makes the minimum at the edges deeper, and starts a barrier at the center of the potential. This leads to the instability of the solution. We want to stress that in this paper the treatment of the BEC is purely classical. Quantum fluctuations around the BEC solution will also play a role in the stability of the BEC, such as losses to the continuum. That will be the subject of a future study. 

The paper is organized as follows. In Section~II, we present the general methodology
to construct exact kovaton solutions to the GPE in any spatial dimension by using
our reverse engineering approach. Then, Sec.~III presents the 1D kovaton solutions
together with their stability analysis results emanating from Derrick's theorem as well as the energy 
landscape as a function of a collective position coordinate for one of the kinks which breaks the parity symmetry.  In Sec.~IV, we consider 2D square and radial kovaton solutions, and similar to Sec.~III, we utilize Derrick's theorem to discuss their stability. The stability analysis results
of Secs.~III and~IV are compared with numerical results that are presented
in Sec.~V. We briefly discuss the generalization of our approach to 3D kovatons in Sec.~VI,
and in Sec.~VII, we state our conclusions.

%
\section{\label{s:reverseE}Finding exact Kovaton solutions by reverse engineering}

The time-dependent, non-linear \Schrodinger\ equation (NLSE) with an 
external potential [or the Gross-Pitaevskii equation (GPE)] is given by
\begin{equation}\label{e:NLSE}
   \{\,
     -
     \laplacian
     +
     g \, |\psi(\vb{r},t)|^{2}
     +
     V(\vb{r}) \,
   \} \, \psi(\vb{r},t)
   =
   \rmi \, \partial_t \, \psi(\vb{r},t) \>, 
\end{equation}
where $\psi(\vb{r},t)\in\mathbb{C}$ is the wave function, and $\nabla^{2}$ is the Laplacian operator in the respective spatial dimension.  The real-valued function $V(\vb{r})$ is the external potential in the NLSE. For this form of the equation, $g>0$ refers to the repulsive case pertinent to the study of most BECs. On the other hand, the case with $g<0$ is the one usually studied in connection with blowup of bright solitons in the NLSE~\cite{sulem}.

Suppose that $u(\vb{r})\in\mathbb{R}$ is the solution to Eq.~\eqref{e:NLSE} at $t=0$. 
If we assume a time-dependent solution for $\psi(\vb{r},t)$ given by  the separation of
variables ansatz:
\begin{equation}\label{e:psi-u}
   \psi(\vb{r},t)
   =
   u(\vb{r}) \, \rme^{-\rmi \, \omega t} \>,
\end{equation}
then Eq.~\eqref{e:NLSE} is written as:
\begin{equation}\label{e:ueq}
   \omega \, u(\vb{r})
   +
   \nabla^{2} u(\vb{r})
   -
   g \, u^2(\vb{r}) \, u(\vb{r})
   =
   V(\vb{r}) \, u(\vb{r}) \>.
\end{equation}
If we have an analytic expression for $u(\vb{r})$, we can then find the 
potential that makes $u(\vb{r})$ an {\it exact} solution to Eq.~\eqref{e:ueq},
and thus to Eq.~\eqref{e:NLSE} through Eq.~\eqref{e:psi-u}. We note in passing
that Eq.~\eqref{e:ueq} can be directly compared with the time-independent GPE for 
the condensate wave function~\cite{Gross1961,Pitaevskii-1961} (see Appendix~\ref{s:units}
for units) given by:
\begin{equation}\label{e:GPEmu}
   \Bigl \{\,
      - \frac{\hbar^2}{2m} \laplacian
      + 
      V(\vb{r})
      + 
      U_0 \, u^2(\vb{r}) \,
   \Bigr \} \,
   u(\vb{r}) 
   = 
   \mu \, u(\vb{r}) \>,
\end{equation}
where $U_0 = 4 \pi \hbar^2 a_s/m$ is the coupling constant, and $a_s$ is the $s$-wave 
scattering length of two interacting bosons. The norm of the wave function, denoted by
$M$ is a constant of motion, and is given by 
\begin{equation}
  M = \int \dd[3]{x} u^2(\vb{r}) \>.
\end{equation}
Thus $\omega$ can be identified with the chemical potential $\mu$, and the norm 
$M$ with the particle number $N$ up to a rescaling. 
Throughout this paper, we will use $M$ and $N$ interchangeably. 

For a given $u(\vb{r})$, exact solutions to Eq.~\eqref{e:ueq} are possible provided 
that we can find a well-behaved external potential function $V(\vb{r})$ such that
\begin{equation}\label{e:Vequation}
   V(\vb{r}) 
   =
   \omega 
   +
   [\, \laplacian u(\vb{r}) \,] / u(\vb{r})
   -
   g \, u^2(\vb{r}) \>.
\end{equation}
In our reversed engineering approach, the density $\rho(\vb{r}) = u^2(\vb{r})$ 
is specified {\it a priori} and does not depend on $\omega$. As a result, this determines 
a $V(\vb{r})$ from Eq.~\eqref{e:Vequation} so that $u(\vb{r})e^{-i \omega t}$ is 
an exact solution of the NLSE. Although changing $\omega$  shifts the potential by a constant, 
this shift has no effect on the stability of the solutions, so for convenience we will set 
$\omega=\omega_0$ in all our plots, where $\omega_0$ is chosen so that $V(\vb{r}) \rightarrow 0$
as $r \rightarrow \infty$. For arbitrary $\omega$, the potential as well as the 
energy per particle gets shifted by $\omega- \omega_0$.

The conserved energy for solutions of Eq.~\eqref{e:NLSE} is given by
\begin{equation}\label{e:energy}
   E[\psi,\psi^{\ast}] 
   =
   \int \dd[d]{x} 
   \bigl \{\,
     \, | \grad{\psi} |^2
      +
      (g/2) \, |\psi|^{4}
      +
      V(\vb{r}) \, | \psi |^2 \, 
   \bigr \} \>,
\end{equation}
and the conserved particle number by
\begin{equation}\label{e:particalnumber}
   M[\psi,\psi^{\ast}]
   =
   \int \dd[3]{x} |\psi|^{2}\>.
\end{equation}
Varying the energy $E[\psi,\psi^{\ast}]$ while holding the normalization 
$M[\psi,\psi^{\ast}]$ constant leads to the time-independent GPE [cf.~Eq.~\eqref{e:GPEmu}] 
with Lagrange multiplier $\mu$.

%
\section{\label{s:1D}Kovatons in one dimension}

In one spatial dimension (1D), we create a kovaton by a combination of kink and anti-kink solutions by assuming the following \ansatz\ for $u(x)$:
\begin{equation}\label{e:1DuAnsatz}
   u(x)
   =
   A\, [\, \tanh(q-x) + \tanh(q+x) \,] \>,
\end{equation}
where $A$ and $2 q$ are its amplitude and width, respectively. 
The conserved particle number is given by
\begin{equation}\label{e:1DMass}
   M
   =
   4 \, A^2 \, [\, 2 q \coth(2q) - 1 \,] \>,
\end{equation}
which fixes $A$ in terms of $M$ and $q$ of the distribution. We thus find
\begin{align}
   \rho(x) 
   &= 
   u^2(x)
   \label{1D:rho} \\
   &=
   \frac{M \sinh ^2(2 q) \text{sech}^2(q-x) \text{sech}^2(q+x)}
        {4 (2 q \coth (2 q)-1)} \>.
   \notag
\end{align}
To determine the 1D potential in this case, we substitute Eq.~\eqref{e:1DuAnsatz} into Eq.~\eqref{e:Vequation}, and obtain:
\begin{align}
   V(x)
   &=
   V_0(x) - g \rho(x)\>,
   \label{e:1DV1-II} \\
   V_0(x)
   &=
   \omega
   + 
   \frac{\cosh(4x) - 2 \cosh(2q) \cosh(2x) - 3}
        { 2\, [\, \cosh^2(q) + \sinh^2(x) \,]^2} \>.
   \notag
\end{align}
Here we choose $\omega = -4$ so that $V(x) \rightarrow 0$ as $x \rightarrow \pm\infty$.  which displays the dependence of  $V$ on the number of particles when $g \neq 0$.

%

\begin{figure}[t]
   \centering
   \hspace{0.1em}
   \subfigure[\ $\rho(x)$]
   {\includegraphics[width=0.75\columnwidth]{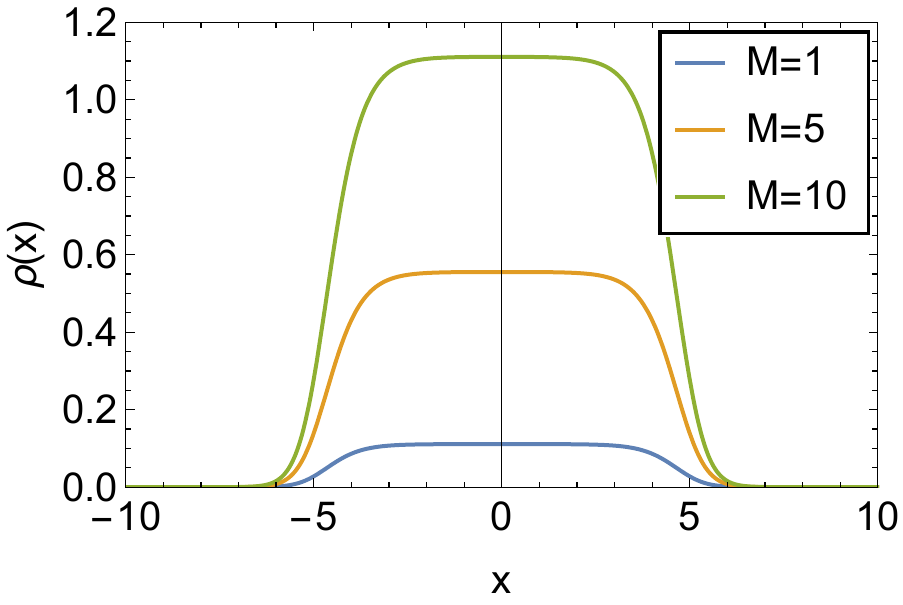}}
   \hspace{0.1em}
   \subfigure[\ $V(x)$ for $g=+1$]
   {\includegraphics[width=0.75\columnwidth]{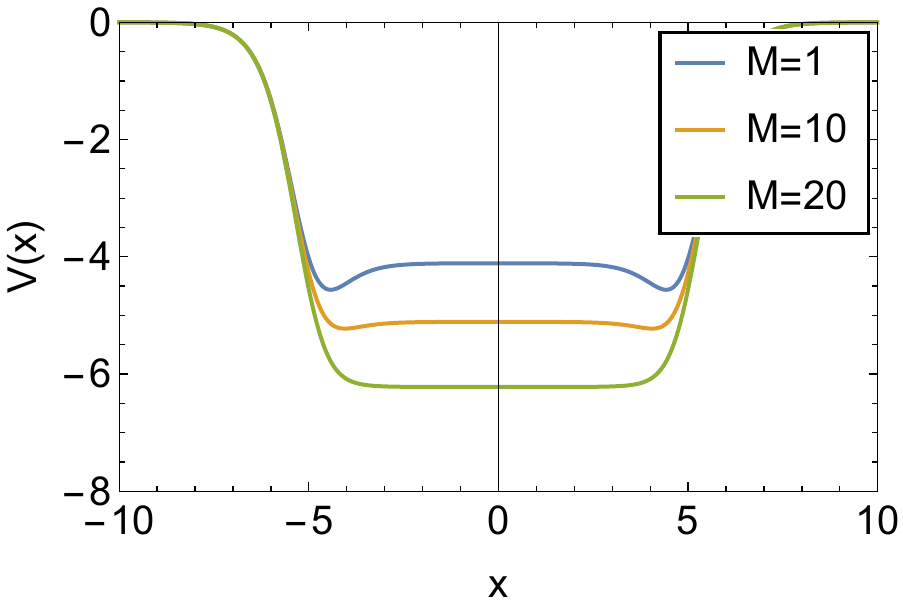}}
   \hspace{0.1em}
   \subfigure[\ $V(x)$ for $g=-1$]
   {\includegraphics[width=0.75\columnwidth]{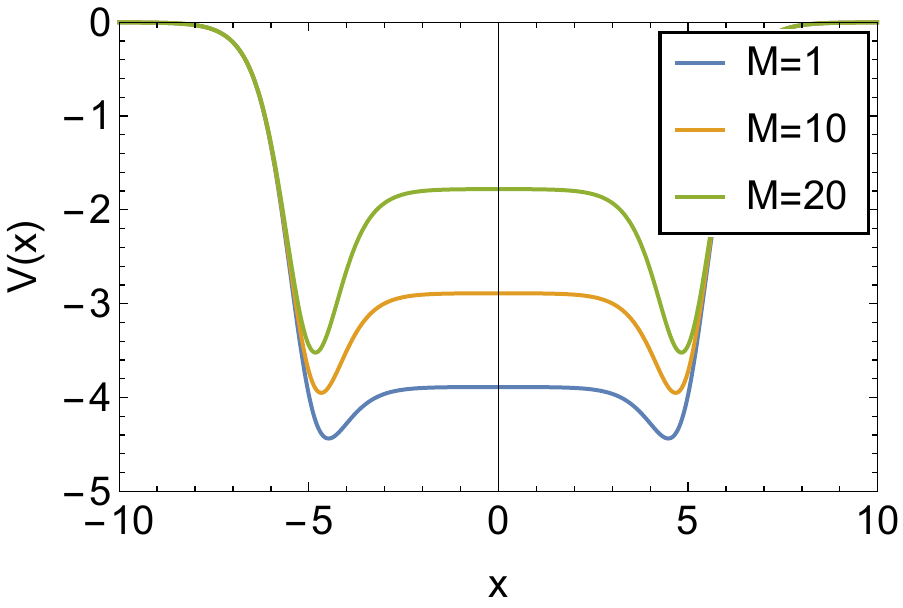}}
   \caption{\label{f:Fig1abc} One-dimensional condensate density $\rho(x)$ and 
   potentials $V(x)$ for $q=5$ and $\omega = -4$ for $g = \pm 1$.}
\end{figure}

In Fig.~\ref{f:Fig1abc}, we summarize our analytical results for the 1D case.
In particular, we present the condensate density $\rho(x) = u^2(x)$ in 
panel (a) of the figure for $M=1,5,10$ with $q=5$. The panels (b) and (c) 
in Fig.~\ref{f:Fig1abc} depict the confining potentials for (b) $g=1$ and 
(c) $g=-1$, respectively, and for various values of the particle number 
$M$ (see the legends therein). We see that for the repulsive case the 
potential is progressively morphed into a finite square well potential, 
whereas for the attractive case the minima near $\pm q$ get deeper and 
a barrier forms in the center. 

%
%

\begin{figure}[ht!]
   \includegraphics[width=0.75\columnwidth]{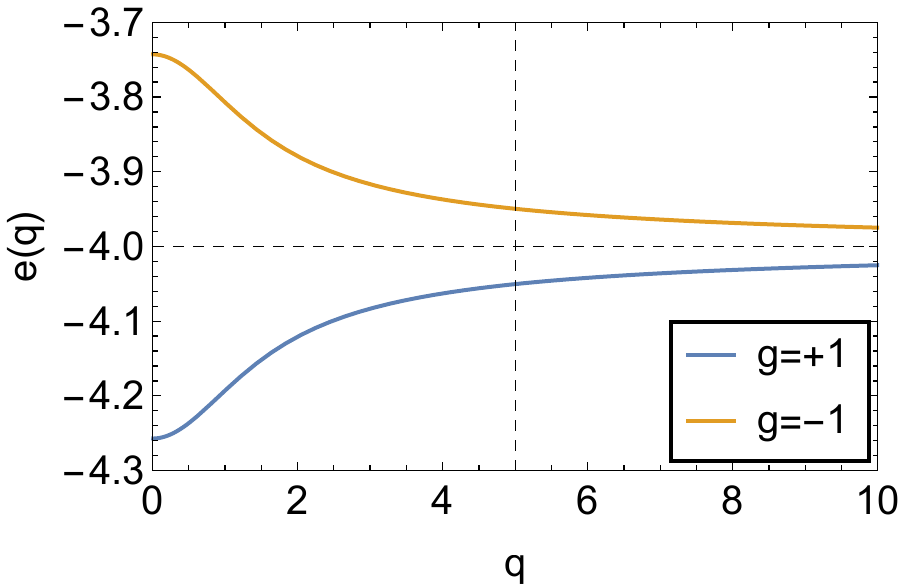}
   \caption{\label{f:Fig2}The 1D energy per particle as a function 
   of $q$ with $\omega = - 4$ and $M = 1$ for $g=\pm 1$.}
\end{figure}

The energy per particle is the sum of three terms: $e(q) = e_1(q) + e_2(q) + e_3(q)$, where
\begin{widetext}
\begin{subequations}\label{1D:allH}
\begin{align}
   e_1(q)
   &=
   \int \dd{x}
   \left(u'(x)\right)^2 / M
   \label{1D:allH1} \\
   &=
   \frac{\csch^3(2q) \, (\, - 24 q \cosh(2q) + 9 \sinh(2q) + \sinh(6q) \,)}
        { 6 \, ( 2 q \coth(2q) - 1 ) } \>,
   \notag \\
   e_2(q,M)
   &=
   \frac{g}{2}
   \int \dd{x}
   u^4(x) / M
   \label{1D:allH2} \\
   &=
   \frac{ g M \csch^3(2q) \, (\, 12 q \, ( 9 \cosh(2q) + \cosh(6q) ) 
          - 27 \sinh(2q) - 11 \sinh(6q) \, ) }
        { 48 \, ( 2 q \coth(2q) - 1 )^2 }\>,
   \notag \\
   e_3(q)
   &=
   \int \dd{x}
   V(x) \, u^2(x) / M
   =
   \int \dd{x}
   \bigl [\,
      \omega \, u^{2} -g \, u^4 + u \, u'' \,   
   \bigr ]/M
   \label{1D:allH3_v1} \\
   &=
   \omega - 2 \, e_2(q) - e_1(q) \>,
   \notag
\end{align}
\end{subequations}
\end{widetext}
where in the last term we have used Eq.~\eqref{e:ueq}, and integrated 
by parts. The resulting energy per particle is then given by
\begin{align}
   e(q)
   &= 
   e_1(q) + e_2(q,M)
   +
   \omega - 2 \, e_2(q,M) - e_1(q)
   \notag \\
   &=
   \omega - e_2(q,M) \>.
   \label{1D:Hresult}
\end{align}
In Fig.~\ref{f:Fig2}, we show the energy per particle as a function 
of $q$ emanating from Eq.~\eqref{1D:Hresult} for values of $M=1$ and $g=\pm1$.
%
%
%
%
%

%
\subsection{\label{ss:1D:Derrick}Stretching instability}

Derrick's theorem~\cite{Derrick-1964} gives a criterion for stability of 
a solution of \Schrodinger's equation under a rescaling $x \rightarrow \beta x$ 
in the soliton wave function keeping the mass $M$ fixed. This transformation is a self-similar transformation.
For all the exact solutions we present her,  we find that for the repulsive case, the energy of 
the stretched (or contracted)  solution is always a minimum at the exact solution value $\beta=1$.
However for the attractive case the energy as a function of $\beta$ shows an 
instability as we increase $M$ in that at  $\beta=1$ the minimum gets progressively shallower 
and the energy has an inflection point near $\beta=1$.  Note that for this problem, where $V$ is 
considered an {\textit{external}} potential, the confining potential is actually different for each value of $M$.

For the streched wave function, we have:
\begin{align}
   &u(x,\beta,M)
   \label{1D:wf} \\
   & \hspace{1em}
   =
   A(\beta,M) \, 
   [\, \tanh(q - \beta x) + \tanh(q + \beta x) \, ] \>,
   \notag
\end{align}
where now
\begin{equation}\label{DerrickA}
   A(\beta,M)
   =
   \sqrt{\frac{\beta M}{ 4 \, [\, 2q \coth(2q) - 1\,]} } \>.      
\end{equation}
The external potential $V(x)$ is held fixed and is given by:
\begin{equation}\label{e:1DV0} 
   V(x)
   =
   V_0(x)
   -
   g \, u_0^2(x)  \>,
\end{equation}
where $V_0(x)$ is given by \eqref{e:1DV1-II}, and $u_0(x)$ fixed by
\begin{equation}\label{1D:u0def}
   u_0(x,M)
   =
   A_0(M) \, [\, \tanh(q-x) + \tanh(q+x) \, ] \>,
\end{equation}
with
\begin{equation}\label{e:A0def}
   A_0(M)
   =
   \sqrt{\frac{M}{ 4 \, [\, 2q \coth(2q) - 1\,]} } \>,      
\end{equation}
and is \emph{independent} of $\beta$. 
Upon using the notation $h_i(\beta,M) = E_i(\beta,M)/M$, 
the energy per particle of the stretched wave function [cf. Eq.~\eqref{1D:wf}] is 
the sum of three terms: $h(\beta,M) = h_1(\beta,M) + h_2(\beta,M) + h_3(\beta,M)$
as before with $\omega_0 = -4$ but now with
\begin{subequations}\label{1D:Hbeta}
\begin{align}
   h_1(\beta,M)
   &=
   \int \dd{x}
   [\, u'(x,\beta,M) \,]^2 / M
   =
   \beta^2 \, e_1(q) \>,
   \label{1D:H1beta}  \\ 
   h_2(\beta,M)
   &=
    \frac{g}{2}
   \int\dd{x}
   u^4(x,\beta,M) / M
   =
    \beta \, e_2(q) \>,
   \label{1D:H2beta} \\
   h_3(\beta,M)
   &=
   \int \dd{x}
   V(x) \, u^2(x,\beta,M) / M
   \notag \\
   &=
   j_1(\beta,M) - j_2(\beta,M) \>,
   \label{1D:allH3}
\end{align}
\end{subequations}
where $e_1(q)$ and $e_2(q)$ are given by Eqs.~\eqref{1D:allH1} and~\eqref{1D:allH2} respectively, and $j_1(\beta,M)$ and $j_2(\beta,M)$ are just numeric and given by the integrals:
\begin{subequations}\label{1D:J12values}
\begin{align}
   j_1(\beta,M)
   &=
   \int_{-\infty}^{\infty} \hspace{-1em} \dd{x}
   V_0(x) \, u^2(x,\beta,M) / M \>,
   \label{1D:j1} \\
   j_2(\beta,M)
   &=
   g \int_{-\infty}^{\infty} \hspace{-1em} \dd{x}
   u_0^2(x,M) \, u^2(x,\beta,M) / M,
   \label{1D:j2}
\end{align}
\end{subequations}
where $V_0(x)$ is given in \eqref{e:1DV1-II}.
%
%
\begin{figure}[t]
   \centering
   \subfigure[\ $g=+1$]
   {\includegraphics[width=0.75\columnwidth]{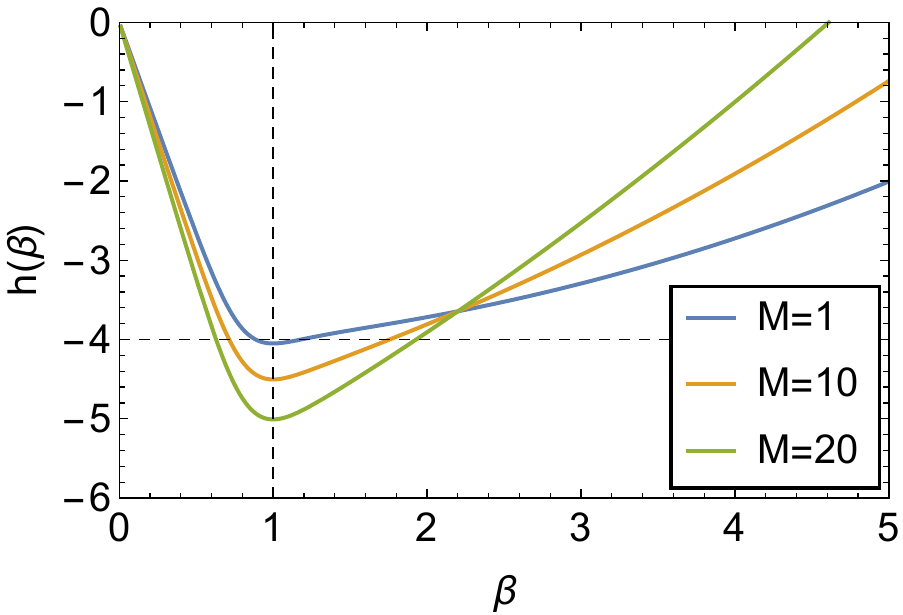}}
   \hspace{0.1em}
   \subfigure[\ $g=-1$]
   {\includegraphics[width=0.75\columnwidth]{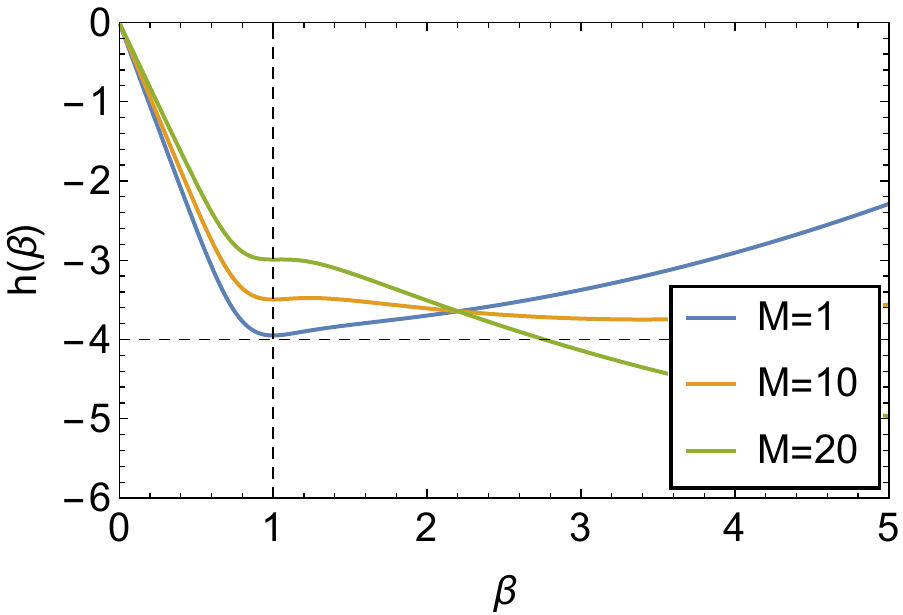}}
   \caption{\label{f:Derrick-1D-Ms}The 1D energy $h(\beta)$ as a function
   of $\beta$ for $g = \pm 1$ and for $M_0 = 1$.} 
\end{figure}
%
%
In Fig.~\ref{f:Derrick-1D-Ms} we present $h(\beta,M)$ as a function of $\beta$
for $q=5$ and $g=\pm 1$ for various values of $M$ (see the legend therein).  
For the repulsive case with $g=1$ shown in Fig.~\ref{f:Derrick-1D-Ms}(b), there is a distinct minimum at $\beta = 1$, and so Derrick's theorem predicts that this system is stable for all values of $M$.  For the attractive self-interaction case with $g=-1$ shown in Fig.~\ref{f:Derrick-1D-Ms}(a), it is not clear that there is a minimum at $\beta = 1$ for large values of $M$.
  It can be discerned from the figure that there is a minimum of the potential for $M=1$ although the minimum gets exceedingly narrow in its width and depth for $M=10$ and $M=20$. For $M=1$, the minimum is at $\beta = 1$ with a
minimum value of $h(1,1) \approx -3.94959$, and the latter agrees with the exact value of the 
energy at $q=5$ and $M=1$. Similarly, for $M = 20$, $h(1,20)\approx -2.99177$, which also agrees 
with the exact energy calculation. Since the second derivative remains positive  for all $M$ (see Appendix B), we cannot use the criterion
of the second derivative vanishing at $\beta=1$ to determine a critical mass $M_c$.  However, from the curves $h(\beta,M) $  it
is clear that even when $M=10$ the solution is unstable to be driven to larger $\beta$ by a small perturbation (i.e. blowup). 

The numerical stability simulations we have performed in Section~\ref{s:Numerical} indicate that for the attractive self interaction ($g=-1$), the kovaton becomes unstable at considerably smaller values of $M$ than we could expect from the energy landscape as a function of $\beta$.  The instability breaks the $x \rightarrow -x$ symmetry and it involves a solution at the 
minimum at $x=q$ for a slight deformation in the  positive $x$ direction.

%
\subsection{\label{ss:1Dtranslation}Translational instability}

%
%
\begin{figure}[pt!]
   \centering
   \subfigure[\ $h_3(a,M)$ for $g=+1$]
   {\includegraphics[width=0.75\columnwidth]{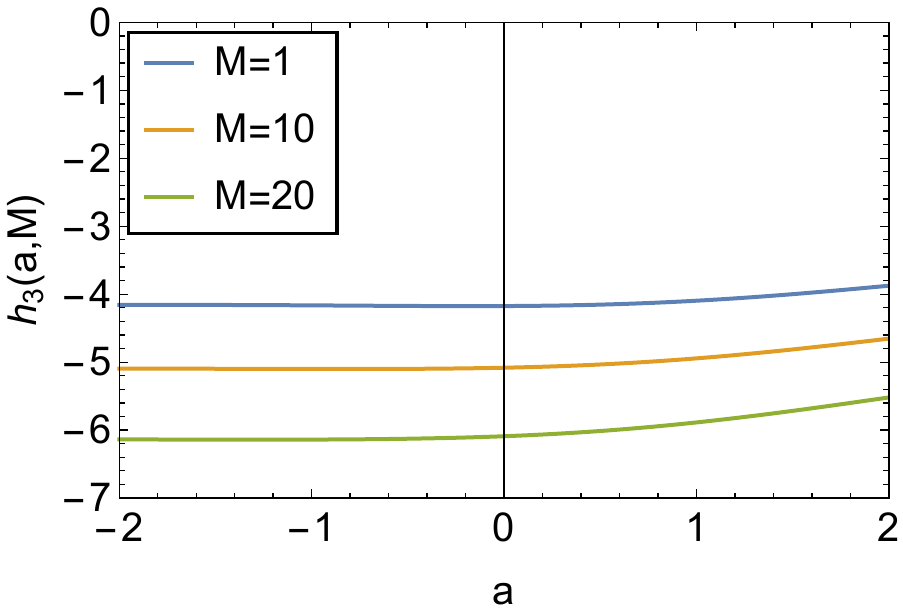}}
   \hspace{0.1em}
    \subfigure[\ $h_3(a,M)$ for $g=-1$]
   {\includegraphics[width=0.75\columnwidth]{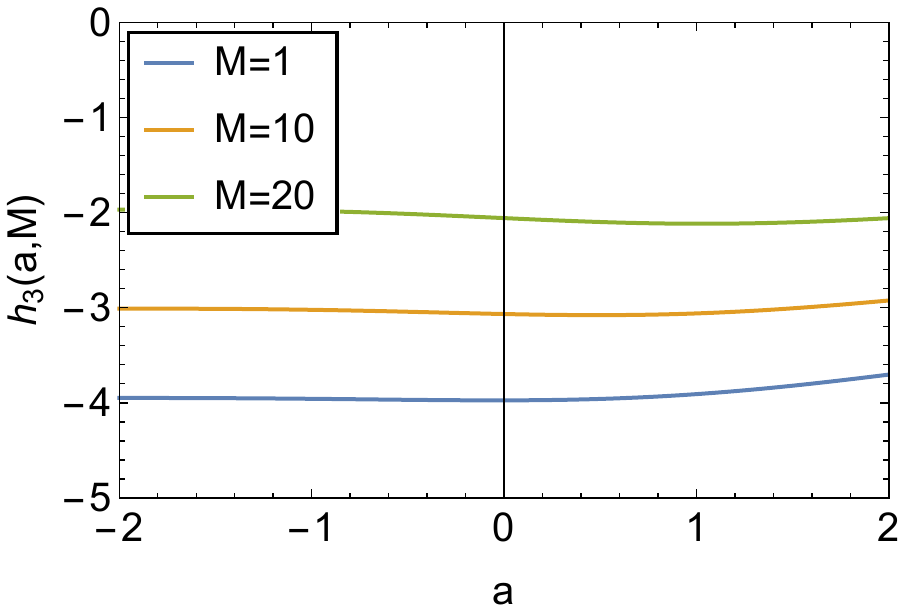}}
   \hspace{0.1em}
  \caption{\label{f:1D-trans-1}Plot of $h_3(a,M)$ for the 1D case
   for $g = \pm 1$.}
\end{figure}
%
%
Because the numerics indicate that there is a parity violating instability we would like to see if the kovaton is stable to an asymmetric translation of the wave function
 \begin{align}
   u(x,q,a,M)
   &=
   A(M) \, 
   [\, \tanh(q+a - x)  + \tanh(q +  x) \, ] \>,
   \notag \\
   A(M)
   &=
   \sqrt{\frac{M}{4\, (2 (q+a/2) \coth (2q+a) -1)}} \>, 
   \label{e:uxatrans}
\end{align}
while keeping the particle number $M$ fixed. We have considered a symmetric version of  this transformation previously in~\cite{Dawson2017, Cooper-2022}, and have shown that the critical particle number $M_c$ found using this method is the same as that found by studying the stability of small 
oscillations in a four collective coordinate approximation to the dynamics of a perturbed wave function, and then setting the oscillation frequency of the translational parameter $q(t)$, i.e. $\omega_q$ to zero.  We now  calculate the energy as a function of $a$ holding $M$ fixed.  The confining potential $V(x)$  is given in \eqref{e:1DV0}.  The energy per particle number $M$ is again the sum of three terms: $h(a,M) = h_1(M) + h_2(M) + h_3(a,M)$ with $h_1(M)$ and $h_2(M)$ being unchanged by the asymmetric shift.  As a result, the dependence on $a$ only involves the $h_3(a,M)$ term:
\begin{align}
   h_3(a,M)
   &=
   \int_{-\infty}^{\infty} \hspace{-1em} \dd{x}
   V(x) \, u^2(x,a,M) / M
   \notag \\
   &=
   j_1(a,M) - j_2(a,M) \>,
   \label{1D:allH3_transl}
\end{align}
where $j_1(a,M)$ and $j_2(a,M)$ and given by the integrals:
\begin{subequations}\label{1D:Jvalues_a}
\begin{align}
   j_1(a,M)
   &=
   \int_{-\infty}^{\infty} \hspace{-1em} \dd{x}
   V_0(x) \, u^2(x,a,M) / M \>, 
   \label{1D:j1_a} \\
   j_2(a,M)
   &=
   g \int_{-\infty}^{\infty} \hspace{-1em} \dd{x}
   u^2_0(x,M) \, u^2(x,a,M) / M,
   \label{1D:j2_a} \\
\end{align}
\end{subequations}
which are determined numerically with $V_0(x)$ given by \eqref{e:1DV1-II}.

\begin{figure}[t]
   \centering
   \includegraphics[width=0.85\columnwidth]{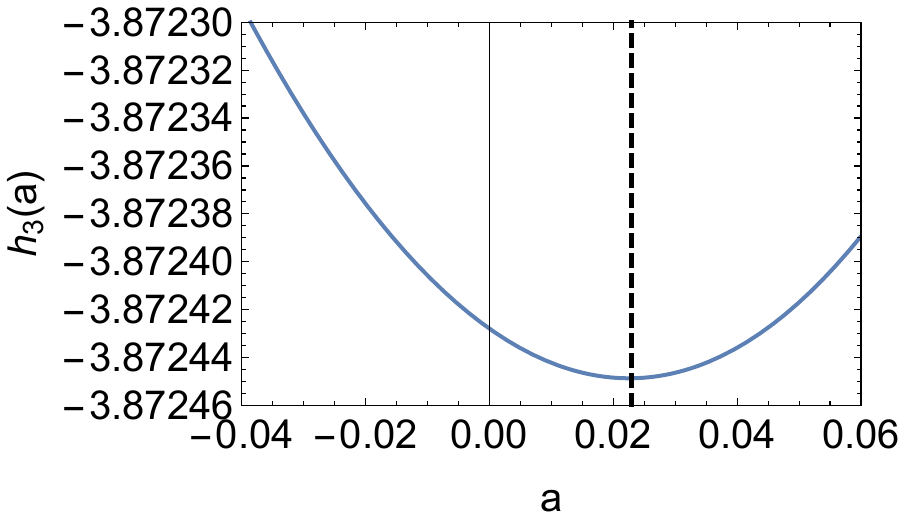}
   \hspace{0.1em}
  \caption{\label{f:1D-trans-2}Plot of $h_3(a)$ for $M=2$ for the 1D case
   for $g = -1$.  The minimum is at $a = 0.023$.}
\end{figure}

The results are shown in Fig.~\ref{f:1D-trans-1}. The kovaton looks unstable for both $g=+1$ and $g=-1$ for large $M$. For the latter case ($g=-1$) there is a critical mass $M$ for which the minimum starts moving away from $a=0$.  For $q=5$ this occurs when $M=1.63$. To show this effect we plot $h_3(a)$ as a function of $a$ at $M=2$, which is shown in Fig.~\ref{f:1D-trans-2}.
For that case we find the minimum occurs at $a=0.023$ showing that the right hand side of the kovaton wants to move to the right.  
%
%
%

Thus we see that if we choose a collective coordinate that shifts just the position of the kink making up the right side of the kovaton (here $a$ is proxy for the position of the kink on the right side of the kovaton) ($x > 0$) then it will start moving to the right once it is perturbed with $M > 1.63$.  So this crude way of taking into account that  numerical simulations show that  the mechanism that determines the onset of instabilities breaks parity invariance.   This type of instability sets in much sooner (as a function of $M$) then does the usual self-similar  blowup instability of the NLSE in the absence of a confining potential.

%
\section{\label{s:2Dkovatons}Kovatons in two spatial dimensions}

In this section we turn our attention to 2D kovaton solutions. The latter appear
in two distribution types: square and radial shapes. 

%
\subsection{\label{s:NLSE-2D-square} The 2D square kovaton}

\begin{figure}[ht!]
   \centering
   \includegraphics[height=.18\textheight]{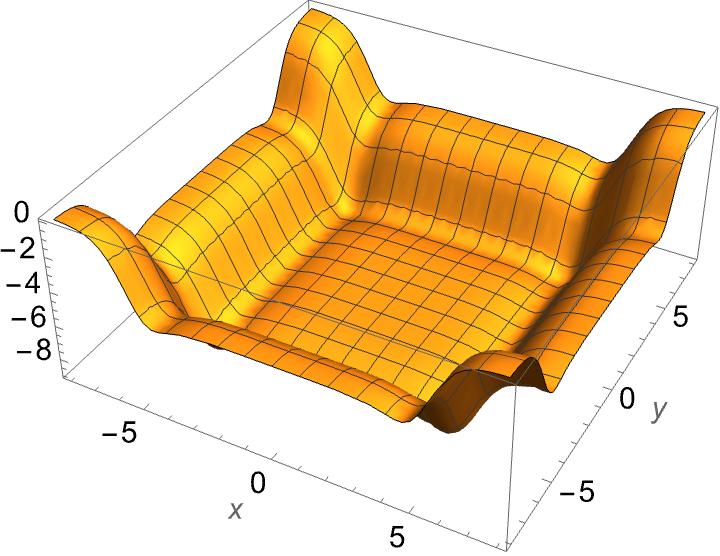}
   \caption{\label{f:V0-2Dsquare} $V_0 (x,y)$ for the square kovaton with $q=5$}
\end{figure}

\begin{figure}[ht!]
   \centering
   \includegraphics[height=.18\textheight]{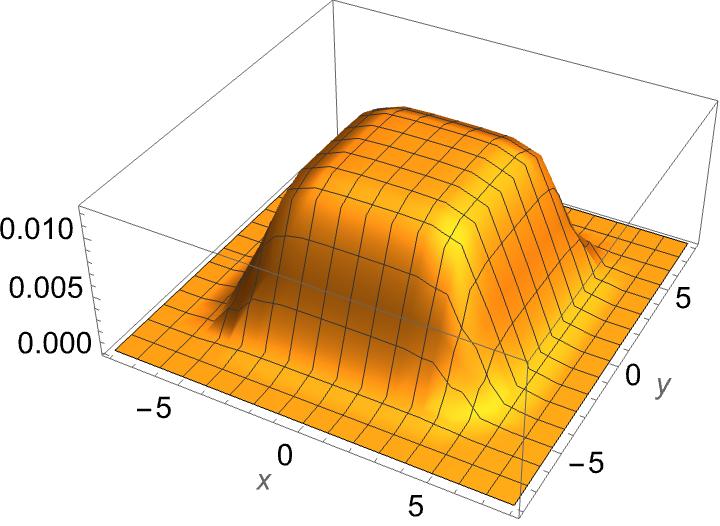}
   \caption{\label{f:rho-2Dsquare}
 $\rho (x,y)$ for the square kovaton with $q=5$, $M=1$
    }
\end{figure}

Motivated by the 1D kovaton solution of Eq.~\eqref{e:1DuAnsatz}, one can generalize
this in 2D to be a square kovaton solution which is the product 
of 1D kovaton solutions in the $x$ and $y$ directions. That is, the wave 
function for the 2D square kovaton solution is given by:
\begin{align}
   u(x,y)
   &=
   A(q) \, 
   [\, \tanh(q-x) + \tanh(q+x) \, ]
   \notag \\
   & \hspace{2.2em}
   \times
   [\, \tanh(q-y) + \tanh(q+y) \, ] \>,
   \label{e:2Dwf}
\end{align}
where the amplitude in terms of $M$ is
\begin{equation}\label{e:2DAq}
   A(q)
   =
   \frac{ \sqrt{M} }{ 4 \, [\, 2q \coth(2q) - 1\,]} \>.
\end{equation}
As in the 1D case, we can reverse engineer a potential in 2D that makes Eq.~\eqref{e:2Dwf} an exact solution. Indeed, the confining potential in question is:
\begin{align}
   V(x,y)
   &=
   V_0(x,y) + \omega - g \, u^2(x,y) \>,
   \label{e:VxyII-b} \\
   V_0(x,y)
   &=
   -
   2 \csch(2 q)
   \notag \\
   & \hspace{1.2em}\times
   \bigl [\,
      \cosh(q + x) \sech(q - x) \tanh(q - x) 
      \notag \\
      & \hspace{2em}
      + 
      \cosh(q - x) \sech(q + x) \tanh(q + x)
      \notag \\
      & \hspace{2em}
      + 
      \cosh(q + y) \sech(q - y) \tanh(q - y) 
      \notag \\
      & \hspace{2em}
      + 
      \cosh(q - y) \sech(q + y) \tanh(q + y) \,
   \bigr ] \>, 
   \notag   
\end{align}
where we select $\omega = -8$ so that $V(x,y) \rightarrow 0$ 
at $|x|, |y| \rightarrow \infty$. 
We display $V(x,y)$ in Fig.~\ref{f:V0-2Dsquare}.
We again see that for the linear \Schrodinger\ equation, the potential needed to confine a kovaton solution is similar to a  finite square well in two dimensions. It is further rounded out at the edges and has its true minimum near the boundary of the well. The density $\rho(x,y)$ for $M=1$ and $q=5$ is shown in Fig.~\ref{f:rho-2Dsquare}.
For the repulsive self-interaction, $g=1$ the interaction term deepens the well, 
whereas for the attractive case again, it causes the minimum of the potential at 
the edges to deepen and a barrier to rise away from the edge. However until $M$ 
gets quite large the self-interaction term is small compared to $V_0(x,y)$.

\begin{figure}[t]
   \includegraphics[width=0.75\columnwidth]{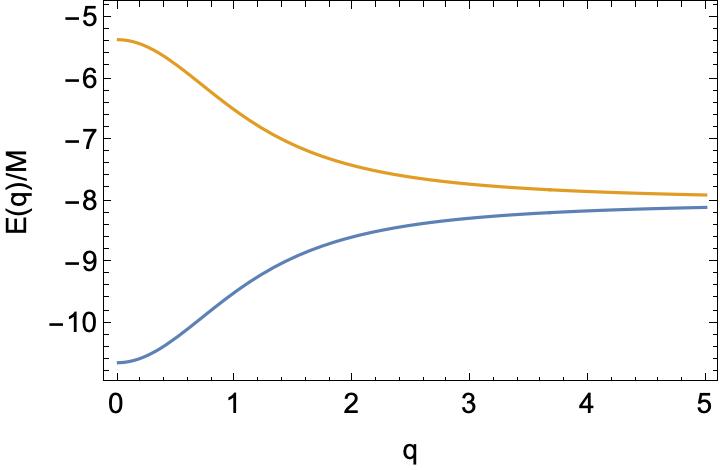}
   \caption{\label{f:Eq2DS} Plot of $h(q)$ for the 2D square case
   with $\omega = - 8$, $M =20$, and $ g = \pm 1$. The red curve is 
   for the attractive case ($g=-1$).}
\end{figure}

We now proceed similar to the 1D case. The energy per particle is the sum of three 
terms: $e(q) = e_1(q) + e_2(q) + e_3(q)$.  We find:
\begin{widetext}
\begin{subequations}\label{2D:E}
\begin{align}
   e_1(q)
   &=
   \iint\dd{x} \dd{y}
   \left[\nabla u(x,y)\right]^2 /M
   \label{2D:e1} \\
   &=  
   \frac{2 \csch^2(2q) (5 + \cosh(4q) - 12 q \coth(2q))}
        { 3 ( 2 q \coth(2q) - 1 ) } \>,   
   \notag \\
   e_2(q,M)
   &=
   \frac{g}{2}
   \iint\dd{x} \dd{y}
   u^4(x,y) /M
   \label{2D:e2} \\
   &=
   \frac{ g M \csch^6(2q)\,(-12 q ( 9 \cosh(2q) + \cosh(6q) ) 
          + 27 \sinh(2q) + 11 \sinh(6q) )^2 }
        { 1152 \, ( 2 q \coth(2q) - 1 )^4 },
   \notag \\
   e_3(q)
   &=
   \iint\dd{x} \dd{y}
   V_0(x,y) \, u^2(x,y) / M
   = 
   \omega - 2 \, e_2(q,M) - e_1(q) \>,
   \label{2D:e3}
\end{align}
\end{subequations}
\end{widetext}
where in the last term, we have used again the equations of motion and 
performed integration by parts. The resulting energy per particle is then 
given by:
\begin{equation}\label{2D:eresult}
   e(q) 
   =
   \omega - e_2(q,M) \>,
\end{equation}
and is plotted in Fig.~\ref{f:Eq2DS} as a function of $q$, and for $g=\pm1$.

%
\subsection{\label{ss:2Dstretching}Derrick's theorem for the 2D square kovaton}

\begin{figure}[t!]
   \centering
   \subfigure[\ $h(\beta,M)$ for $g=1$]
   {\label{f:Derrick2Da}
    \includegraphics[width=0.75\columnwidth]{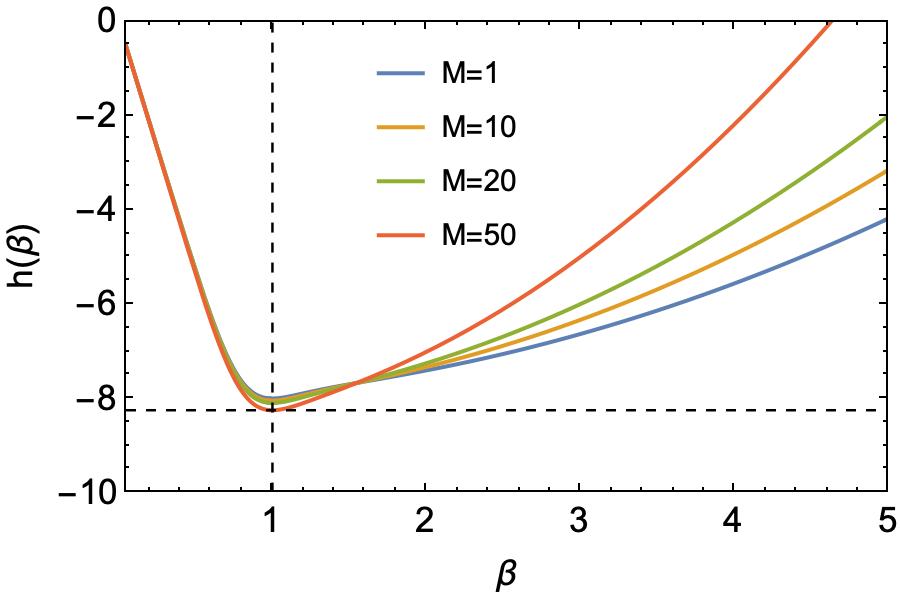}}
   \subfigure[\ $h(\beta,M)$ for $g=-1$]
   {\label{f:Derrick2Db}
    \includegraphics[width=0.75\columnwidth]{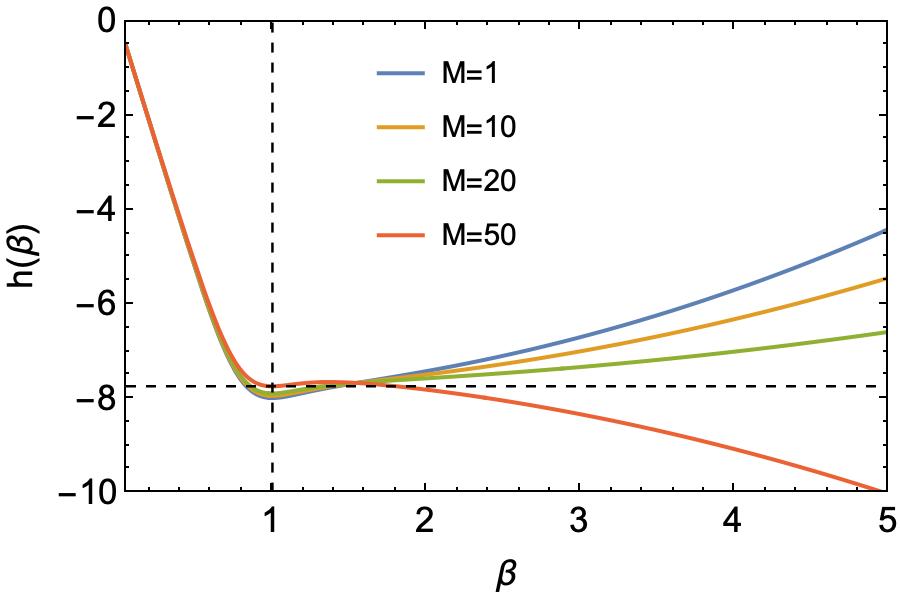}}
   \caption{\label{f:Derrick2D}Plot of $h(\beta,M)$ for the 2D square case
   with $g = \pm 1$, $\omega_0 = -8$, and $M = 1,10,20,50$.}
\end{figure}

For Derrick's theorem in the 2D square case, we consider the energy for the self-similar solution 
with  $x_i \rightarrow \beta x_i$ while keeping the mass $M$ fixed.  We get the same general picture for the 
$h(\beta,M)$ for $g= \pm1$ as for the 1D case.   For the repulsive interactions, $\beta=1$ is a minimum, whereas for the attractive case
as we increase $M$ an inflection point develops at $\beta >1$.  This is seen in  Fig.~\ref{f:Derrick2D}. 
%
%
%
%
Since Derrick's theorem does not give a reliable value for $M_c$ we will not discuss this further. 

%
\subsection{\label{sss:TwoD-Radial} Radially-symmetric kovatons in 2D}

\begin{figure}[ht!]
   \centering
   \includegraphics[width=0.75\columnwidth]{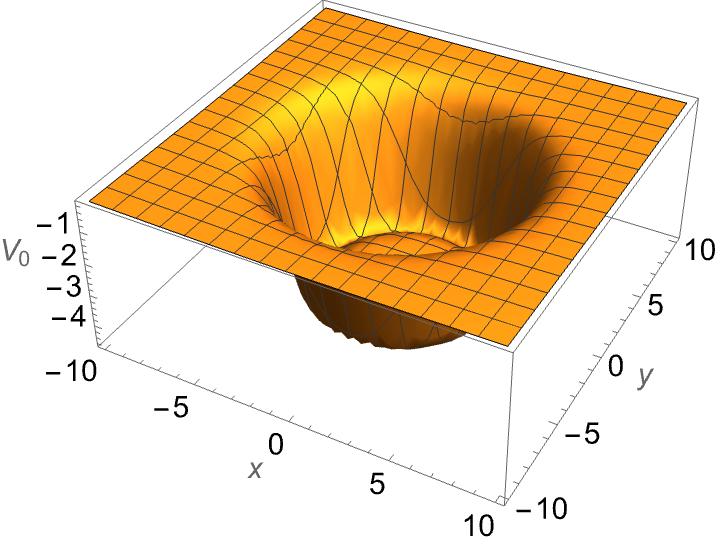}
   \caption{\label{f:V0-2Dround}$V_0(x,y)$ for $q=5$.}
\end{figure}

\begin{figure}[t]
   \centering
   \subfigure[\ $\rho(r)$]
   {\label{f:2DRrho}
    \includegraphics[width=0.75\columnwidth]{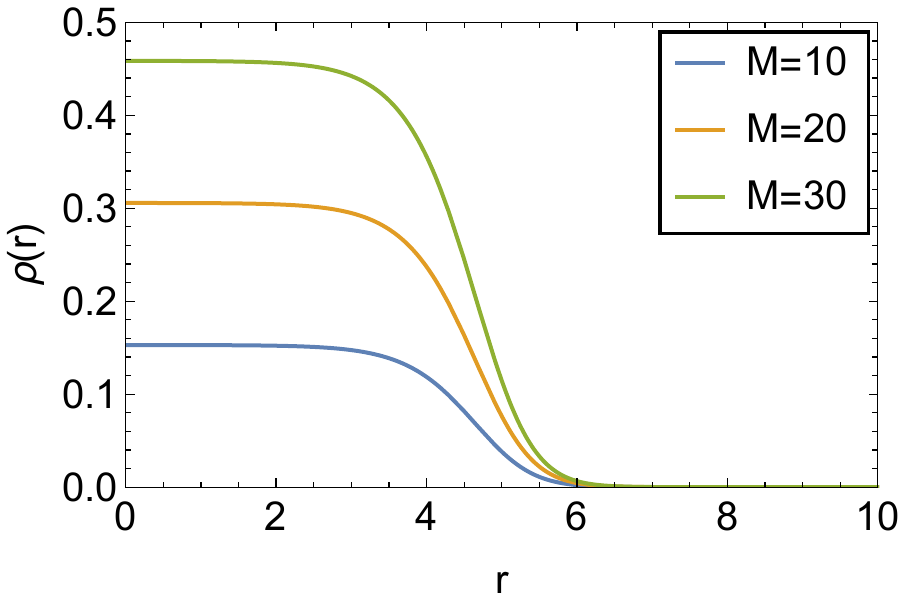} }
   \hspace{0.1em}
   \subfigure[\ $g=+1$]
   {\label{f:2DRVgp}
    \includegraphics[width=0.75\columnwidth]{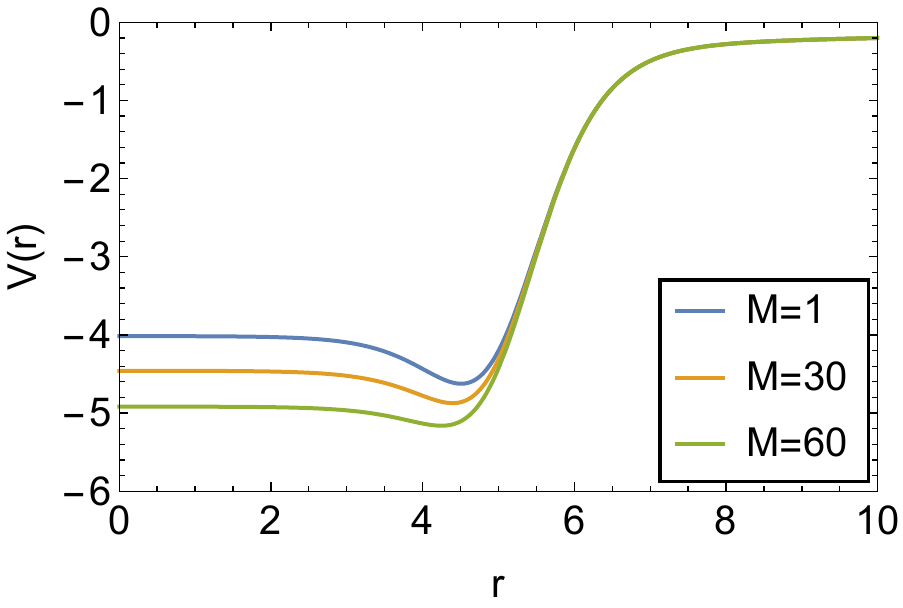} }
   \hspace{0.1em}
   \subfigure[\ $g=-1$]
   {\label{f:2DRVgm}
   \includegraphics[width=0.75\columnwidth]{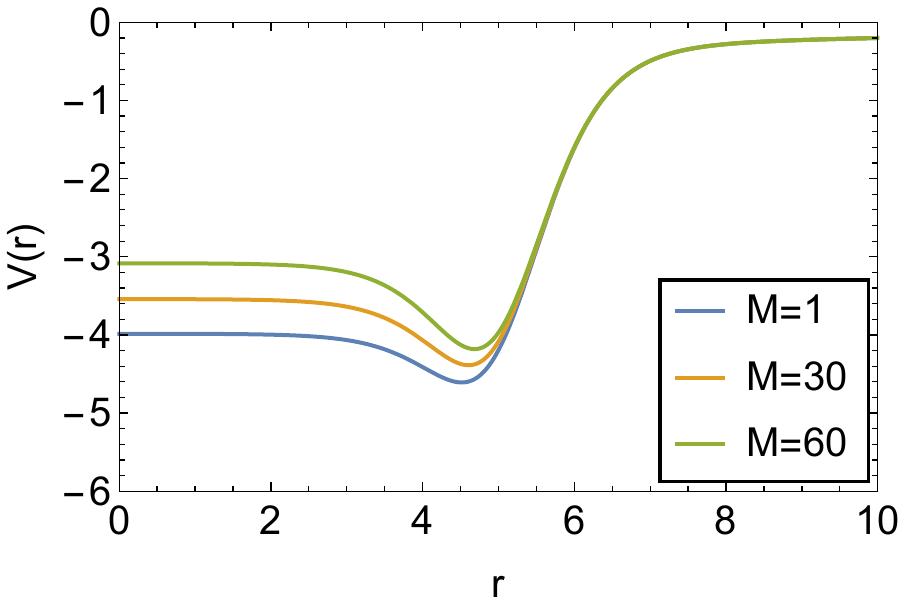} }
   \caption{\label{f:2drhor} The radial density $\rho(r)$ and potentials $V(r)$ 
   for $g=\pm 1$ for the 2D radial case with $M=20$ and $q = 5$.} 
%
%
\end{figure}

\begin{figure}[t]
   \includegraphics[width=0.75\columnwidth]{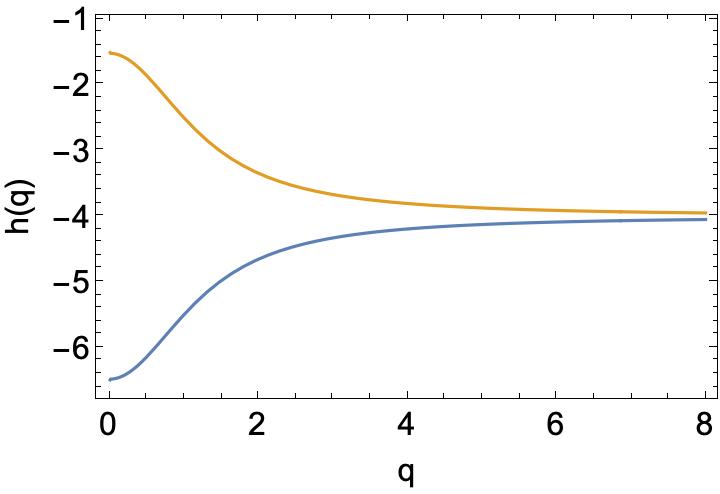}
   \caption{\label{f:D2Renergy} Plot of $e(q)$ for the 2D radial case
     with $\omega = - 4$, $q = 5$, and $M =20$. 
     The repulsive case ($g=+1$) is given by the blue curve,
     the attractive case ($g=-1$) is given by the red curve.}
\end{figure}

Another possibility for a 2D kovaton is a radially-symmetric kovaton solution
of the form:
\begin{equation}\label{e:2DR-wf}
   \psi(r,\theta,t) = u(r) \, \rme^{-\rmi \, \omega t}
   \qc
   u(r) \in \mathbb{R} \>,
\end{equation}
where
\begin{equation}\label{e:2DRu}
   u(r)
   =
   A(M,q) \, [\, \tanh(q-r) + \tanh(q+r) \, ] \>.
\end{equation}
In this case, its density is given by $\rho(r) = u^2(r)$, and the particle number $M$ is given by
\begin{widetext}
\begin{equation}\label{e:2DRmass}
   M
   =
   2 \pi \int_{0}^{\infty} \hspace{-0.5em} r \dd{r} \rho(r)
   =
   4 \pi \, A^2 \,
   \{\,
      -
      \Li_2[-e^{2 q}] \coth (2 q)
      -  
      (\, q^2 + (\pi^2/12) \, ) \coth(2 q)
      -
      \log( \rme^{2q} + 1 ) + q \,
   \} \>,
\end{equation}
where $\Li_{n}[x]$ is the PolyLog function of degree $n$~\cite{polylog}.  Solving for $A^2(M,q)$, we find:
\begin{equation}\label{e:AMq}
   A^2(M,q)
   =
   \frac{M}
   { 4\pi \,    
   \bigl \{\,
      -
      \Li_2[-e^{2 q}] \coth (2 q)
      -  
      (\, q^2 + (\pi^2/12) \, ) \coth(2 q)
      -
      \log( \rme^{2q} + 1 ) + q \,
   \bigr \} } \>.
\end{equation}
Substitution of Eq.~\eqref{e:2DRu} into \eqref{e:Vequation} gives:
%
\begin{align}
   V(r)
   &=
   V_0(r) - g \, \rho(r) \>,
   \label{e:2DR-V} \\
   V_0(r)
   &= 
   \omega 
   -
   \frac{ 
      \sech^2(q - r) \, [\, 1 + 2 \, r \tanh(q - r) \,] 
      - 
      \sech^2(q + r) \, [\, 1 - 2 \, r \tanh(q + r) \,] }
        {r \, [\,\tanh(q - r) + \tanh(q + r) \,]} \>,
   \notag
\end{align}
\end{widetext}
with $\omega = -4$.
In Fig. \ref{f:V0-2Dround}  we show the potential $V(x,y)$ for $q=5$. 
We see it is a round  waste-basket like potential which has a slightly deeper minimum near 
the boundary at $r=5$.
%
%
%
%
%
Again for the repulsive case $g=1$ the full potential gets deeper as we increase $M$, 
where for the attractive case $g=-1$ the potential develops deeper minima near $r=q$ 
as well as a barrier in the middle. 
The plot of $\rho(r)$ for $q=5, M=20$ is shown in the left panel of Fig.~\ref{f:2drhor}.
The middle and right panels of the figure depict the potential $V(r)$ as a function 
of $r$ for the case when $g  =  \pm 1$ with $q=5$.
%
%
The energy per particle of the round kovaton is the sum of three terms: 
$e(q) = e_1(q) + e_2(q) + e_3(q)$.  We find:
\begin{widetext}
\begin{subequations}\label{e:2DRee}
\begin{align}
   e_1(q)
   &=
   2 \pi \int_0^{\infty} \hspace{-0.5em} r \dd{r} \left[\nabla u(r)\right]^2 /M
   \label{e:2DRe1} \\
   &=
   -\frac{16 \pi  e^{6 q}}{3 \left(e^{4 q}-1\right)^3} \frac{A^2(M,q)}{M} \, 
   \Bigl \{\,
      -
      12 \Li_2\left( -e^{2 q} \right) \cosh (2 q)
      -
      \left( 12 q^2 + \pi ^2 \right) \cosh (2q)
      \notag \\
      & \hspace{3em}
      +
      \left(q-\log \left(e^{2 q}+1\right)\right) (9 \sinh (2 q)+\sinh (6 q))
      +
      8 \sinh^3(q) \cosh(q) \,
   \Bigr \} \,,
   \notag \\
   e_2(q)
   &=
   g \, \pi \int_0^{\infty} \hspace{-0.5em} r \dd{r} u^4(r) /M 
   \label{e:2DRe2} \\
   &=
   -
   \frac{4 \pi A^4(M,q) \, g \, e^{6 q}}{3 M \left(e^{4 q}-1\right)^3} \,
   \Bigl \{\, 
      9 \left(12 q^2+\pi ^2\right) \cosh (2 q)
      + 
      12 \Li_2\left(-e^{2 q}\right) (9 \cosh (2 q) + \cosh (6 q))
      \notag \\
      & \hspace{1em}
      +
      \left(12 q^2+\pi^2\right) \cosh (6 q) 
     -
     2 \left(q-\log \left(e^{2 q}+1\right)\right) (27 \sinh (2 q)+11 \sinh (6 q))
     \notag \\
     & \hspace{1em}
     -
     16 \sinh^3(q) (2 \cosh (q)+3 \cosh (3 q)) \,
   \Bigr \} \,, 
   \notag \\
   e_3(q)
   &=
   2 \pi \int_0^{\infty} \hspace{-0.5em} r \dd{r} 
   V(r) \, u^2(r) / M
   \label{e:2DRe3} \\
   &=
   2 \pi \int_0^{\infty} \hspace{-0.5em} r \dd{r} 
   \bigl \{\,
      \omega_0 \, u^2(r) -  g \, u^4(r) + u(r) \, [\, \laplacian u(r) \,] \,   
   \bigr \}/M
   \notag \\
   &=
   \omega - 2 \, e_2(q) - e_1(q) \>,
   \notag
\end{align}
\end{subequations}
\end{widetext}
where in the last term we have used the equations of motion and integrated by 
parts, and $A^2(M,q)$ is given by \eqref{e:AMq}. The resulting energy per 
particle is then:
\begin{equation}\label{2DR:etotal}
   e(q) 
   = 
   e_1(q) + e_2(q) 
   +
   \omega - 2 \, e_2(q) - e_1(q)
   =
   \omega - e_2(q) \>,
\end{equation}   
and is plotted in Fig.~\ref{f:D2Renergy}.
%
%

%
\subsubsection{\label{ss:2DRstretching}Derrick's theorem for the 2D radial kovaton}

\begin{figure}[t]   
   \centering
   \subfigure[\ $g = +1$]
   {\label{f:2DR-Derrick-gp}
    \includegraphics[width=0.75\columnwidth]{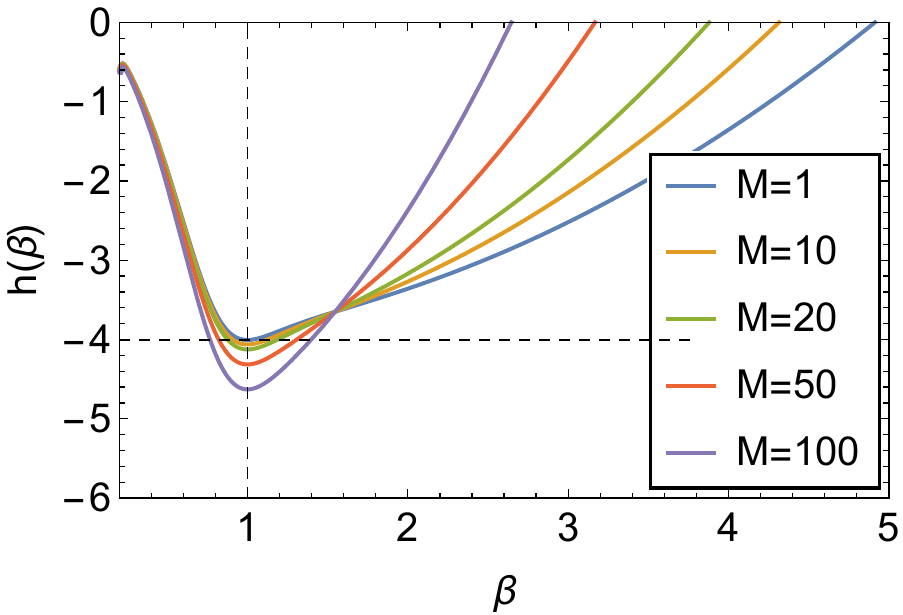} }
   \hspace{0.1em}
   \subfigure[\ $g = -1$]
   {\label{f:2DR-Derrick-gm}
    \includegraphics[width=0.75\columnwidth]{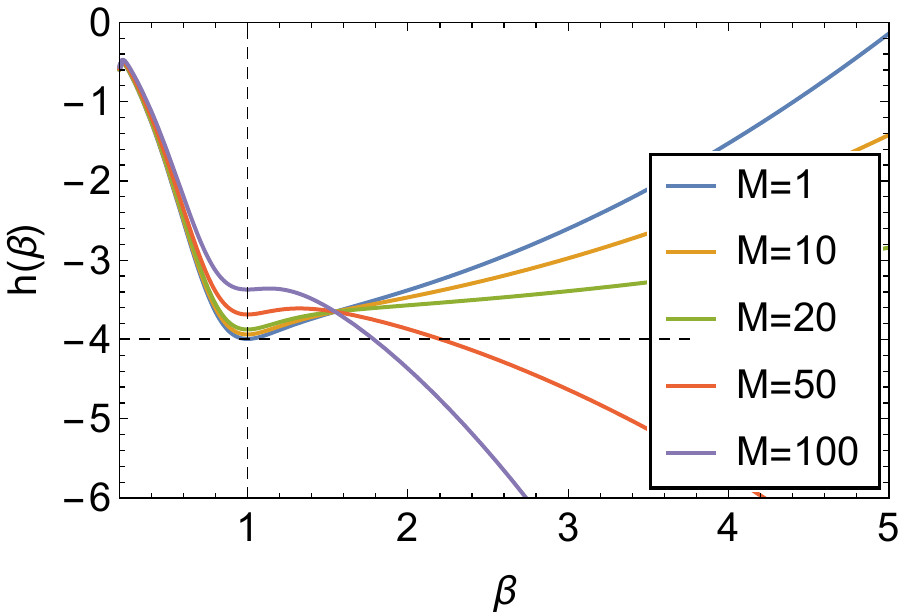} }
   \caption{\label{f:Derrick2Dr} The energy of the stretched round 2D 
   kovaton $h(\beta)$ as a function of $\beta$.}
\end{figure}

For Derrick's theorem in 2D for the round case, we perform the transformation 
$r \rightarrow \beta r$ keeping the mass $M$ fixed. That is, we calculate the energy 
as a function of $\beta$ and at a given $M$ when the wave function has the form:
\begin{widetext}
\begin{align}
   u(r,\beta,M)
   &=
   A(\beta,M) \, 
   [\, \tanh(q - \beta r) + \tanh(q + \beta r) \, ] \>,
   \label{e:2Dwf_r} \\
   A^2(\beta,M)
   &=
   \frac{M \, \beta^2}
   { 4\pi \,    
   \bigl \{\,
      -
      \Li_2[-e^{2 q}] \coth (2 q)
      -  
      \bigl [\, q^2 + \frac{\pi^2}{12} \, \bigr ] \coth(2 q)
      -
      \log( e^{2q} + 1 ) + q \,
   \bigr \} } \>. 
   \notag  
\end{align}
\end{widetext}
The potential is fixed to be the potential of the problem with $\beta=1$. The 
results for the energy of the stretched kovaton as a function of $\beta$ 
for different $M$ are shown in Fig.~\ref{f:Derrick2Dr}.
%
%
%
%
Again we see the same qualitative  behavior of $h(\beta)$. For the repulsive 
case $\beta=1$  is a minimum for all $M$ whereas there is a critical value 
of $M$ which is signaled by there being an inflection point developing near $\beta=1$ as we increase the mass $M$. 

%
\section{\label{s:Numerical}Numerical Analysis and Results for the 1D and 2D GPEs}

In this section, we discuss the existence, stability and selective cases
on the dynamics of kovaton solutions in 1D and 2D. In doing so, we consider
first the steady-state problem, i.e., the GPE of Eq.~\eqref{e:ueq}. The 
physical domains in 1D and 2D, i.e., $\mathbb{R}$ and $\mathbb{R}^{2}$
are truncated respectively into finite ones: $\Omega_{1D}=[-L,L]$ and 
$\Omega_{2D}=[-L,L]^{2}$. We then introduce a finite number of equidistant 
grid points in both cases with lattice spacing $\Delta x=0.04$ (with $L=40$)
for the 1D GPE, and $\Delta x=0.06$ (with $L=15$) for the 2D one. The 
Laplacian that appears in Eq.~\eqref{e:ueq} (and equivalently in Eq.~\eqref{e:NLSE}) 
is replaced by fourth-order accurate, finite differences, where we impose 
zero Dirichlet boundary conditions (BCs) at the edges of the computational
domain, i.e., $u\big|_{\partial \Omega_{1D,2D}}=0$. With this approach, we
want to identify the numerically exact, kovaton solutions on the above computational
grid in order to perform a spectral stability analysis followed by direct dynamical
simulations. It should be noted that one may use directly the exact solution we 
presented in this work for performing a spectral stability analysis but the calculation
will suffer from local truncation errors. The latter are avoided by finding the 
numerically exact kovaton solutions.

We identify numerically exact solutions (with strict tolerances of $10^{-12}$ 
on the convergence and residual errors) by using Newton's method where the associated 
Jacobian matrix of the pertinent nonlinear equations is explicitly supplied therein. 
We note in passing, that the potential $V(\vb{r})$ we consider for our numerical  
simulations is given by Eq.~\eqref{e:Vequation}, and the $u(\vb{r})$ that appears 
therein is replaced by the 1D and 2D (either square or radial) kovaton solutions of 
Eq.~\eqref{e:1DuAnsatz} as well as Eqs.~\eqref{e:2Dwf} and~\eqref{e:2DRu}, respectively. 
The amplitude $A$ of the solution is expressed in terms of the mass $M$, rendering the 
potential to be a function of $M$ (the values of $g$, $\omega$, and $q$ are fixed). Then 
for fixed $M$, we use the exact waveforms of Eqs.~\eqref{e:1DuAnsatz},~\eqref{e:2Dwf} 
and Eq.~\eqref{e:2DRu} as initial guesses to the Newton solver. Upon convergence, we 
perform a sequential continuation over $M$, and trace branches of kovaton solutions 
whose spectral stability analysis is carried out next.


To do so, we consider the perturbation \ansatz:
\begin{align}
   \psi(\vb{r},t)
   &= \psi_0(\vb{r},t) + \varepsilon \, \psi_1(\vb{r},t) + \dotsb
   \label{pertr_ansatz} \\
   &=
   \rme^{- \rmi \omega_0\,t}
   \bigl \{\,
      u_0(\vb{r})
      +
      \varepsilon \, 
      \bigl [ \,
         a(\vb{r}) \, \rme^{\lambda t} + b^{\ast}(\vb{r}) \, \rme^{\lambda^{\ast} t} \,
      \bigr ]
   \bigr \}
   +
   \dotsb
   \notag
\end{align}
where $\varepsilon\ll 1$ and where $u_0(\vb{r})$ satisfies the time-independent Gross-Pitaevskii equation \eqref{e:ueq} with $\omega \rightarrow \omega_0$.  
Upon plugging Eq.~\eqref{pertr_ansatz} into Eq.~\eqref{e:NLSE}, to ${\mathcal{O}}(\varepsilon)$ we arrive at the eigenvalue problem:
\begin{gather}
   \mathcal{A}(\vb{r}) \, \mathcal{V}(\vb{r})
   =
   \rmi \lambda \, \mathcal{V}(\vb{r})
   \label{stab_problem} \\
   \mathcal{V}(\vb{r}) =
   [\, a(\vb{r})\,\,\,b(\vb{r}) \,]^{T}
   \in
   \mathbb{C}^{2}
   \qc 
   \lambda\in\mathbb{C} \>.
\end{gather}
where $\mathcal{A}(\vb{r})$ is the $2\times 2$ matrix
\begin{align}
\mathcal{A}(\vb{r}) 
=
\begin{pmatrix}
   \mathcal{A}_{11}(\vb{r}) & \mathcal{A}_{12}(\vb{r}) \\
   -\mathcal{A}_{12}^{\ast }(\vb{r}) & -\mathcal{A}_{11}(\vb{r})
\end{pmatrix},
\end{align}
and the matrix blocks are given by
\begin{subequations}
\begin{align}
   \mathcal{A}_{11}(\vb{r}) 
   &=
   - \laplacian + 2g \, u_{0}^{2}(\vb{r}) + V(\vb{r}) - \omega_0, \\
   \mathcal{A}_{12}(\vb{r}) 
   &=
   g \, u_{0}^{2}(\vb{r}) \>. 
\end{align}
\end{subequations}
Then, the eigenvalue problem of Eq.~\eqref{stab_problem} is solved
by using the contour-integral based FEAST eigenvalue solver~\cite{doi:10.1137/15M1026572}
(see also~\cite{CHARALAMPIDIS2020105255,PhysRevA.105.053303} for its 
applicability to relevant yet higher dimensional problems too). A
steady-state kovaton solution $u_0(\vb{r})$ is deemed stable
if all the eigenvalues $\lambda=\lambda_{r}+\mathrm{i}\lambda_{i}$ 
have zero real part, i.e., $\lambda_{r}=0$. On the other hand, if 
there exists an eigenvalue with non-zero real part ($\lambda_{r}\neq 0$),
this signals an instability, and thus the solution is deemed linearly
unstable.

\onecolumngrid\
\begin{figure}[pt!]
\begin{center}
\begin{overpic}[height=.16\textheight, angle =0]{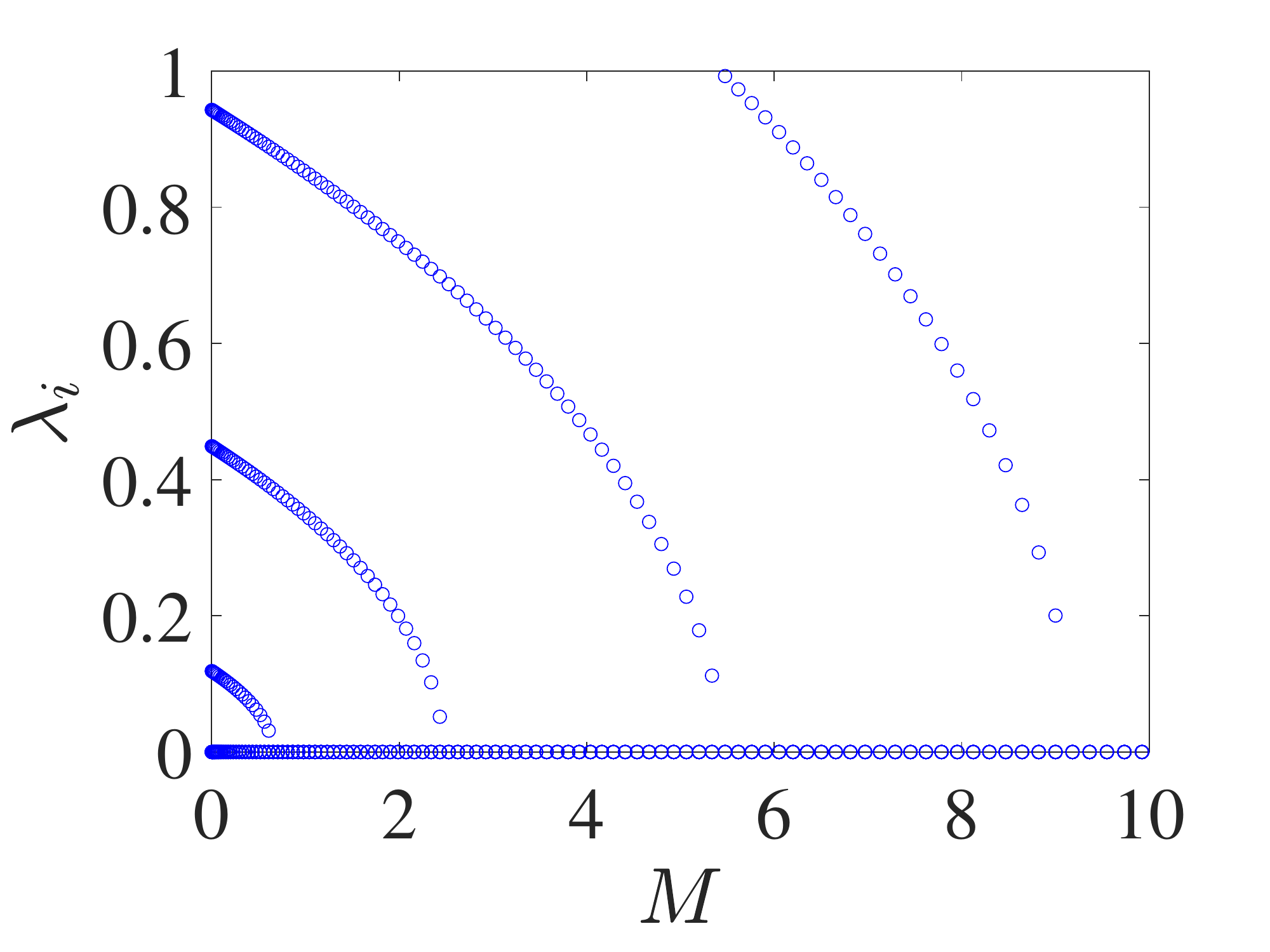}
\put(35,63){$(a)$}
\end{overpic}
\includegraphics[height=.16\textheight, angle =0]{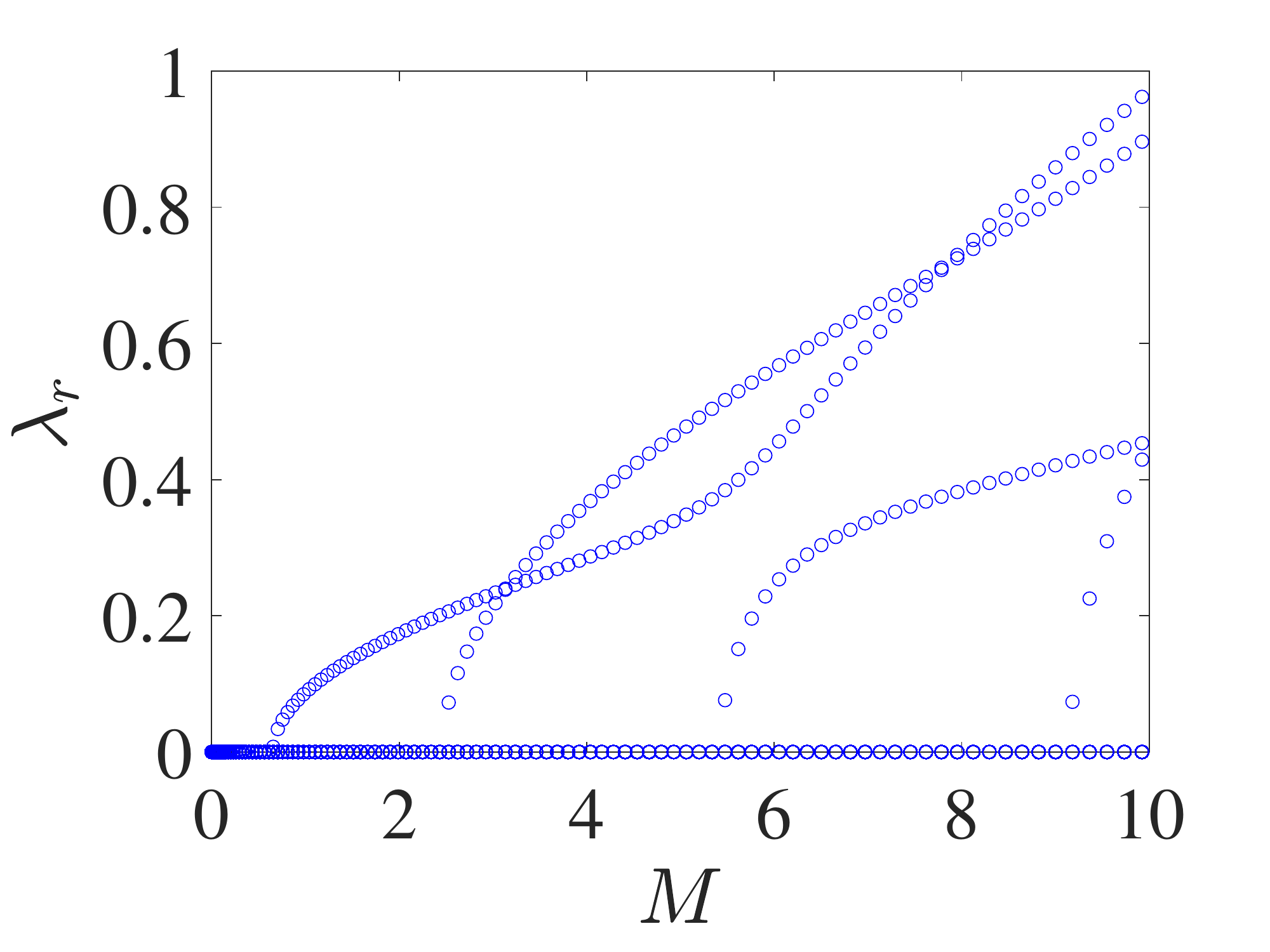}\\
\begin{overpic}[height=.16\textheight, angle =0]{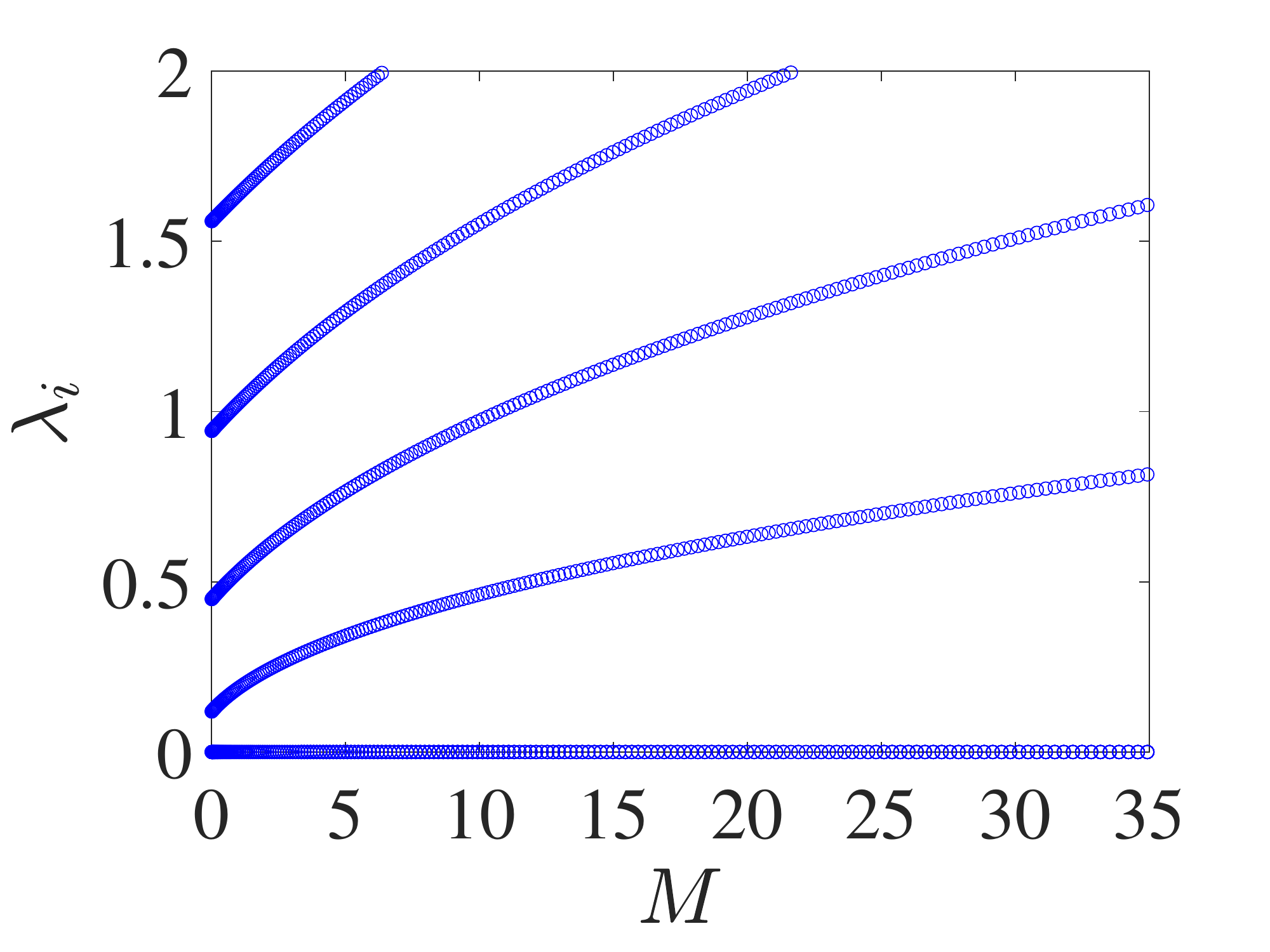}
\put(35,63){$(b)$}
\end{overpic}
\includegraphics[height=.16\textheight, angle =0]{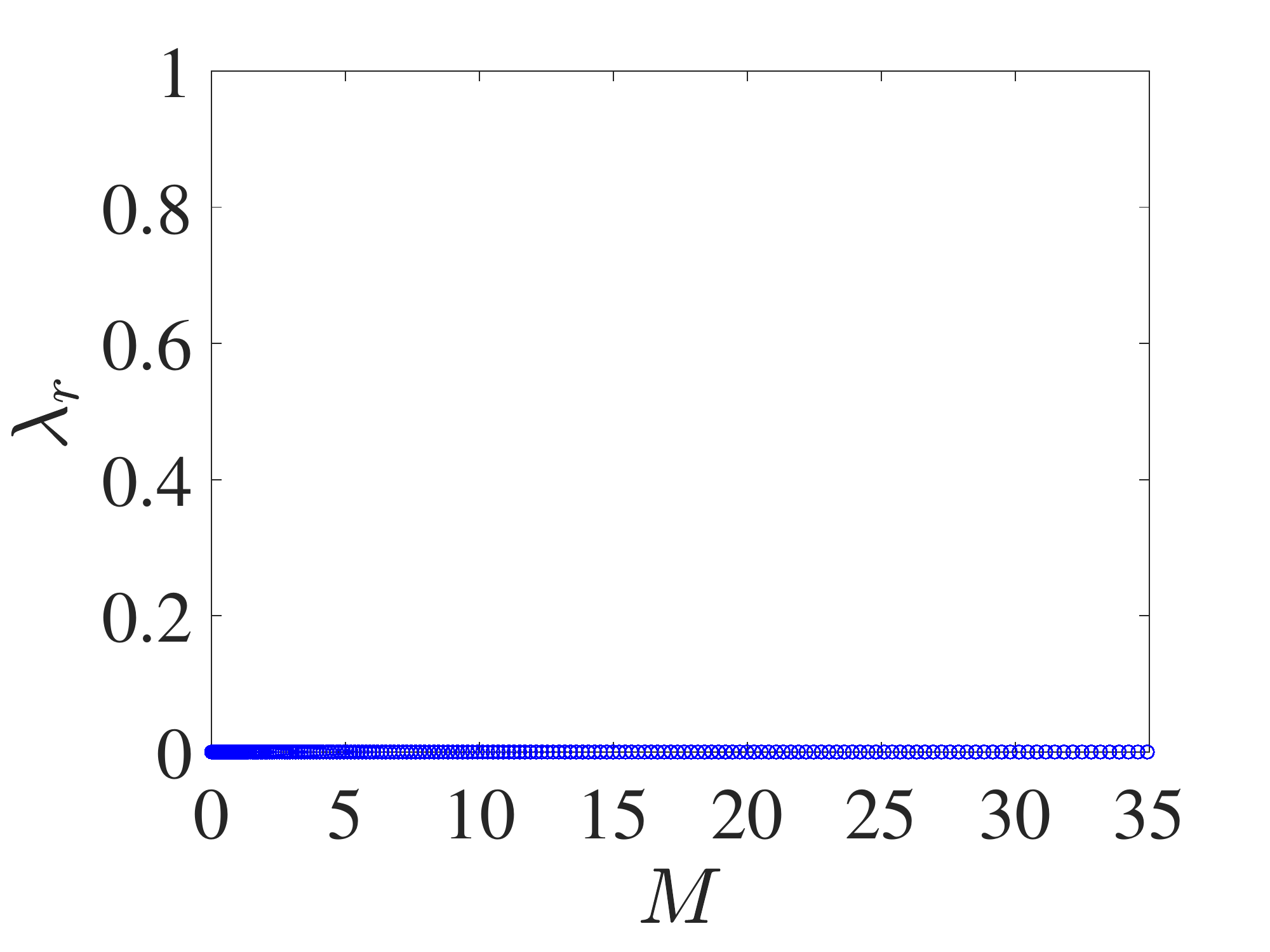}
\end{center}
\caption{\label{fig1}
Spectral stability analysis of 1D kovaton solutions for (a) $g=-1$ (attractive) 
and (b) $g=1$ (repulsive), respectively. The left and right columns depict respectively 
the imaginary and real parts of the eigenvalues of the stability problem of 
Eq.~\eqref{stab_problem}. The parameter values here are $\omega=-4$ and $q=5$.
Note that the parameter $M$ herein coincides with the mass (or $l_{2}$-norm) 
of the kovaton solution via Eq.~\eqref{e:1DMass}.
}

\end{figure}
\twocolumngrid\
%

%
\subsubsection{Numerical Results for the 1D GPE}

\begin{figure}[pt!]
\begin{center}
\includegraphics[height=.16\textheight, angle =0]{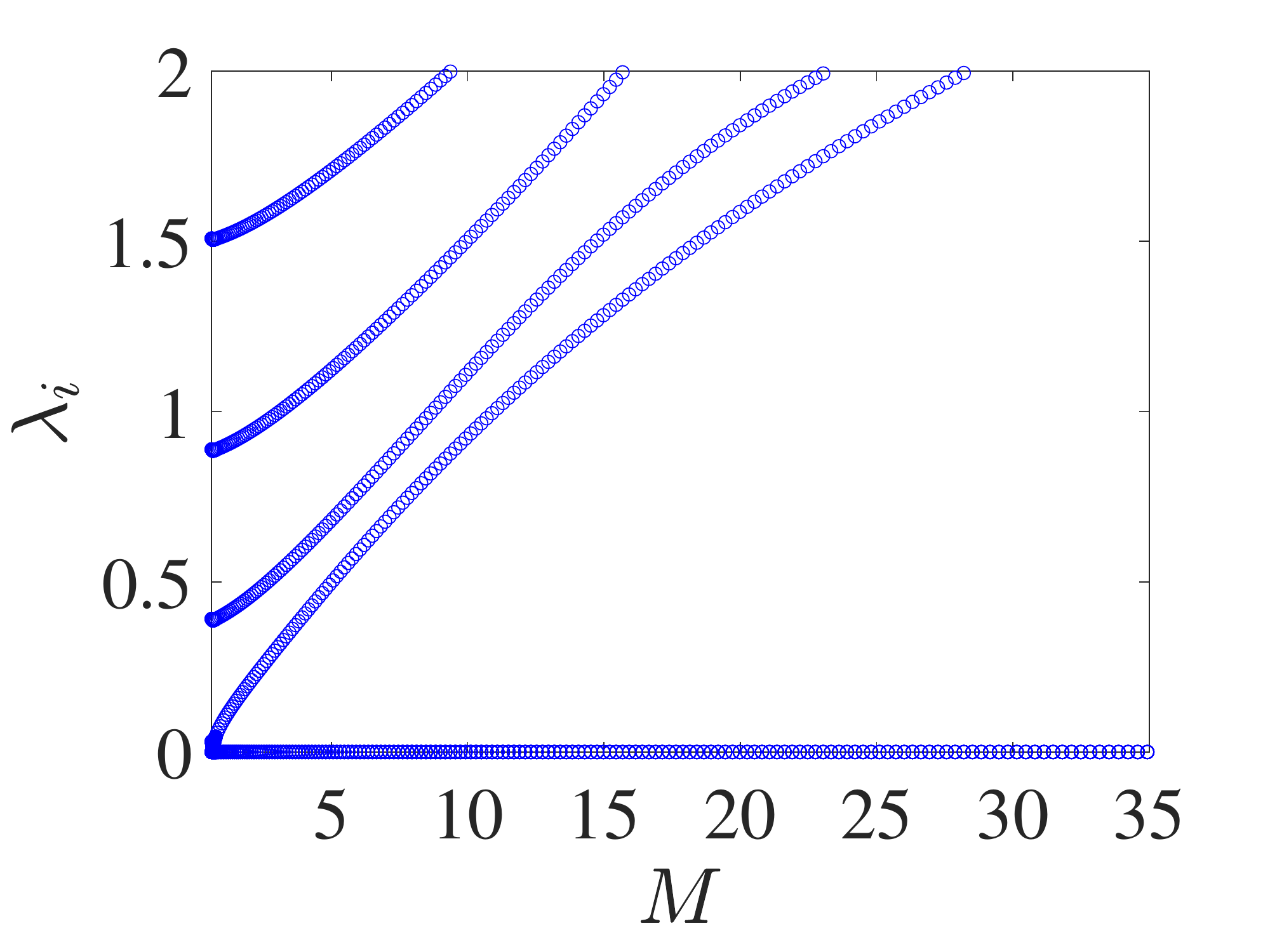}
\includegraphics[height=.16\textheight, angle =0]{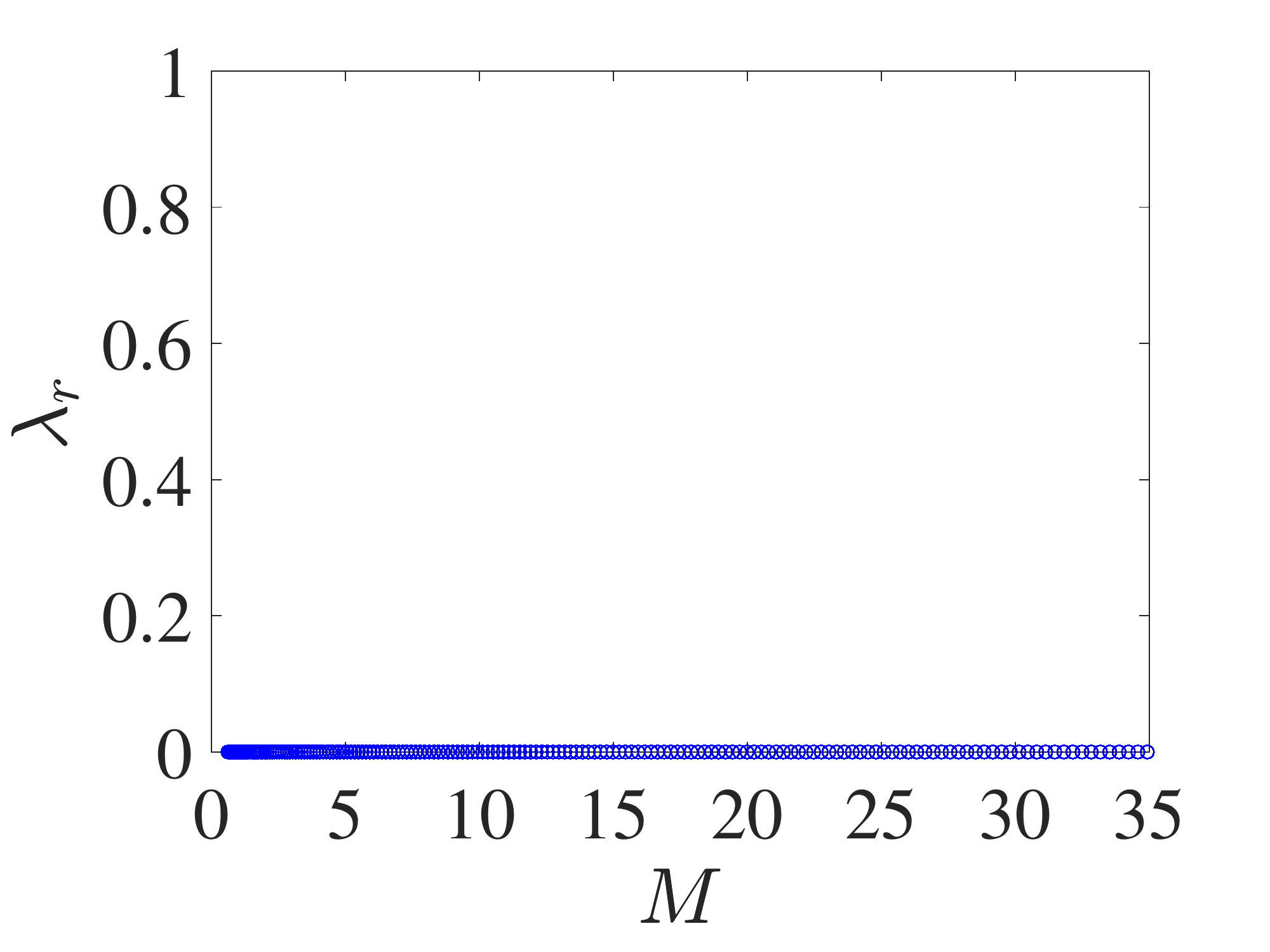}
\includegraphics[height=.16\textheight, angle =0]{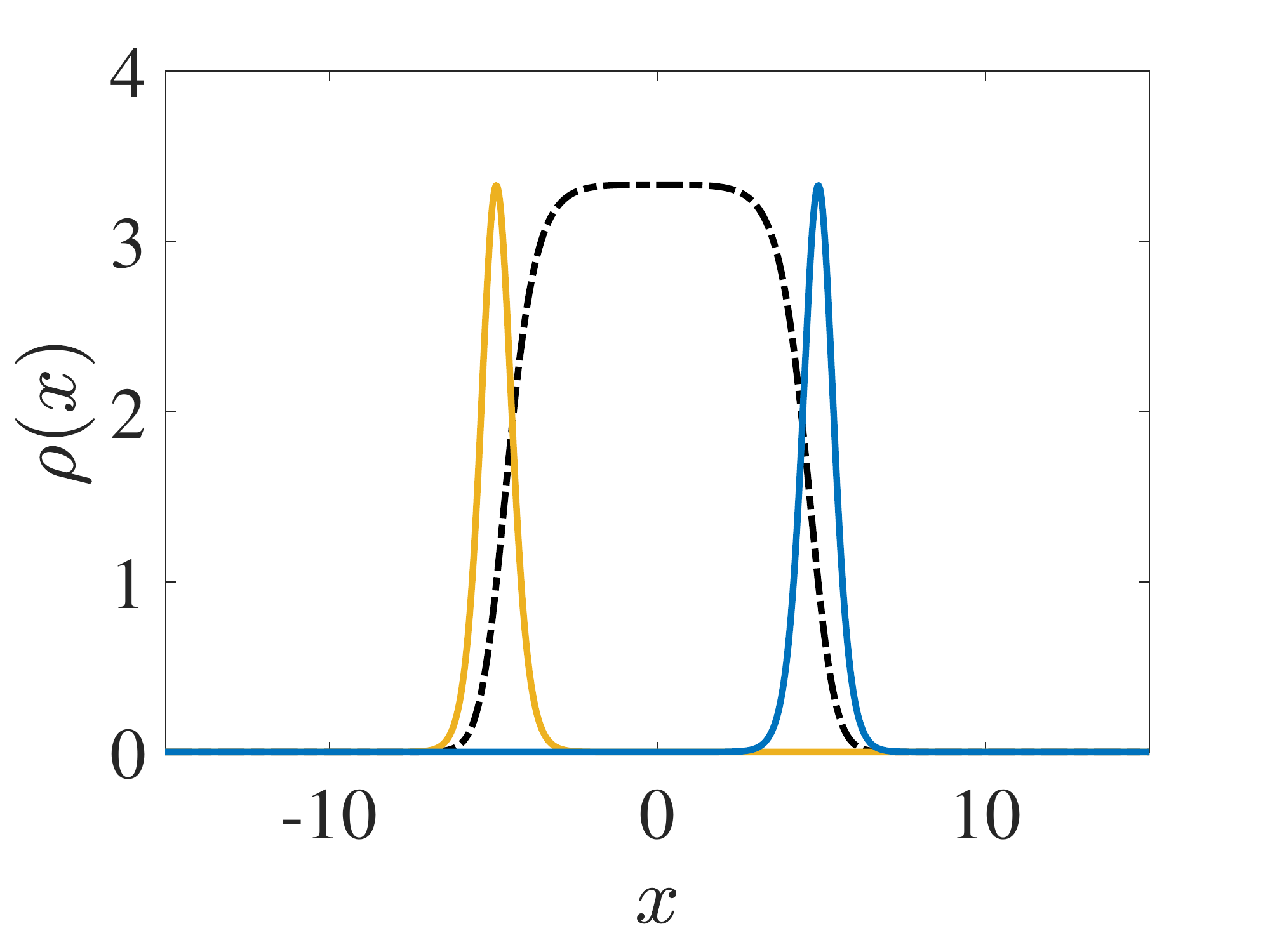}
\includegraphics[height=.16\textheight, angle =0]{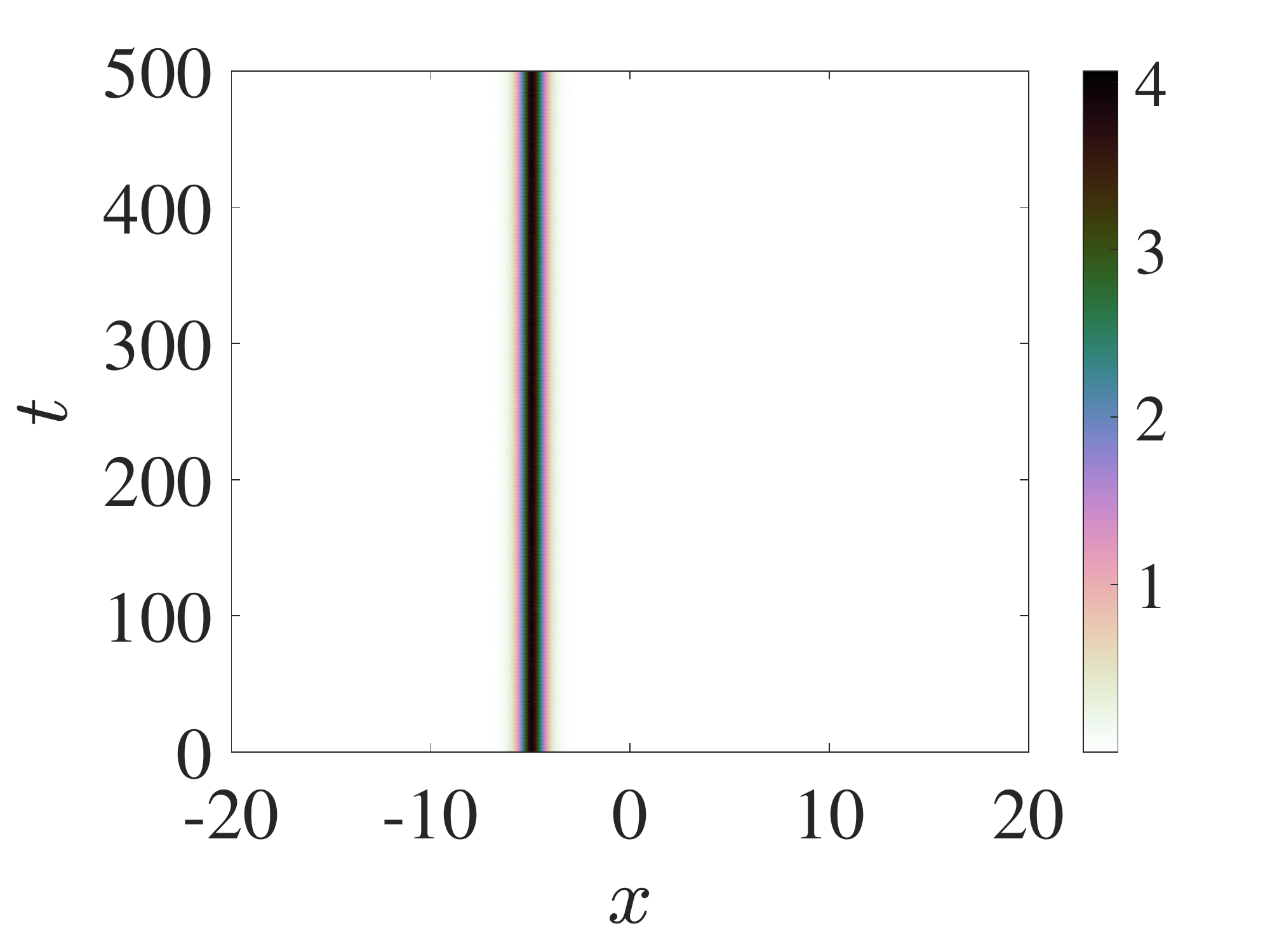}
\includegraphics[height=.16\textheight, angle =0]{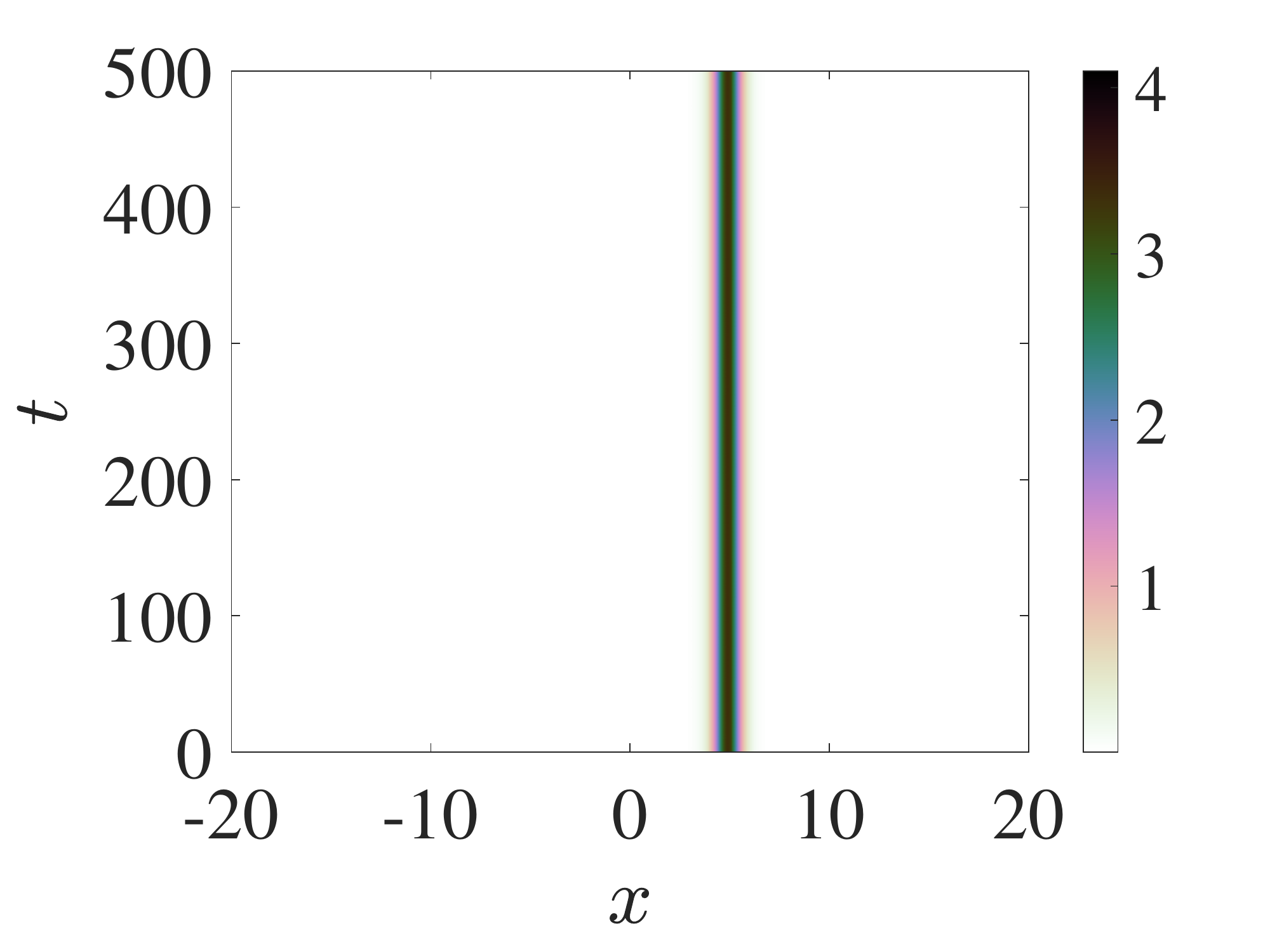}
\includegraphics[height=.16\textheight, angle =0]{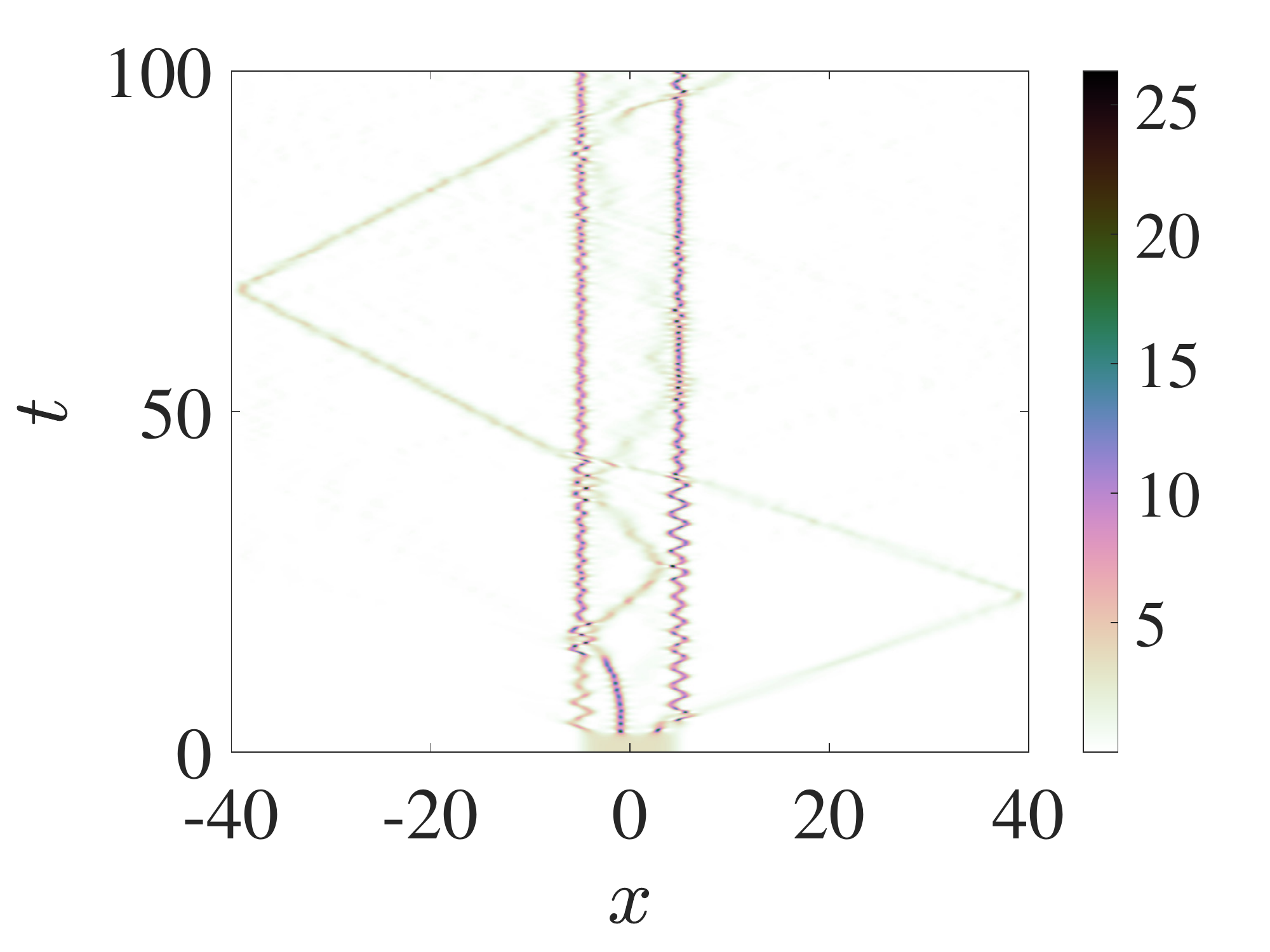}
\end{center}
\caption{\label{fig2}
\textit{Top panels}: Spectral stability analysis and existence results of bifurcating branches 
emanating from the kovaton solution in 1D with $g=-1$ (and $q=5$ as well as $\omega=-4$). The 
left and middle panels depict $\lambda_{i}$ and $\lambda_{r}$ as functions of $M$ (the same 
spectral picture is obtained for the other branch that has the same norm). Note that the bifurcating 
branch is spectrally stable due to the absence of real eigenvalues (see the middle panel). The 
right panel depicts spatial profiles of the density of the bifurcating branches for $M=30$. Note 
that the density of the kovaton solution for $M=30$ is plotted too in the panel with dashed-dotted 
black lines for comparison. \textit{Bottom panels}: Spatio-temporal evolution of densities $\rho(x)$
for the bifurcating branches is shown in the left and middle panels with $M=30$, as well as the kovaton
solution (for the same $M$) in the right panel. For the stable steady-states, we perturbed the initial
condition with a random perturbation (of $10^{-3}\times\mathrm{max}(|u^{(0)}|)$ amplitude) whereas for 
the unstable kovaton solution of the right panel, we perturbed the initial condition by considering 
the eigenvector corresponding to the most unstable eigendirection. 
}
\end{figure}

We begin our discussion on the numerical results by considering first the 1D kovaton 
solution and its spectra as a function of $M$. It should be noted that the parameter
$M$ that appears in the potential coincides with the actual mass (or $l_{2}$-norm) 
of the kovaton solution via Eq.~\eqref{e:1DMass}. The respective results on the stability 
are summarized in Fig.~\ref{fig2} which showcases the dependence of $\lambda_{i}$ and 
$\lambda_{r}$ on the (bifurcation parameter or) mass $M$ for the attractive case with
$g=-1$ (see, Fig.~\ref{fig2}(a)) and repulsive one with $g=1$ (see, Fig.~\ref{fig2}(b)). 
It can be discerned from panel (a), that the kovaton solution is spectrally stable from 
its inception (i.e., $M\ll 1$) to $M_{c}\approx 0.65$ whereupon the solution becomes (spectrally) 
unstable, and the growth rate of the instability increases with $M$. On the other hand, 
and for the repulsive case of $g=1$, the kovaton solutions are spectrally stable throughout 
the parameter interval in $M$ that we consider therein.

It is worth pointing out in Fig.~\ref{fig2}(a) that the emergence of the instability 
is due to the fact that a pair of imaginary eigenvalues cross the origin, and give 
birth to the unstable mode at $M_{c}\approx 0.653$. Moreover, this ``zero crossing'' 
of the pertinent eigenvalues signals the emergence of a pitchfork (or symmetry-breaking) 
bifurcation~\cite{jyang_2012} around that point in the parameter space. Although such 
bifurcations are important in their own right (in fact, and in the present setup, there 
exist more such bifurcations at $M\approx 2.528$, $5.475$, and $M\approx 9.18$), we do
not pursue them all. Such bifurcating branches can be obtained by using Newton's method 
where the solver is fed by the steady-state kovaton solution at the value of $M$ (where
such a zero crossing happens) perturbed by the eigenvector corresponding to that unstable
eigendirection.

Illustratively, we briefly discuss the emergence of two ``daughter'' branches of solutions
at $M_{c}\approx 0.653$, i.e., at the point where the ``parent'' kovaton solution branch
undergoes a symmetry-breaking bifurcation. Indeed, in the top row of Fig.~\ref{fig2}, we
present our results on this bifurcation. In particular, the top left and middle panels
showcase the $\lambda_{i}$ and $\lambda_{r}$ both as functions of $M$ of the bifurcating
branch (the other one has exactly the same spectrum), and the (top) right panel presents
the spatial distribution of the densities, i.e., $\rho(x)$ of two profiles at $M=30$. In
addition, the density of the kovaton solution (emanating from the parent branch)
for the same value of the bifurcation parameter $M$ is included too in the figure, and shown 
with dashed-dotted black lines for comparison. It can be discerned from the middle panel of 
the figure that the daughter branches are spectrally stable all along, i.e., over the parameter
window in $M$ that we considered therein). At the bifurcation point $M_{c}\approx 0.653$, the 
daughter branch ``inherits'' the stability of the parent branch whereas the latter becomes 
(spectrally) unstable past that point, i.e., pitchfork bifurcation. From the top right panel 
of the figure, we further note that the bifurcating solutions resemble solitary yet \textit{shifted} 
pulses.

In the bottom panels of Fig.~\ref{fig2} we corroborate our stability analysis results
by performing time evolution of perturbed steady-states. In particular, the bottom left and 
middle panels of Fig.~\ref{fig2} depict the spatio-temporal evolution of the density $\rho(x)$ 
for the stable bifurcating solutions of the top right panels of Fig.~\ref{fig2}. We added a random 
perturbation with a strong amplitude of $10^{-3}\times\mathrm{max}(|u^{(0)}|)$ to the localized 
pulse. It can be discerned from these two panels that the bifurcating branches are indeed stable 
solutions. On the other hand, the kovaton solution, i.e., the parent branch, is spectrally unstable, 
whose dynamics is shown in the bottom right panel of the figure. We initialized the dynamics therein
by perturbing the steady-state solution with the eigenvector corresponding to the most unstable
eigendirection (essentially, utilizing Eq.~\eqref{pertr_ansatz} for $t=0$ with $\varepsilon$ being 
$10^{-3}\times\mathrm{max}(|u^{(0)}|)$). This way, we feed the instability of the pertinent solution.
It can be discerned from that panel that the solution oscillates in the presence of the potential 
while simultaneously interpolating between the two stable (bifurcating) solutions of the top right 
panel of the figure.

\onecolumngrid\
\twocolumngrid\

\begin{figure}[pt!]
\begin{center}
\begin{overpic}[height=.16\textheight, angle =0]{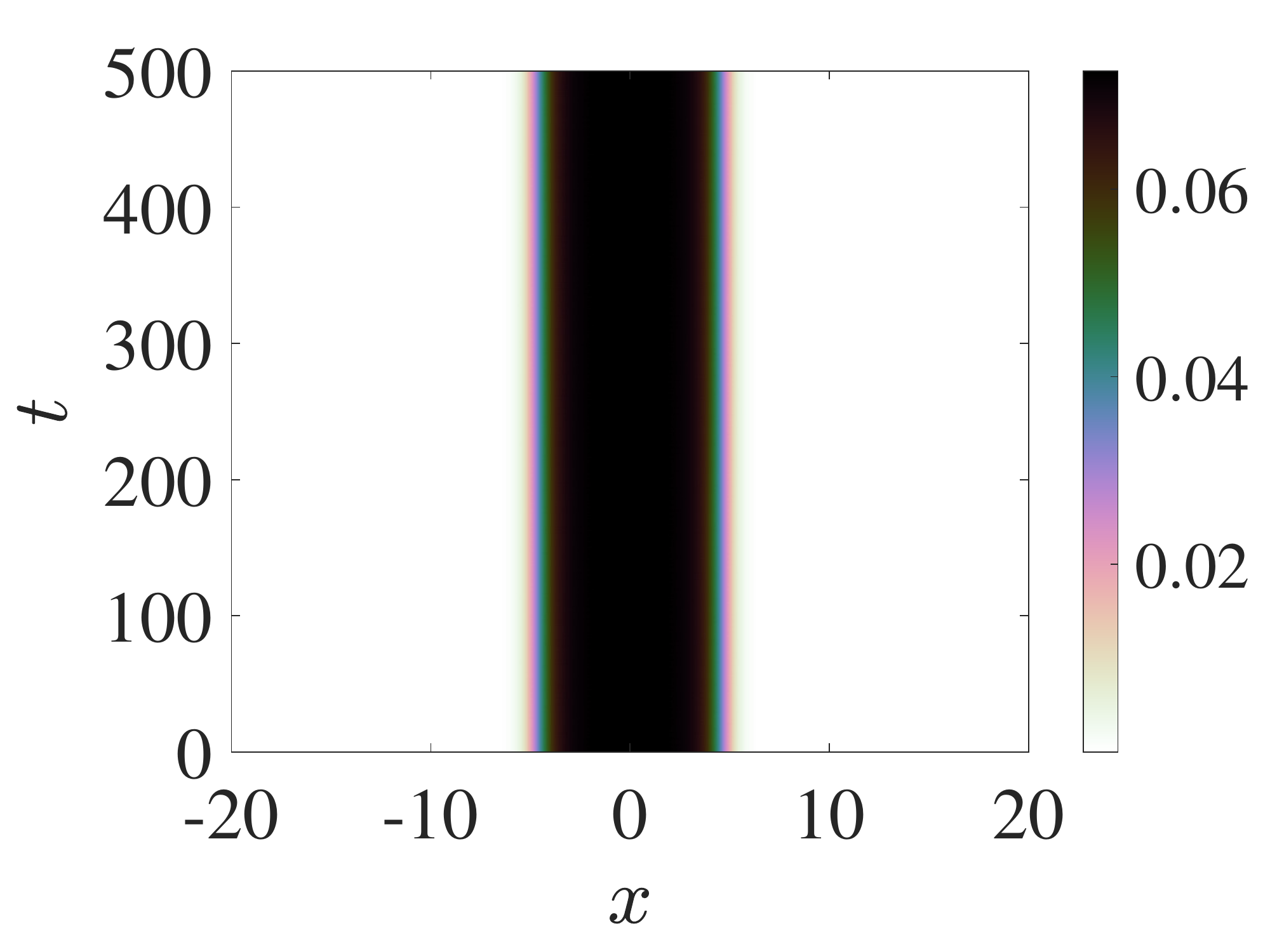}
\put(20,63){$(a)$}
\end{overpic}
\begin{overpic}[height=.16\textheight, angle =0]{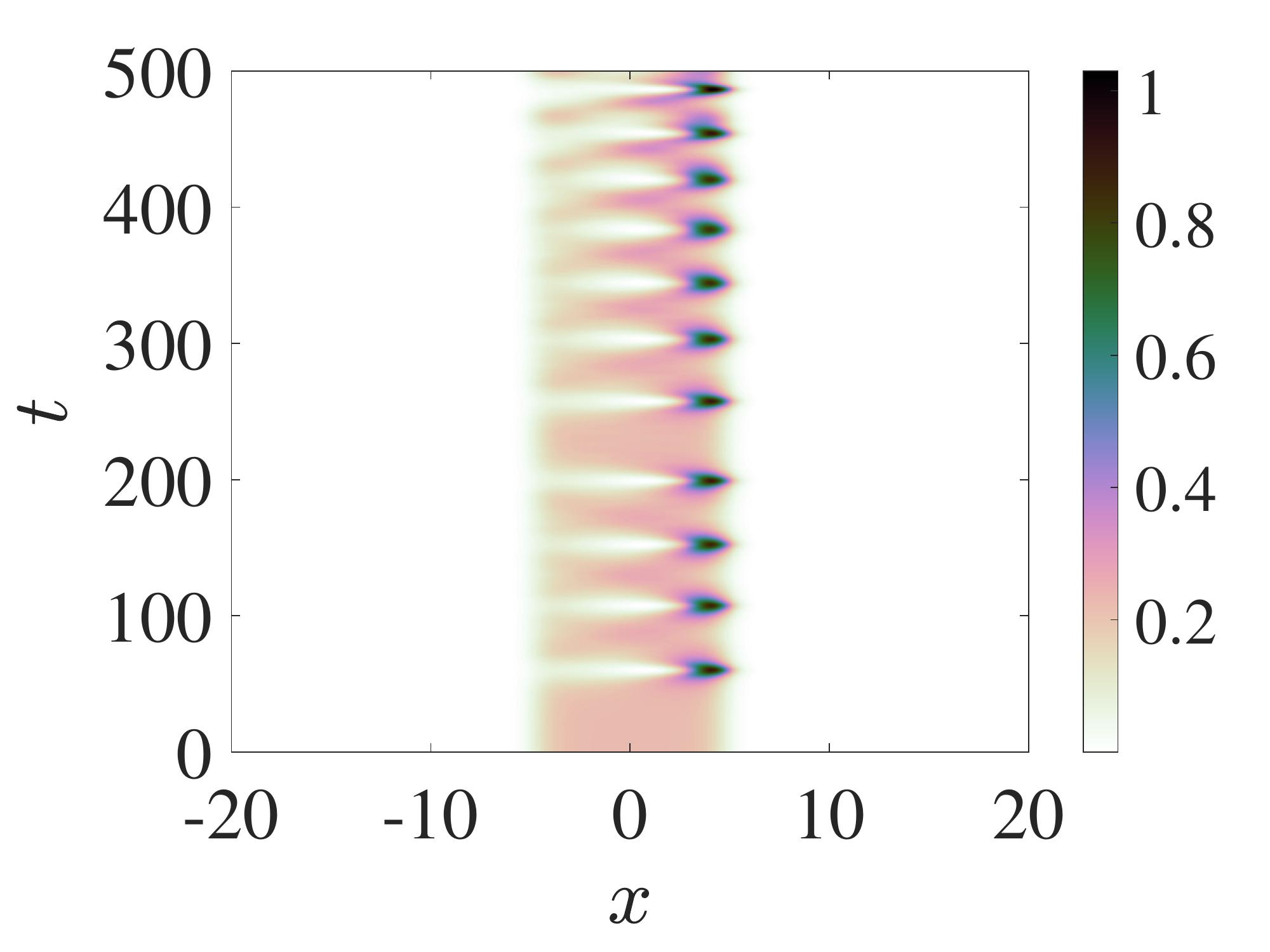}
\put(20,63){$(b)$}
\end{overpic}
\begin{overpic}[height=.16\textheight, angle =0]{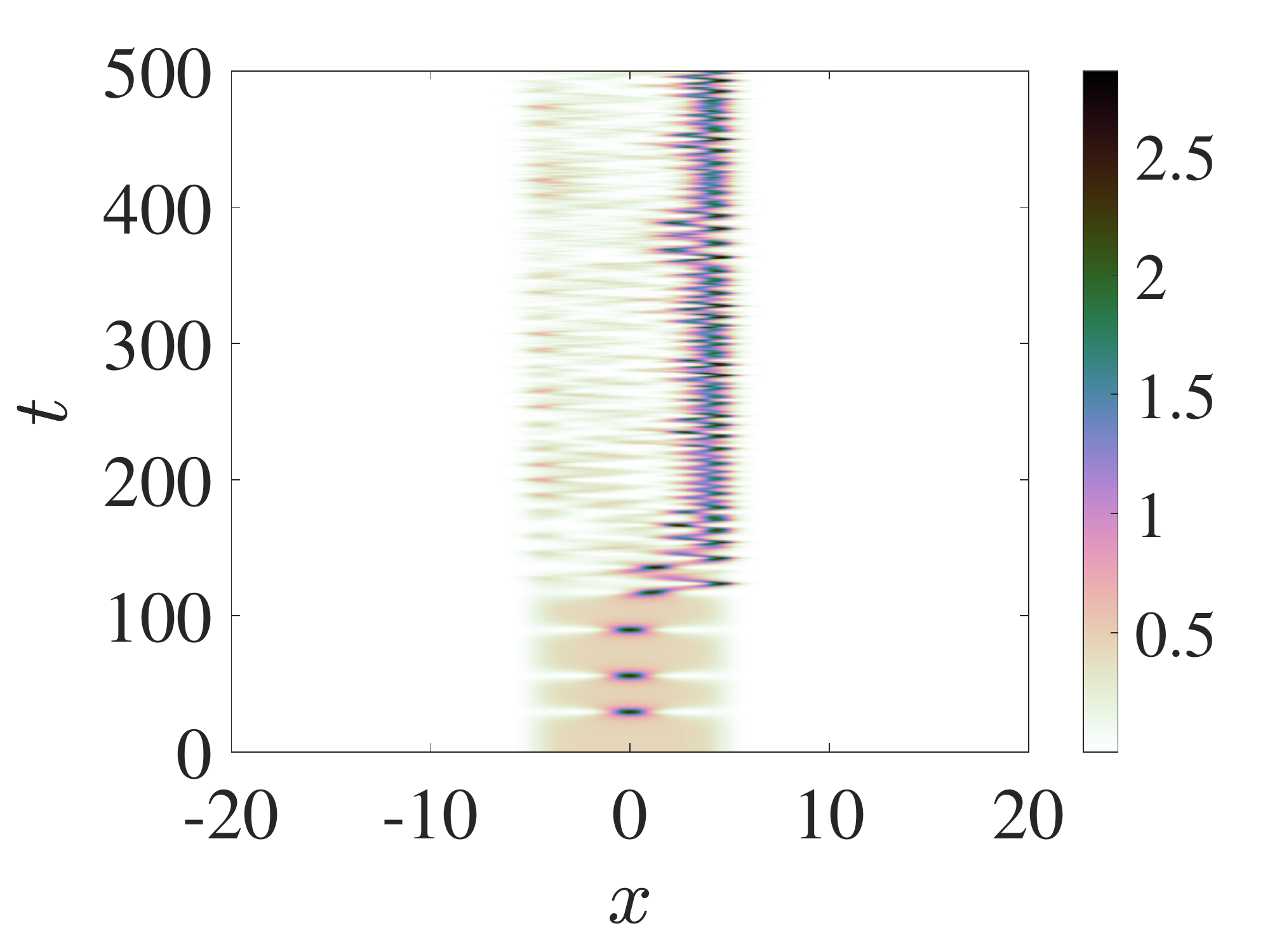}
\put(20,63){$(c)$}
\end{overpic}
\end{center}
\caption{
Spatio-temporal evolution of the density $\rho(x)$ for a perturbed kovaton solution 
for (a) $M=0.65$, (b) $M=2$, and (c) $M=4$, with $g=-1$, $q=5$, and $\omega=-4$. 
For the stable steady-state of panel (a), a random perturbation with amplitude 
$10^{-3}\times\mathrm{max}(|u^{(0)}|)$ was added to the localized pulse whereas 
for the unstable states of panels (b)-(d), the initial condition was perturbed by 
the most unstable eigendirection (and with the same amplitude for the pertinent 
cases).
}
\label{fig3}
\end{figure}

We now move to Fig.~\ref{fig3} which corroborates further our stability analysis 
results for the kovaton solutions themselves by presenting the spatio-temporal 
evolution of the density $\rho(x)$ for a perturbed kovaton solution with (a) $M=0.65$, 
(b) $M=2$, and (c) $M=4$, respectively. Based on Fig.~\ref{fig1}(a), the kovaton 
solution for $M=0.65$ is deemed spectrally stable, and its perturbed dynamics (upon 
adding a random perturbation to the localized region of the kovaton) is shown in 
Fig.~\ref{fig3}(a). It can be clearly discerned from the figure that the kovaton 
solution is dynamically stable. On the other hand, and for panels (b)-(c), the kovaton 
solutions are unstable for $M=2$ and $M=4$ (see, Fig.~\ref{fig1}(a)). We investigate this 
finding dynamically in these panels by furnishing an initial condition corresponding 
to the stationary kovaton solution plus a perturbation added on top of the localized
region of the pulse (as we did before in the bottom right panel of Fig.~\ref{fig2}). 
In Fig.~\ref{fig3}(b), we observe that after a short time interval, the kovaton 
solution starts oscillating in the confining potential featuring a beating pattern 
whose temporal period decreases as time passes by, thus effectively approaching the 
stationary yet stable solitary pulse shown in the top right panel of Fig.~\ref{fig2} 
(see the one depicted with solid blue line). This is not surprising due to the fact 
that the branch associated with this pulse is spectrally stable, thus creating a basin 
of attraction in the dynamics. This is also evident in Fig.~\ref{fig3}(c). Indeed, 
after a transient period of time, featuring a solitary pulse mounted on top of a kovaton 
solution, these oscillations have a progressively smaller period, and the dynamics start 
approaching the stationary state of the top right panel of Fig.~\ref{fig2}. 

We finalize our discussion on the 1D GPE by briefly reporting the stability of kovaton 
solutions with $g=1$, i.e., the repulsive case. We performed dynamical simulations of perturbed 
kovaton solutions in that case, and we corroborated the stability results of Fig.~\ref{fig1}(b)
(the results on the dynamics are not shown). Having finalized a detailed exposure on the 
existence, stability (and bifurcations), as well as dynamics for the 1D GPE, we move now
to the 2D GPE case next.

%
\subsubsection{Numerical Results for the 2D GPE}

\begin{figure}[pt!]
\begin{center}
\begin{overpic}[height=.16\textheight, angle =0]{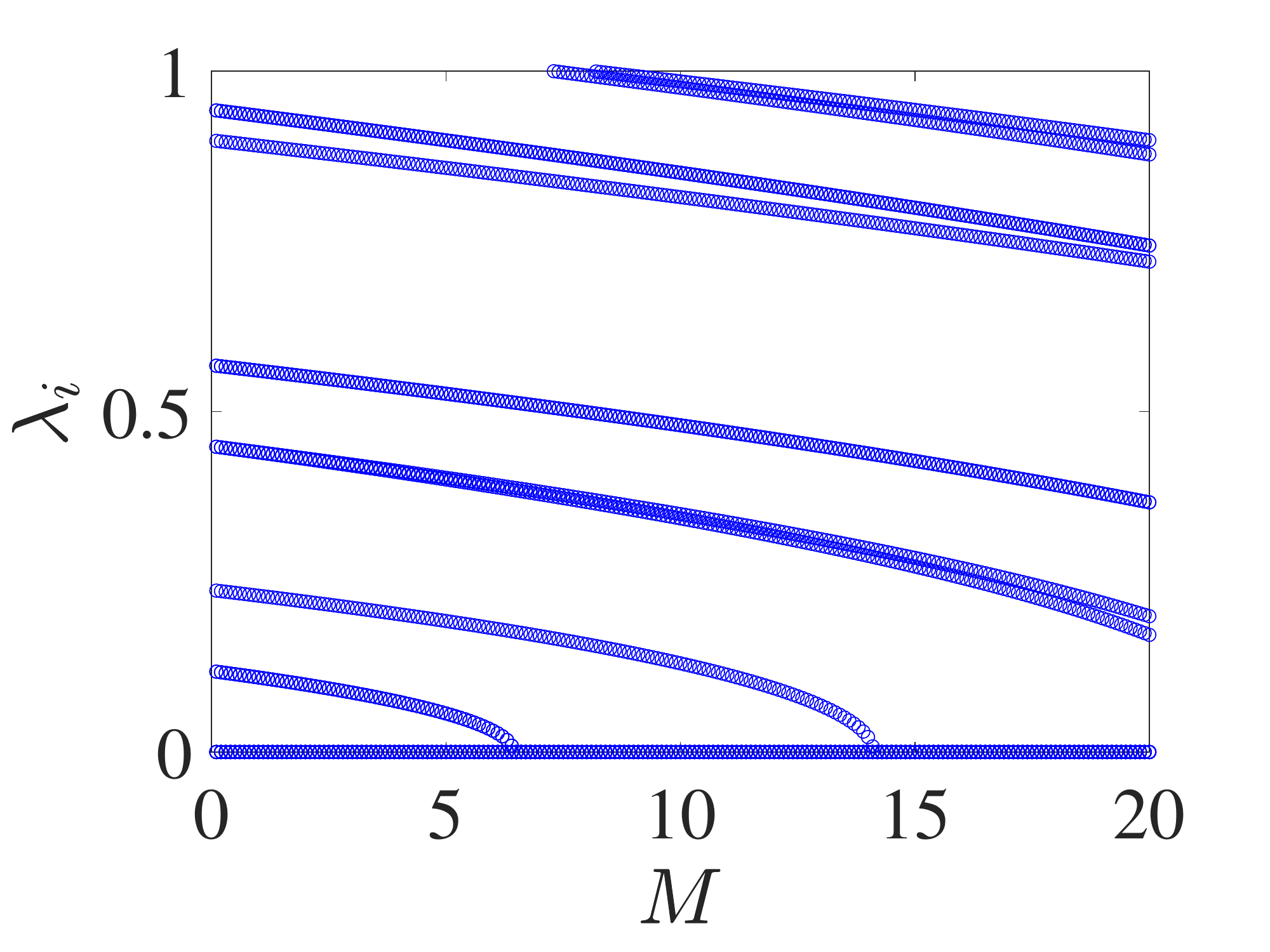}
\put(20,57){$(a)$}
\end{overpic}
\includegraphics[height=.16\textheight, angle =0]{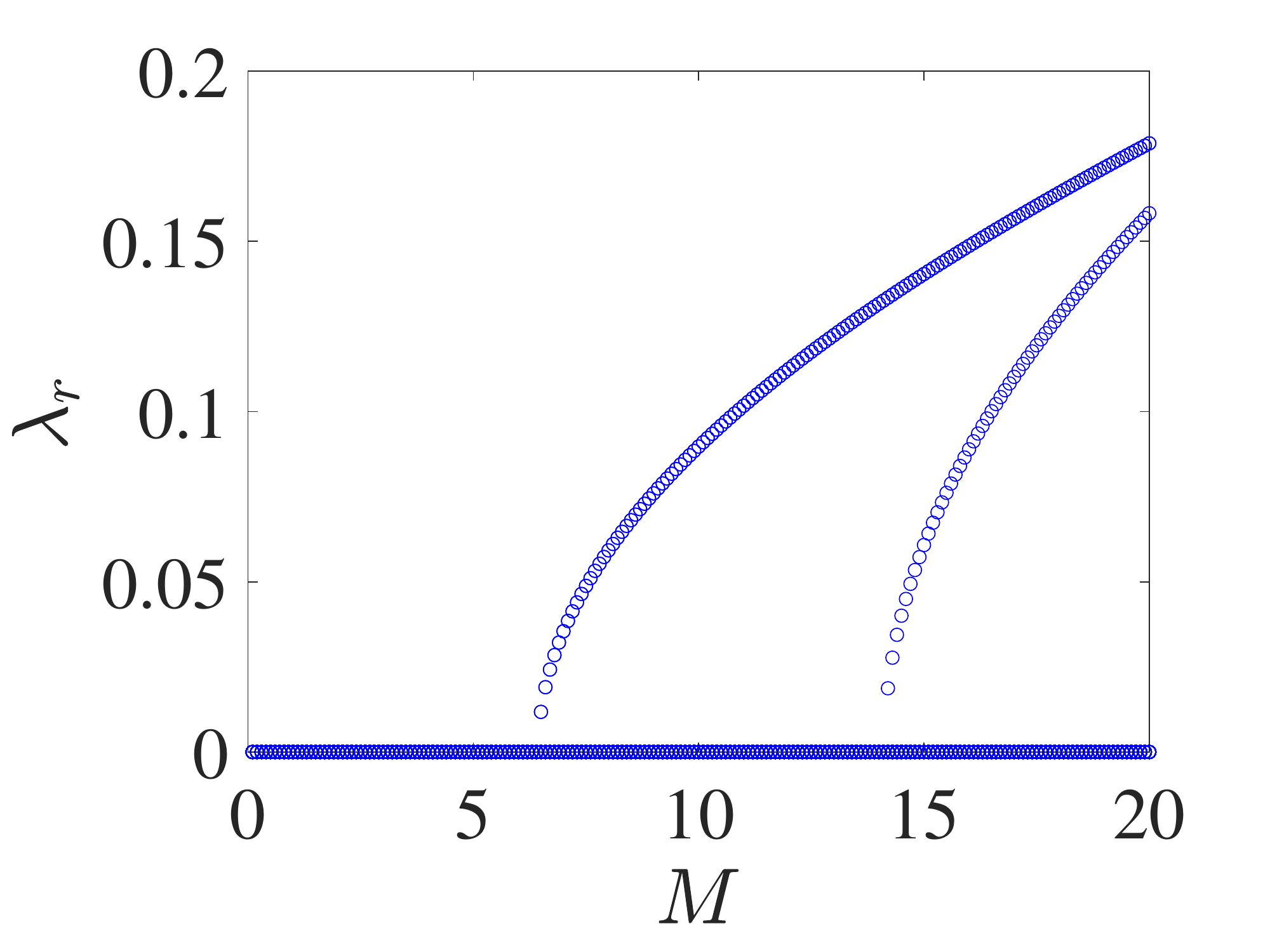}\\
\begin{overpic}[height=.16\textheight, angle =0]{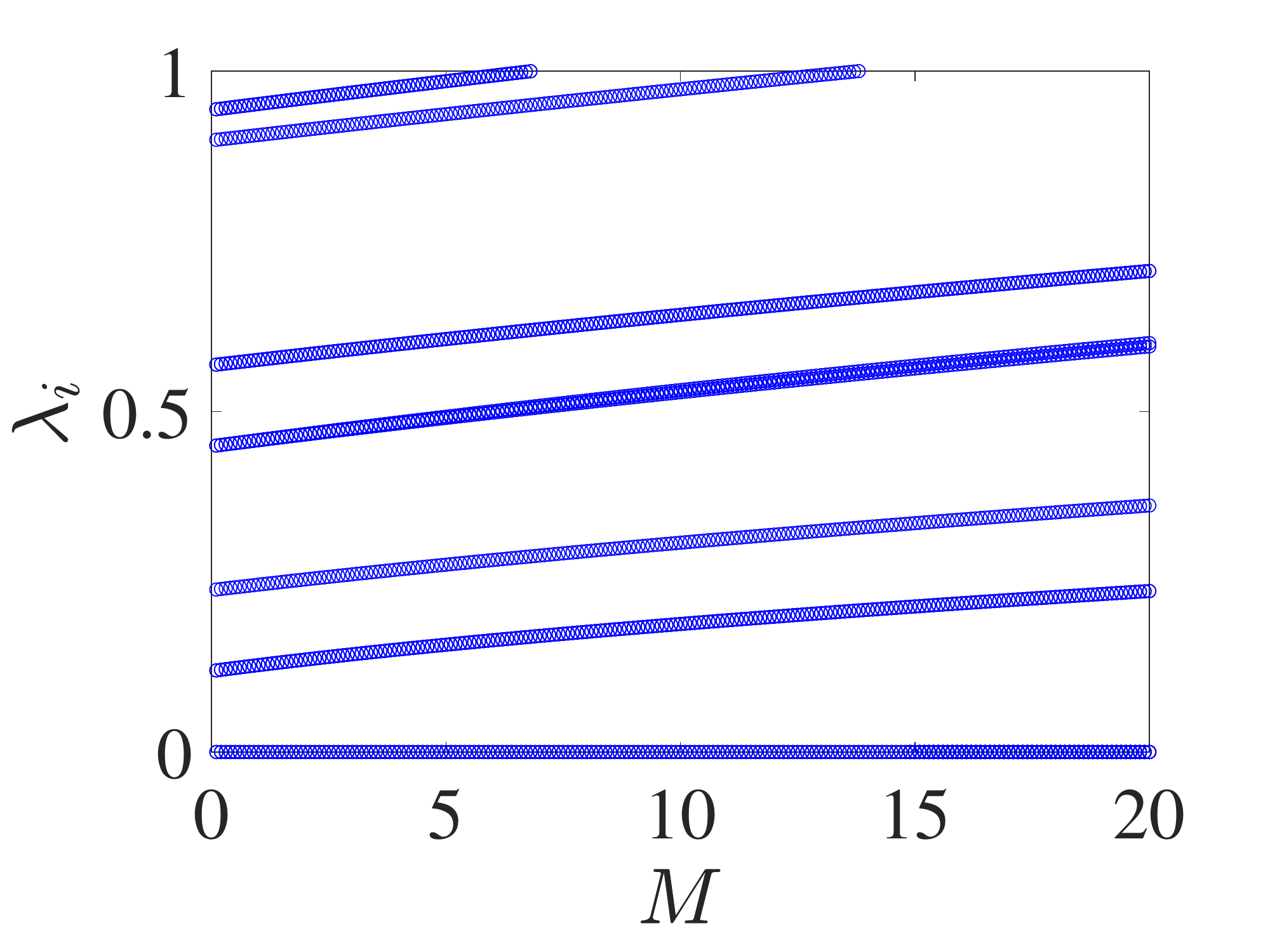}
\put(20,57){$(b)$}
\end{overpic}
\includegraphics[height=.16\textheight, angle =0]{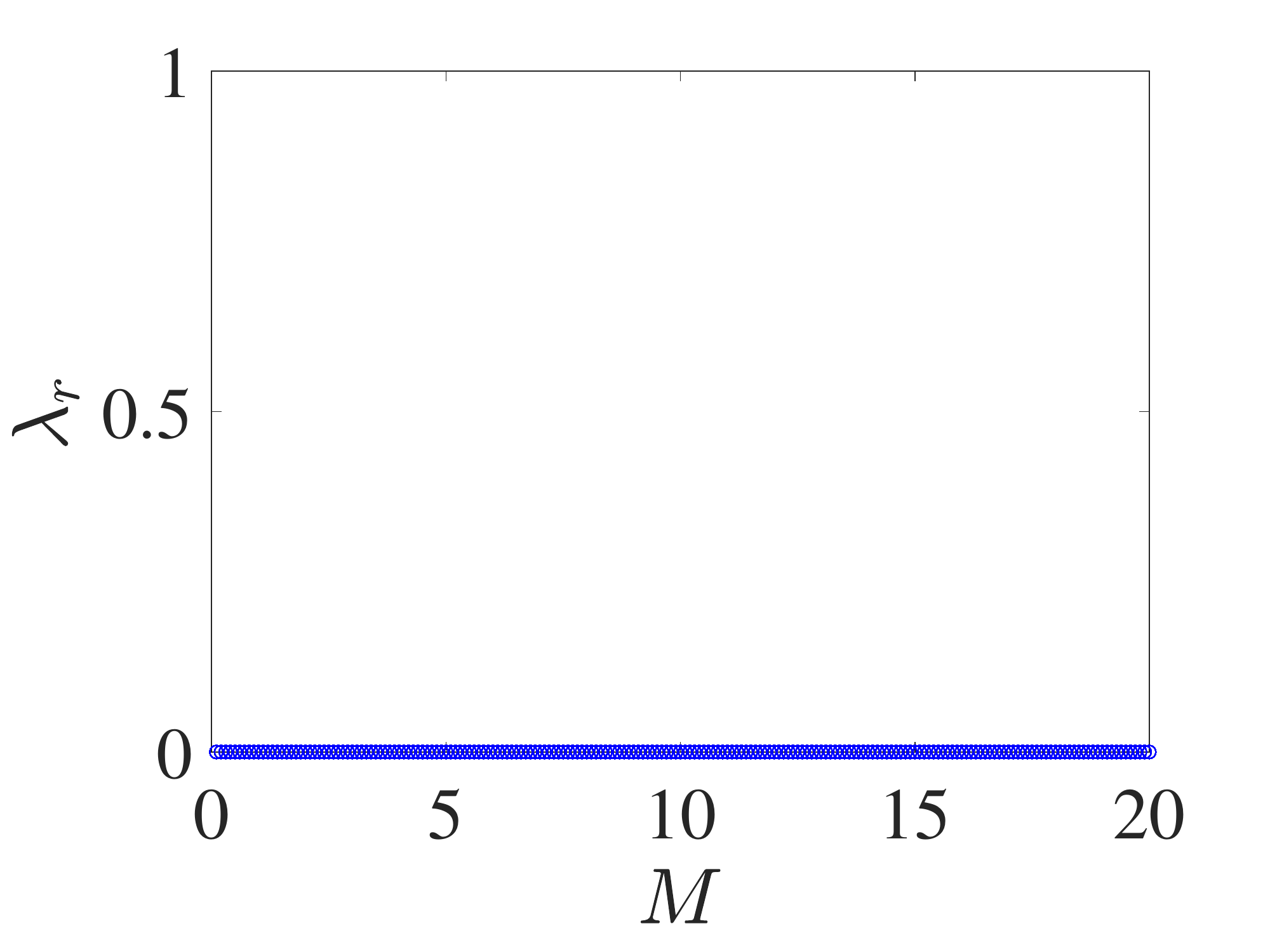}
\end{center}
\caption{
Spectral stability analysis of 2D square kovaton solutions for (a) $g=-1$ (attractive) 
and (b) $g=1$ (repulsive). The format of the figure is the same as of Fig.~\ref{fig1}.
The parameter values here are $\omega=-8$ and $q=5$.
}
\label{fig4}
\end{figure}

\begin{figure}[pt!]
\begin{center}
\begin{overpic}[height=.16\textheight, angle =0]{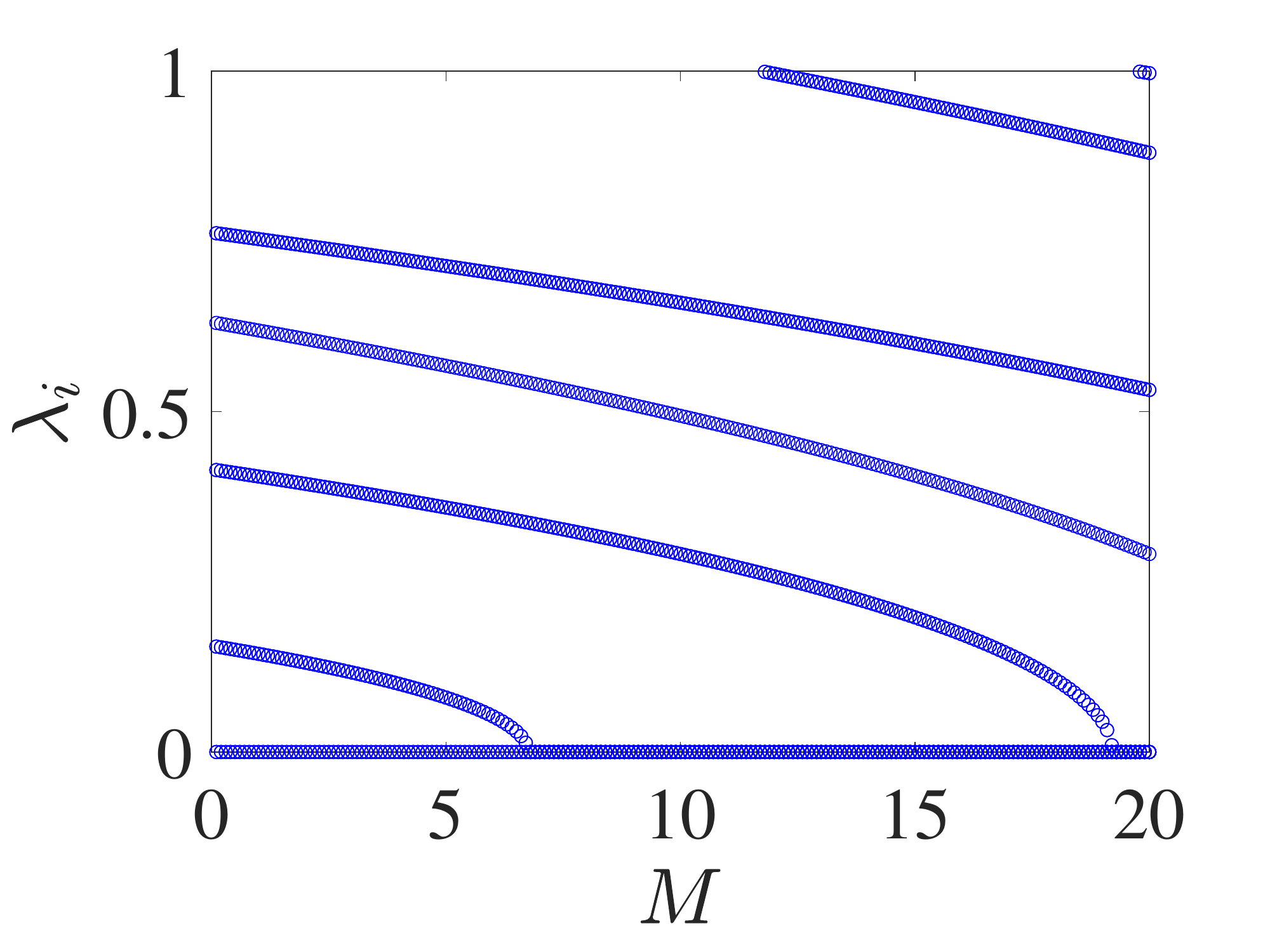}
\put(20,63){$(a)$}
\end{overpic}
\includegraphics[height=.16\textheight, angle =0]{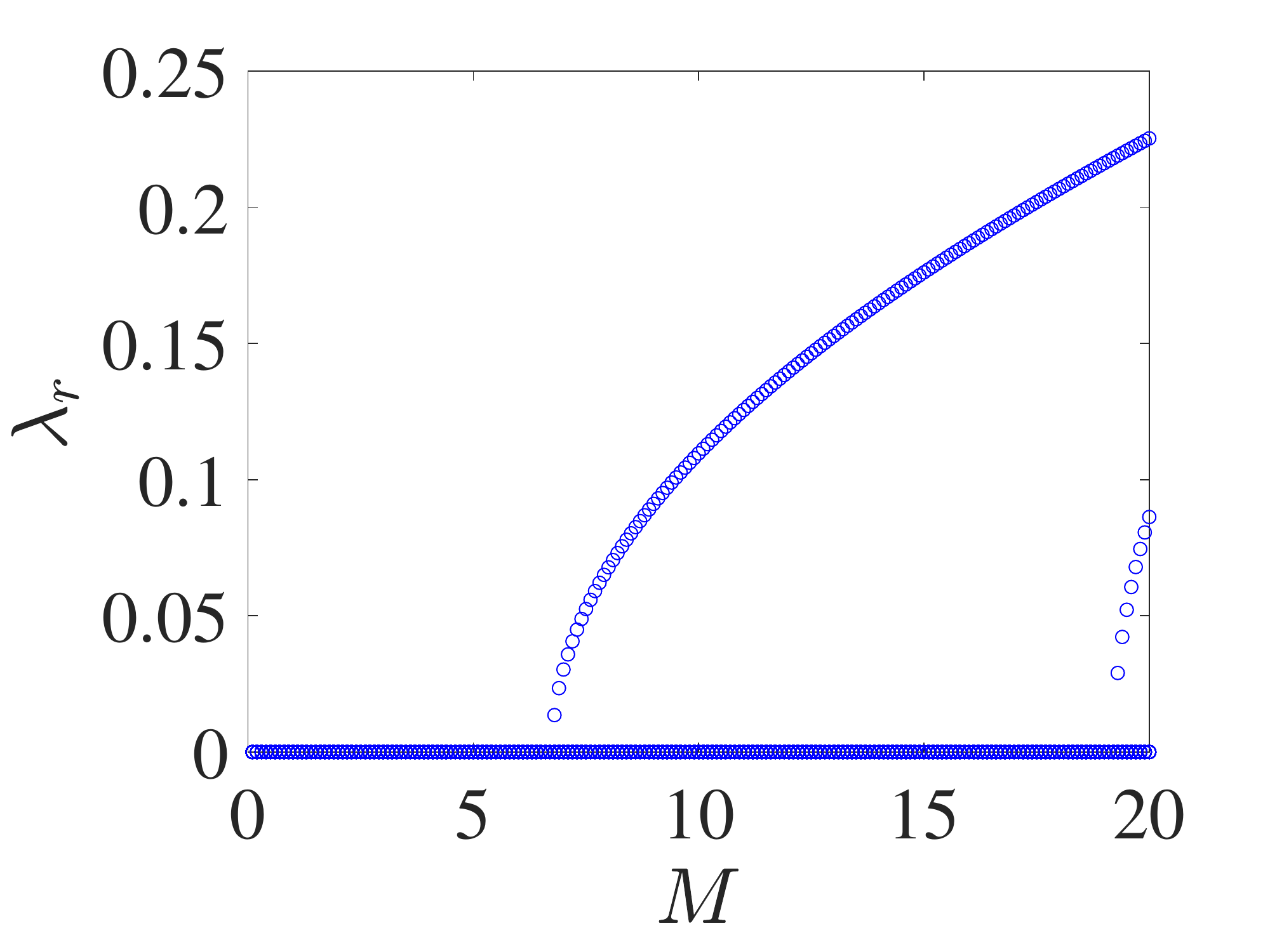}\\
\begin{overpic}[height=.16\textheight, angle =0]{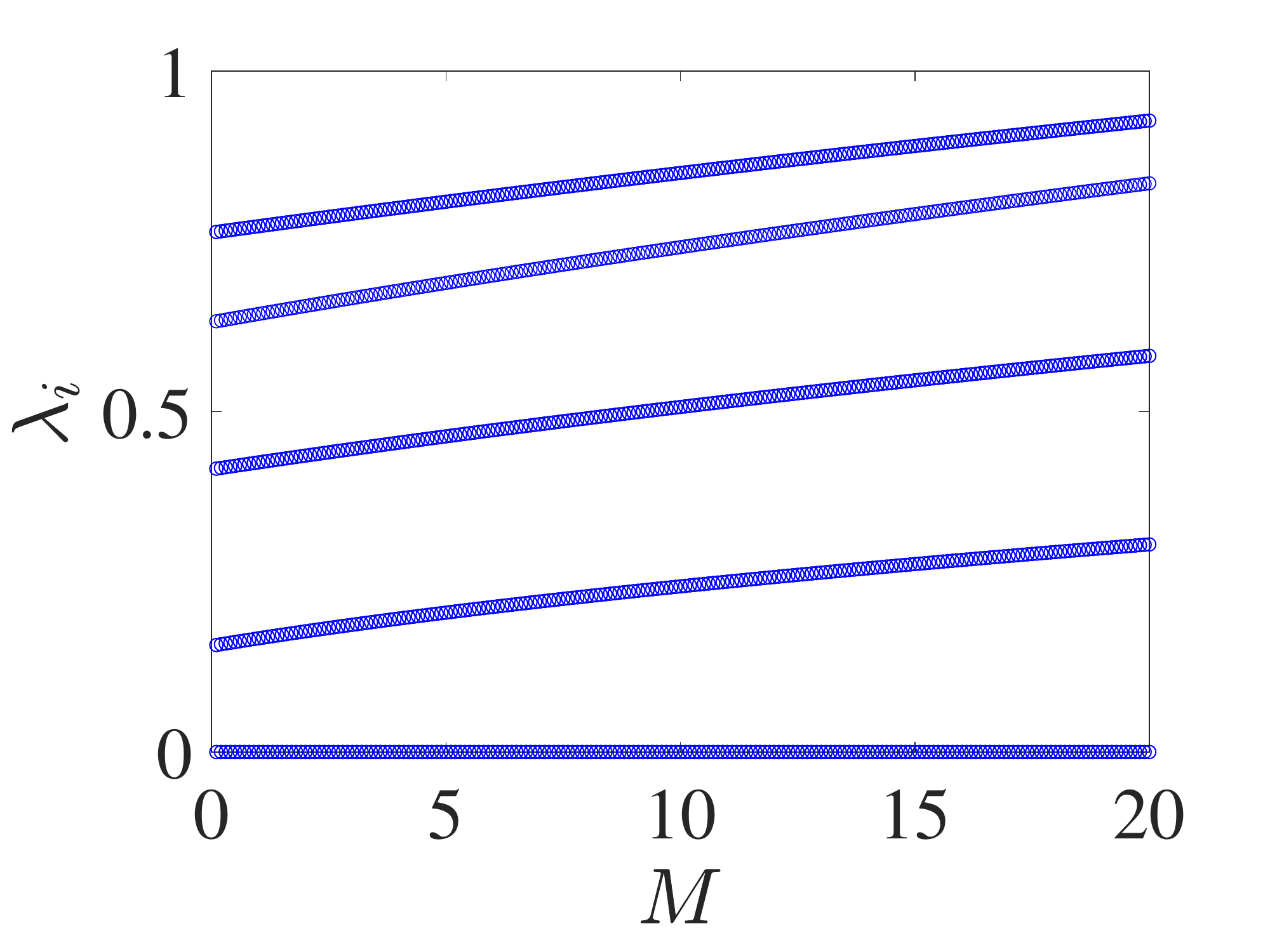}
\put(20,63){$(b)$}
\end{overpic}
\includegraphics[height=.16\textheight, angle =0]{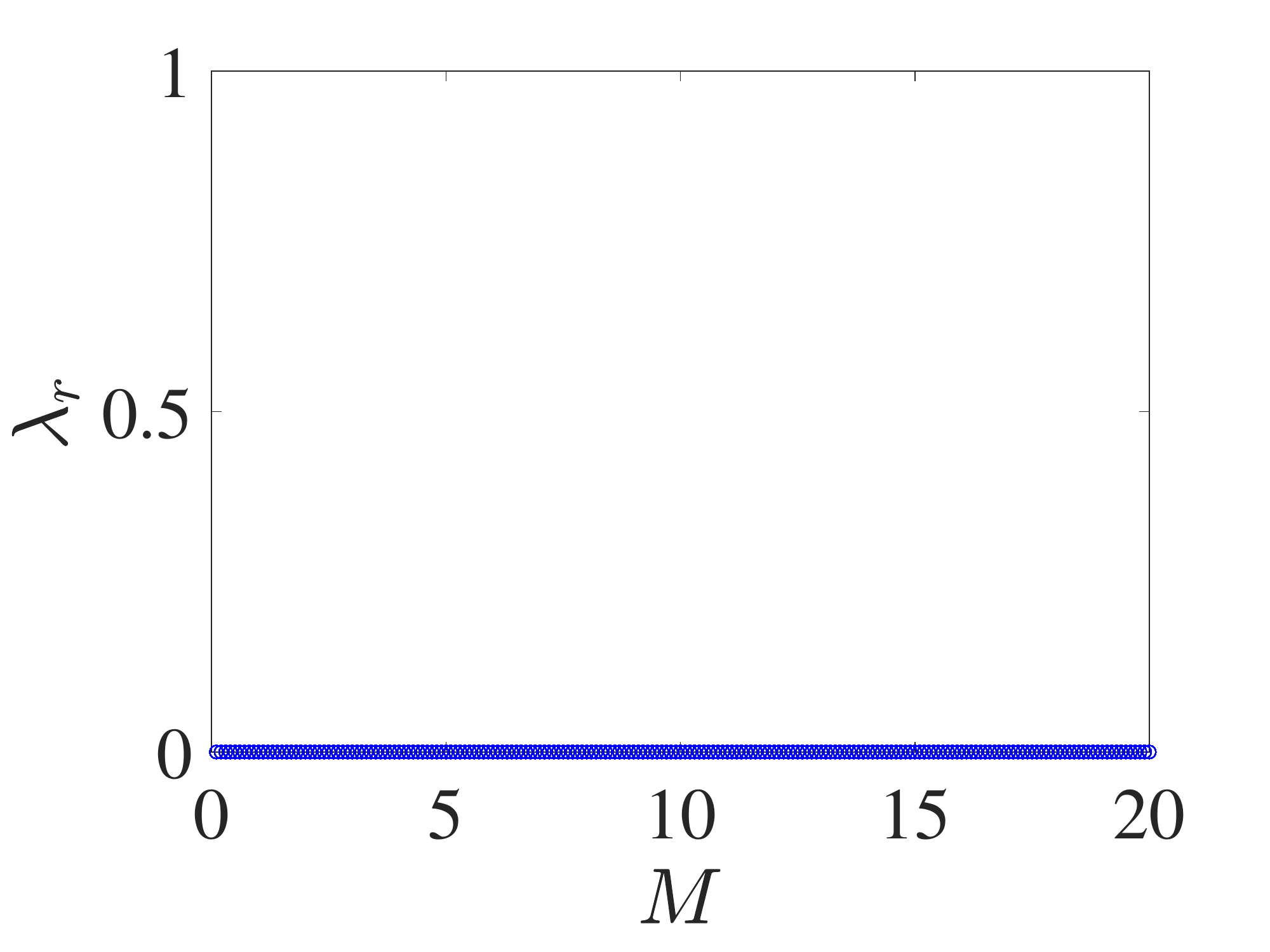}
\end{center}
\caption{
Same as Fig.~\ref{fig4} but for the 2D radial kovaton solutions with (a) $g=-1$ (attractive) 
and (b) $g=1$ (repulsive). The format of the figure is the same as of Fig.~\ref{fig1}.
The parameter values here are $\omega=-4$ and $q=5$.
}
\label{fig5}
\end{figure}

Similar to the 1D case, we present in Figs.~\ref{fig4} and~\ref{fig5}
our spectral stability analysis results for the 2D square and radial kovaton 
solutions, respectively, that emanate from the solution of the eigenvalue 
problem of Eq.~\eqref{stab_problem}. We consider both the attractive case 
with $g=-1$ (see panels (a) in the figures) and the repulsive case with $g=1$ 
(see panels (b) in the figures), where we set $q=5$ for both cases, and 
$\omega=-8$ and $\omega=-4$ for the square and radial kovaton cases, 
respectively. It can be discerned from Fig.~\ref{fig4}(a) that the square
kovaton solution with $g=-1$ is spectrally stable from its inception until
$M_{c}\approx 6.5$. At that value of $M$, we notice a zero crossing of a pair 
of eigenvalues that give birth to an unstable mode whose growth rate increases
with $M$ (see, the top right panel of the figure). Similar to the 1D case, this
signals the fact that the parent square kovaton branch undergoes a pitchfork 
bifurcation at that point although we do not pursue them here. In addition, 
a secondary unstable mode emerges at $M\approx 14.2$ from the same mechanism,
i.e., a zero crossing of a pair of eigenvalues (see, also the top left panel
in the figure). On the other hand, and for the repulsive case, i.e., $g=1$,
the square kovaton solutions are deemed stable over the parameter interval 
in $M$ we considered herein. This is clearly evident in Fig.~\ref{fig4}(b)
(see, in particular, the right panel showcasing $\lambda_{r}$ as a function 
of $M$). A similar result is obtained for the radial kovaton, and is shown
in Fig.~\ref{fig5} where in panels (a) and (b) we present our spectral stability
analysis results for $g=-1$ and $g=1$, respectively. The 2D radial kovaton 
solution with $g=-1$ is stable from its inception and becomes unstable at 
$M_{c}\approx 6.8$, i.e., slightly above the square case. This instability emerges
again from a zero crossing of a pair of eigenvalues (see the left panel therein).
The secondary unstable mode appears at a larger value of $M$ (in contrast to the 
square case), and in particular at $M\approx 19.3$. For the repulsive case of
$g=1$, the 2D radial kovaton is spectrally stable over the interval in $M$ that
we consider in the figure.

\begin{figure}[pt!]
\begin{center}
\includegraphics[height=.16\textheight, angle =0]{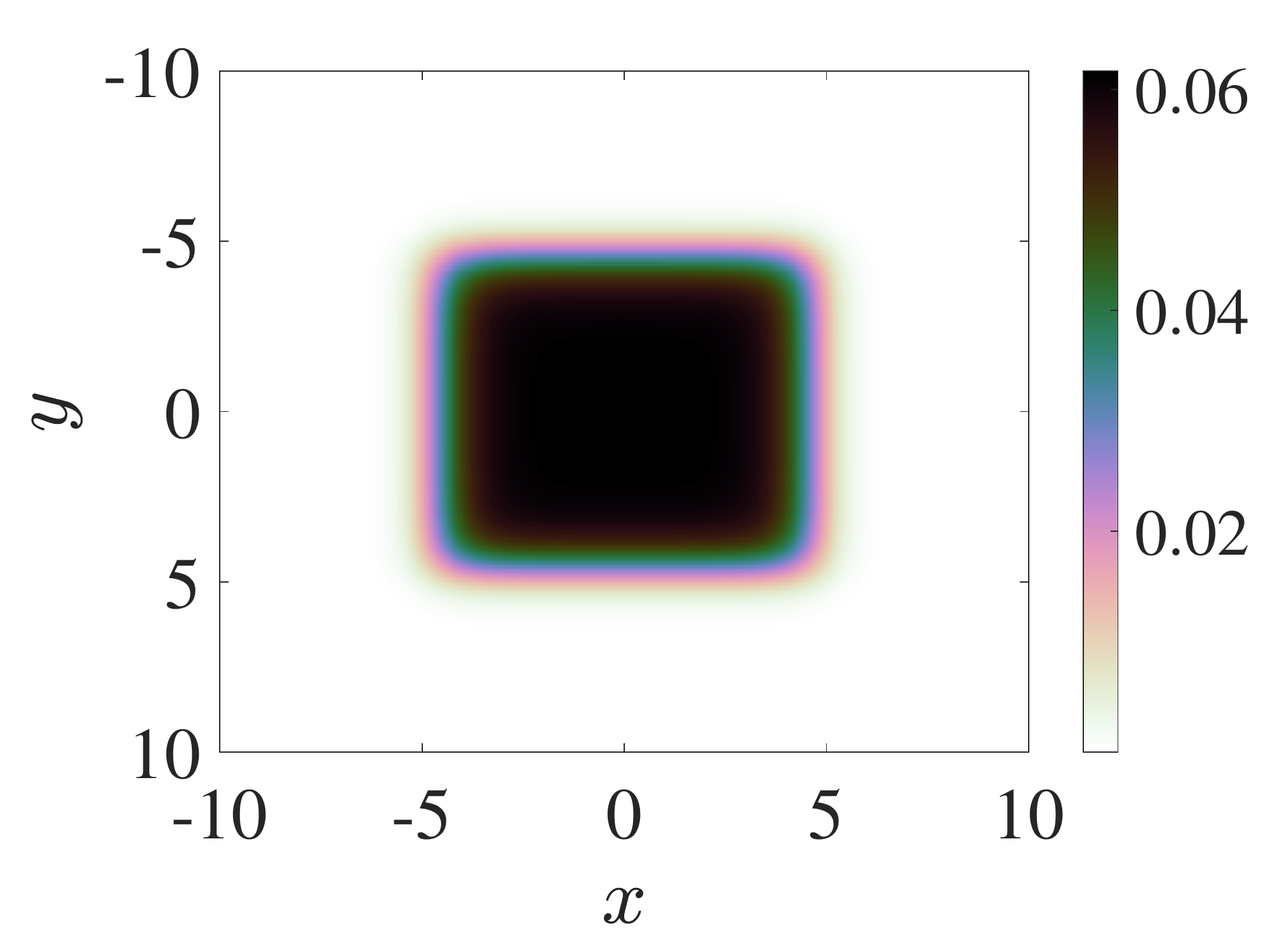}
\includegraphics[height=.16\textheight, angle =0]{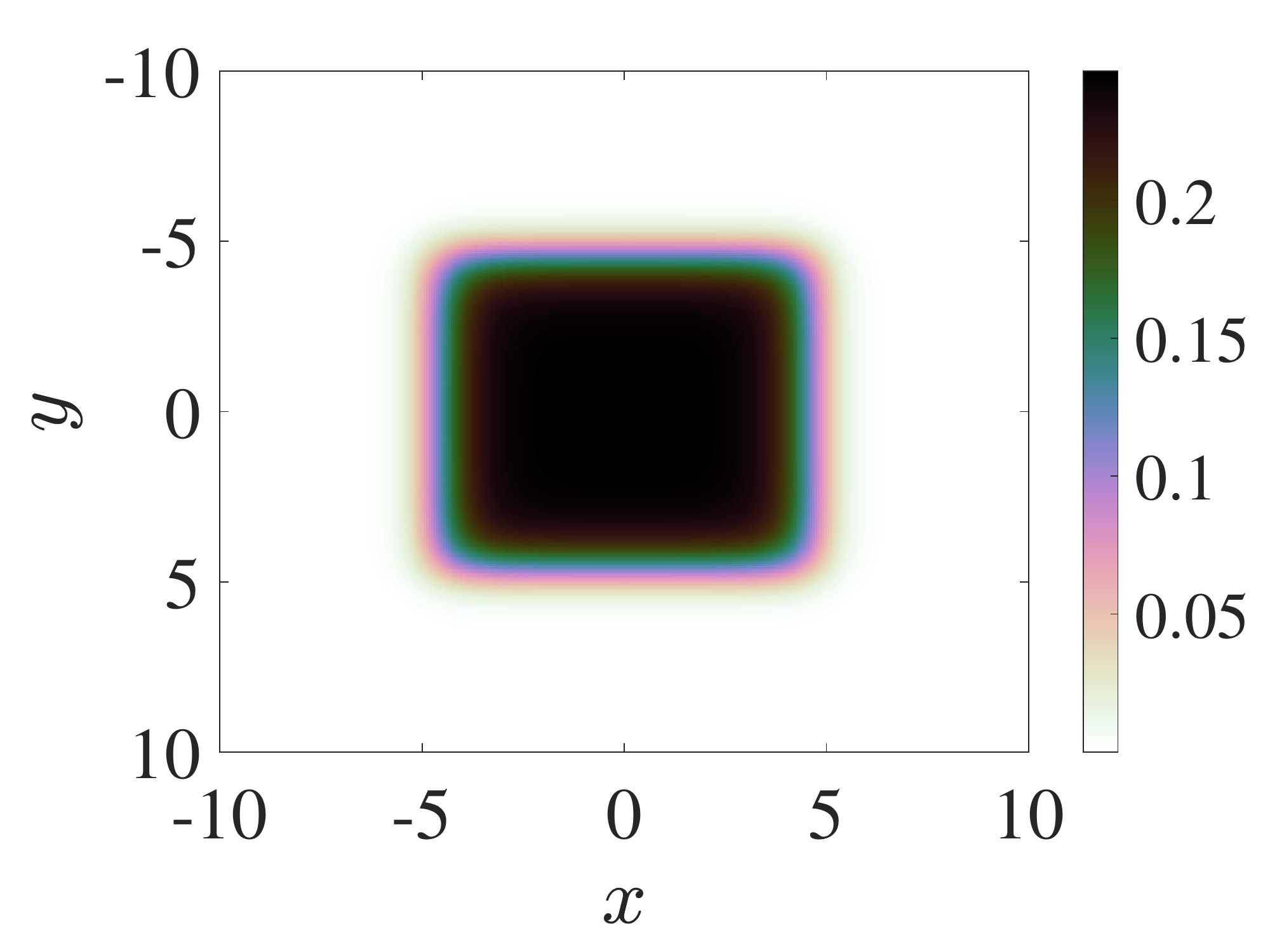}\\
\includegraphics[height=.16\textheight, angle =0]{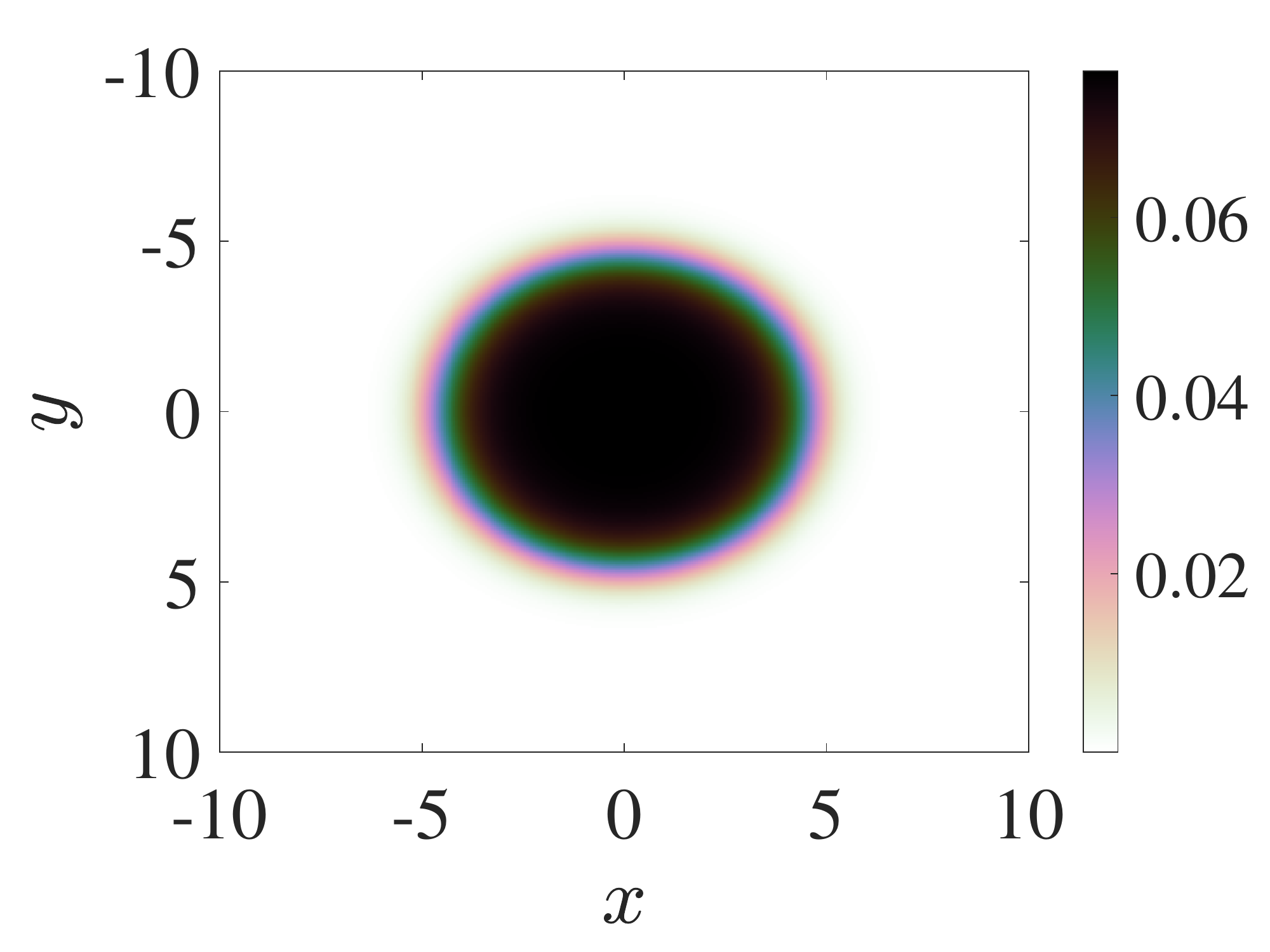}
\includegraphics[height=.16\textheight, angle =0]{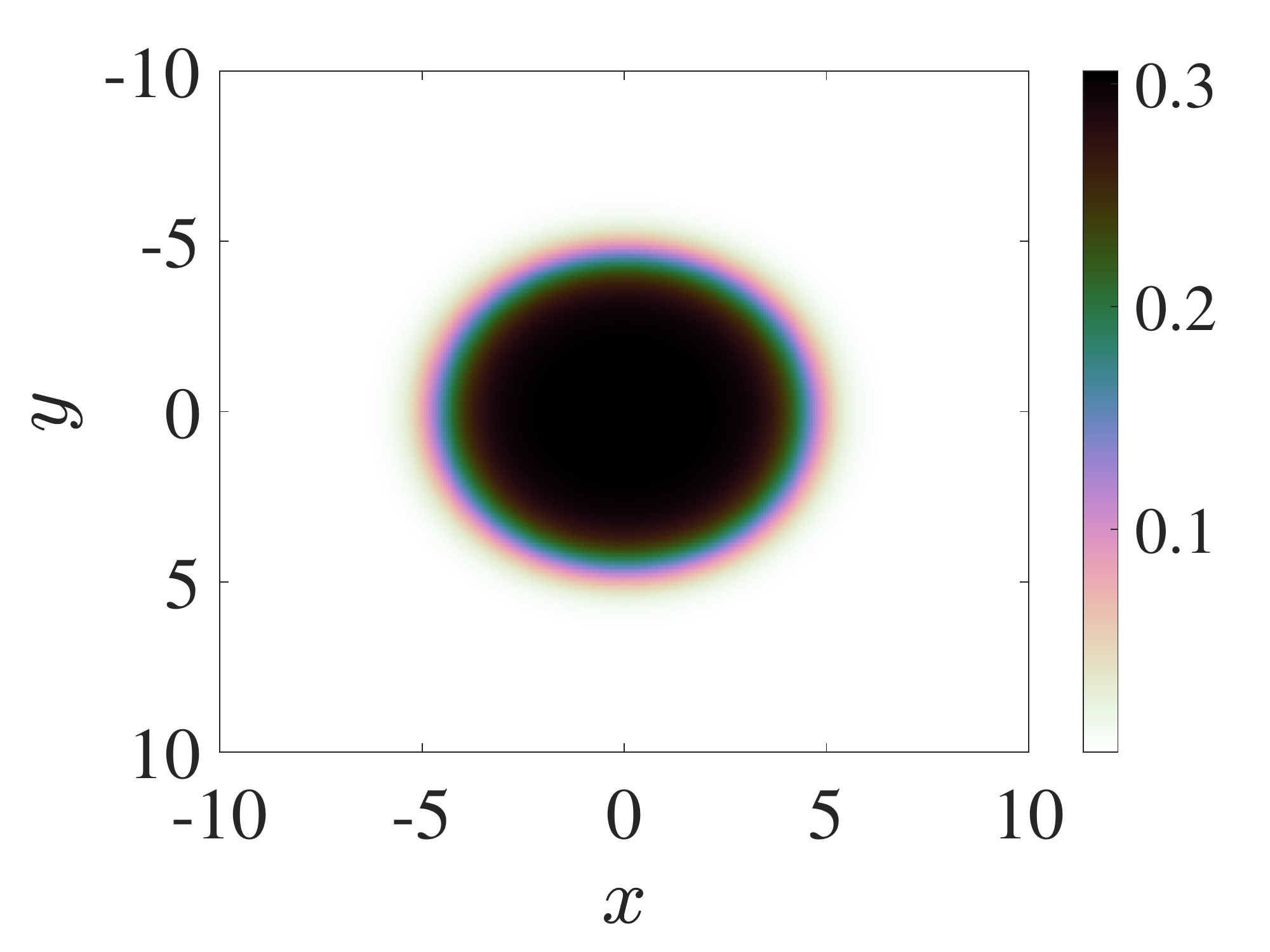}
\caption{
Spatial distribution of the density $\rho(x,y)$ at $t=500$ corresponding to perturbed 
square (top panels) and radial (bottom panels) kovaton solutions for $g=-1$ (left column)
and $g=1$ (right column). The densities shown in the left and right columns correspond to
$M=5$ and $M=20$, respectively, i.e., at values of $M$ where the solutions are linearly 
stable (see, Figs.~\ref{fig4} and~\ref{fig5}). For the square kovatons, $\omega=-8$ whereas 
$\omega=-4$ for the radial ones (with $q=5$ in both cases).
\label{fig6}
}
\end{center}
\end{figure}

\begin{figure}[pt!]
\begin{center}
\begin{overpic}
[height=.14\textheight, angle =0]{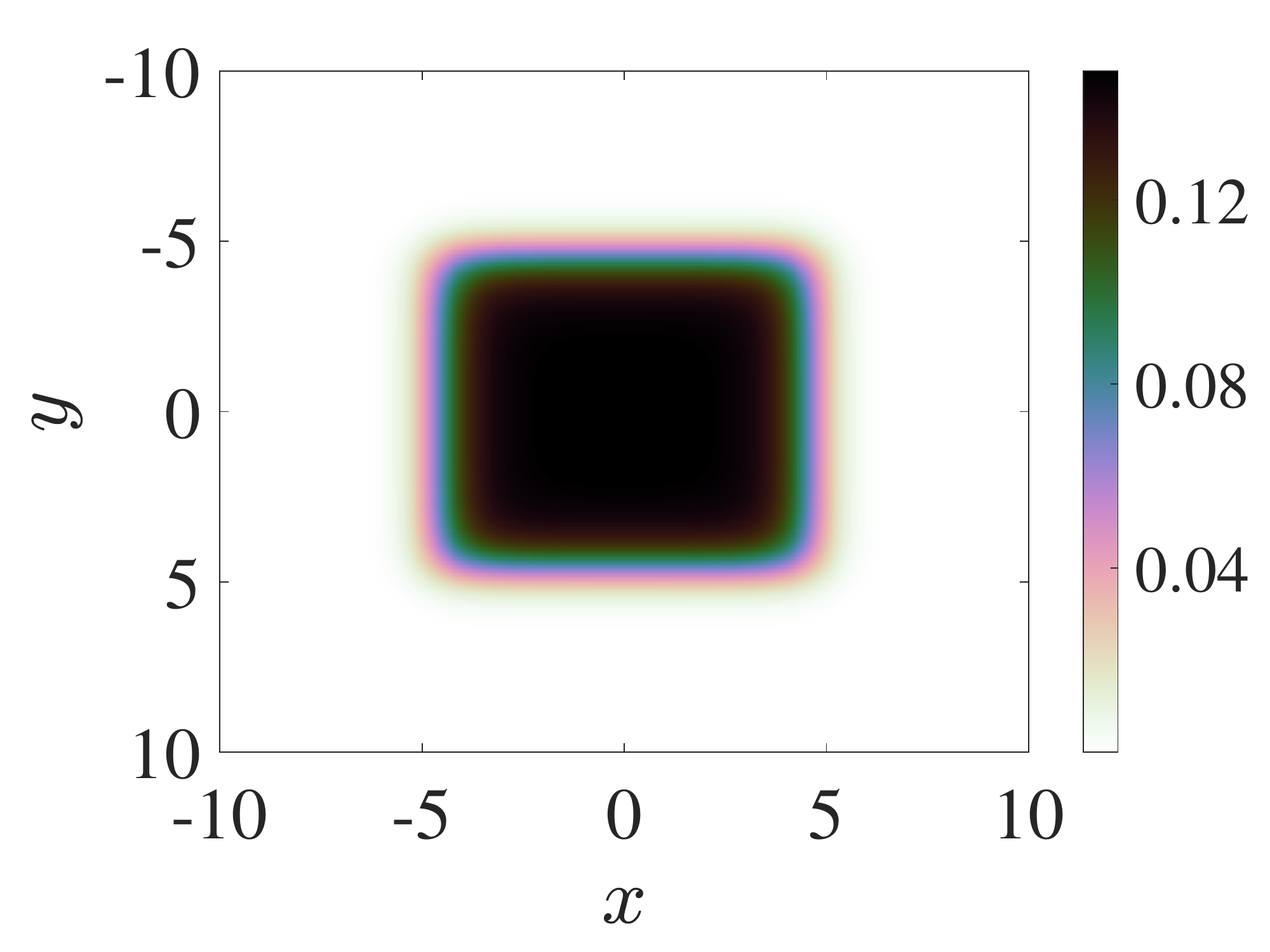}
\put(20,61){$t=0$}
\end{overpic}
\begin{overpic}
[height=.14\textheight, angle =0]{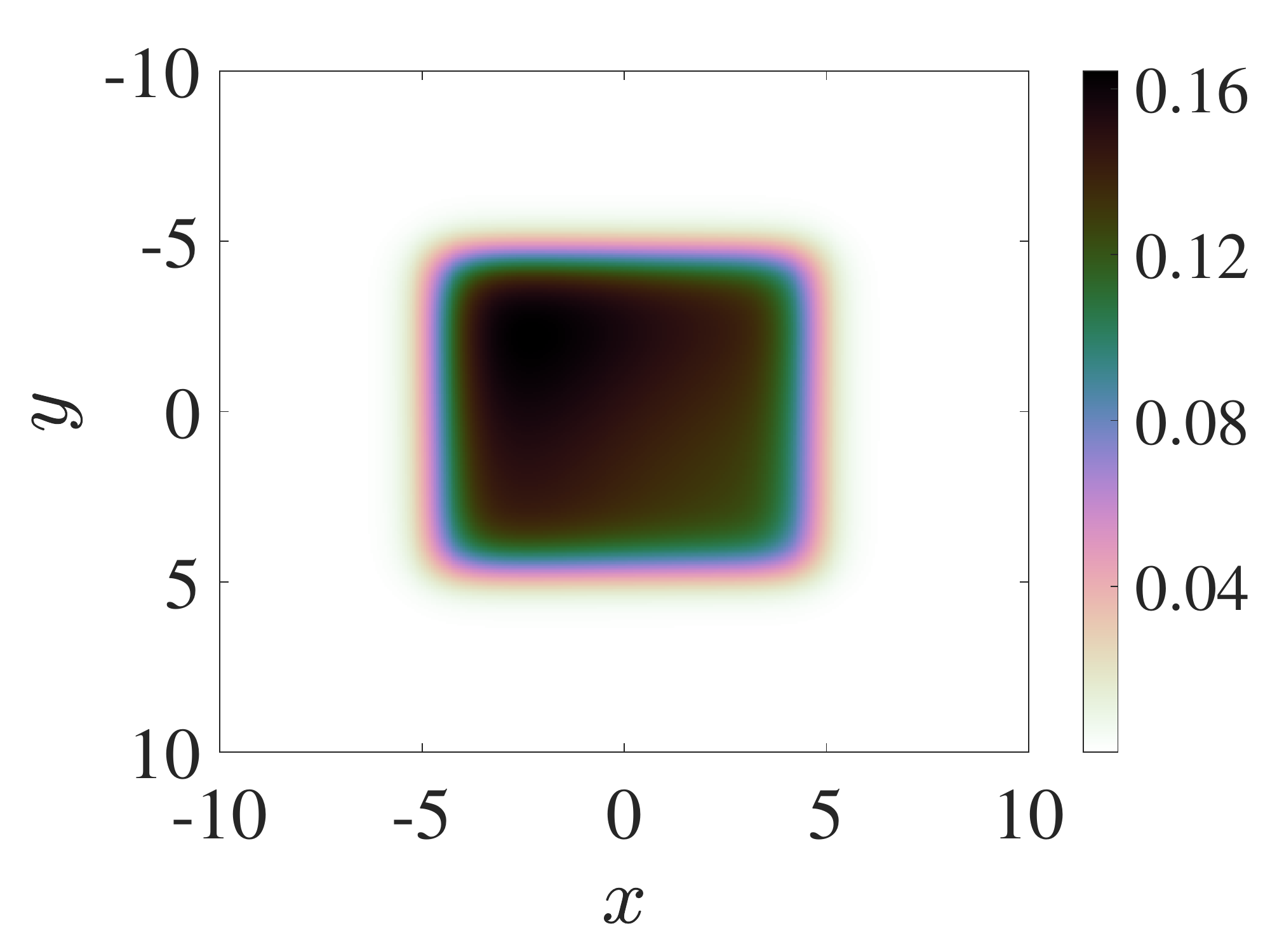}
\put(20,61){$t=285$}
\end{overpic}
\begin{overpic}
[height=.14\textheight, angle =0]{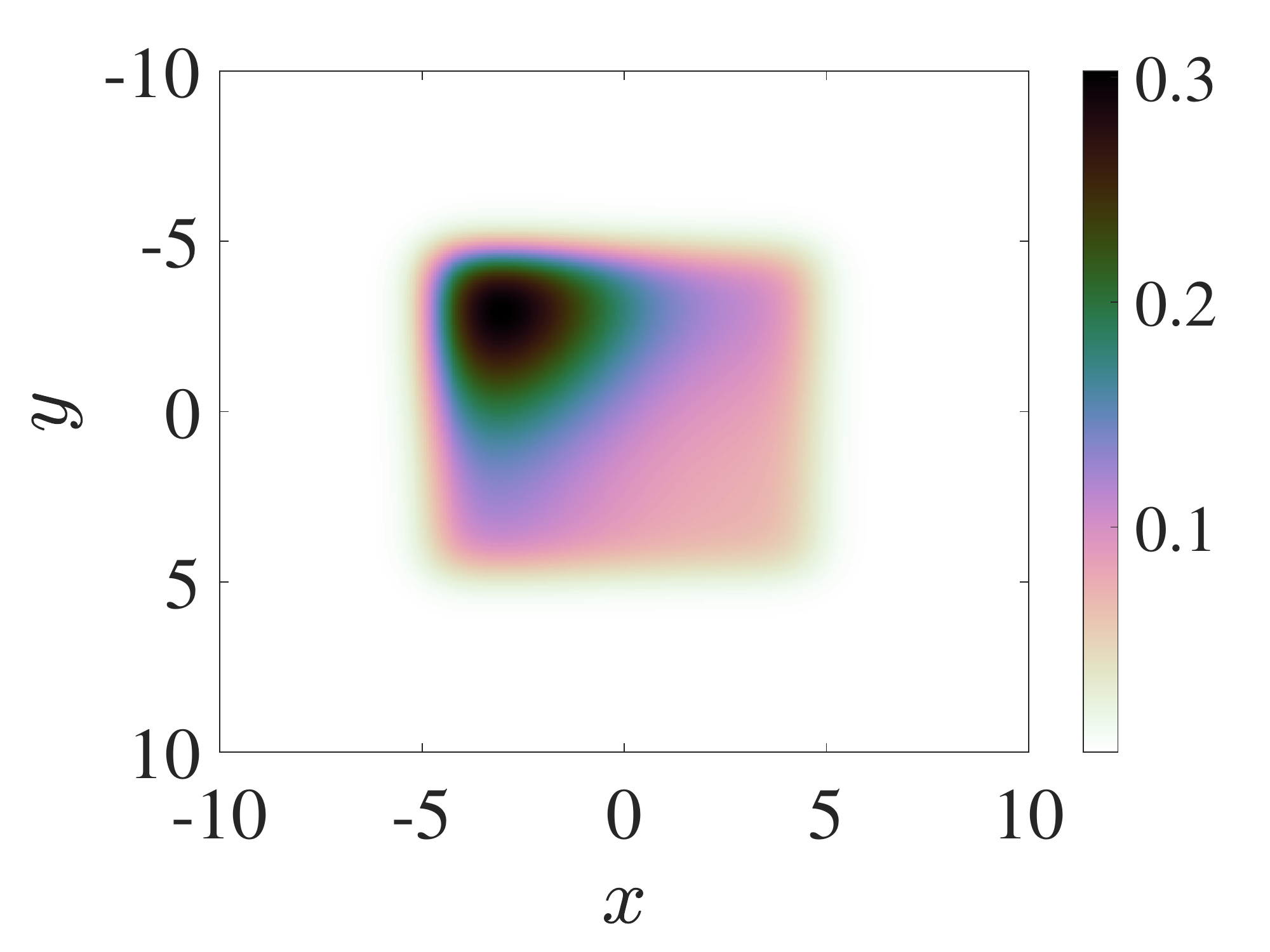}
\put(20,61){$t=300$}
\end{overpic}
\begin{overpic}
[height=.14\textheight, angle =0]{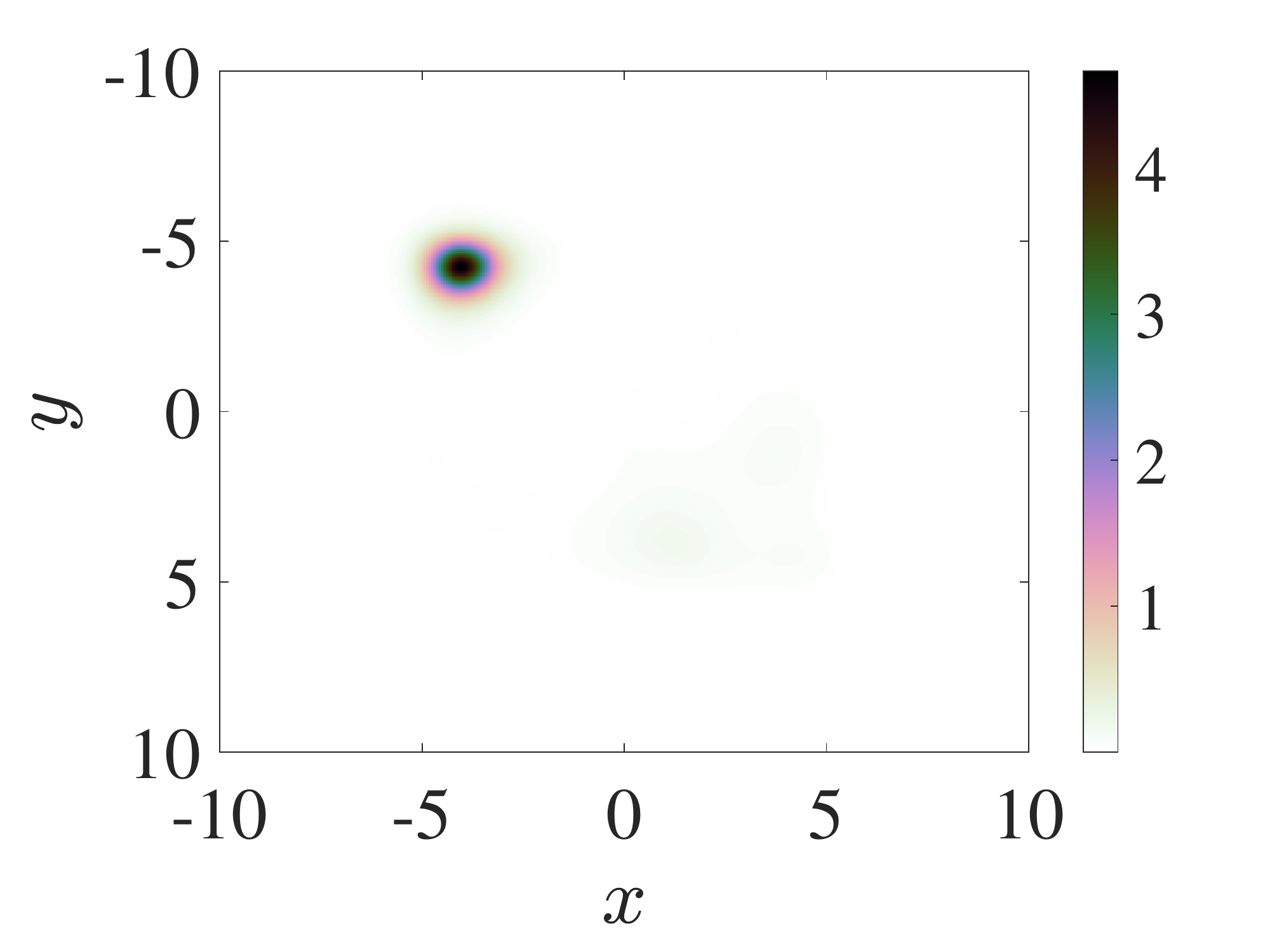}
\put(20,61){$t=500$}
\end{overpic}\\
\begin{overpic}
[height=.14\textheight, angle =0]{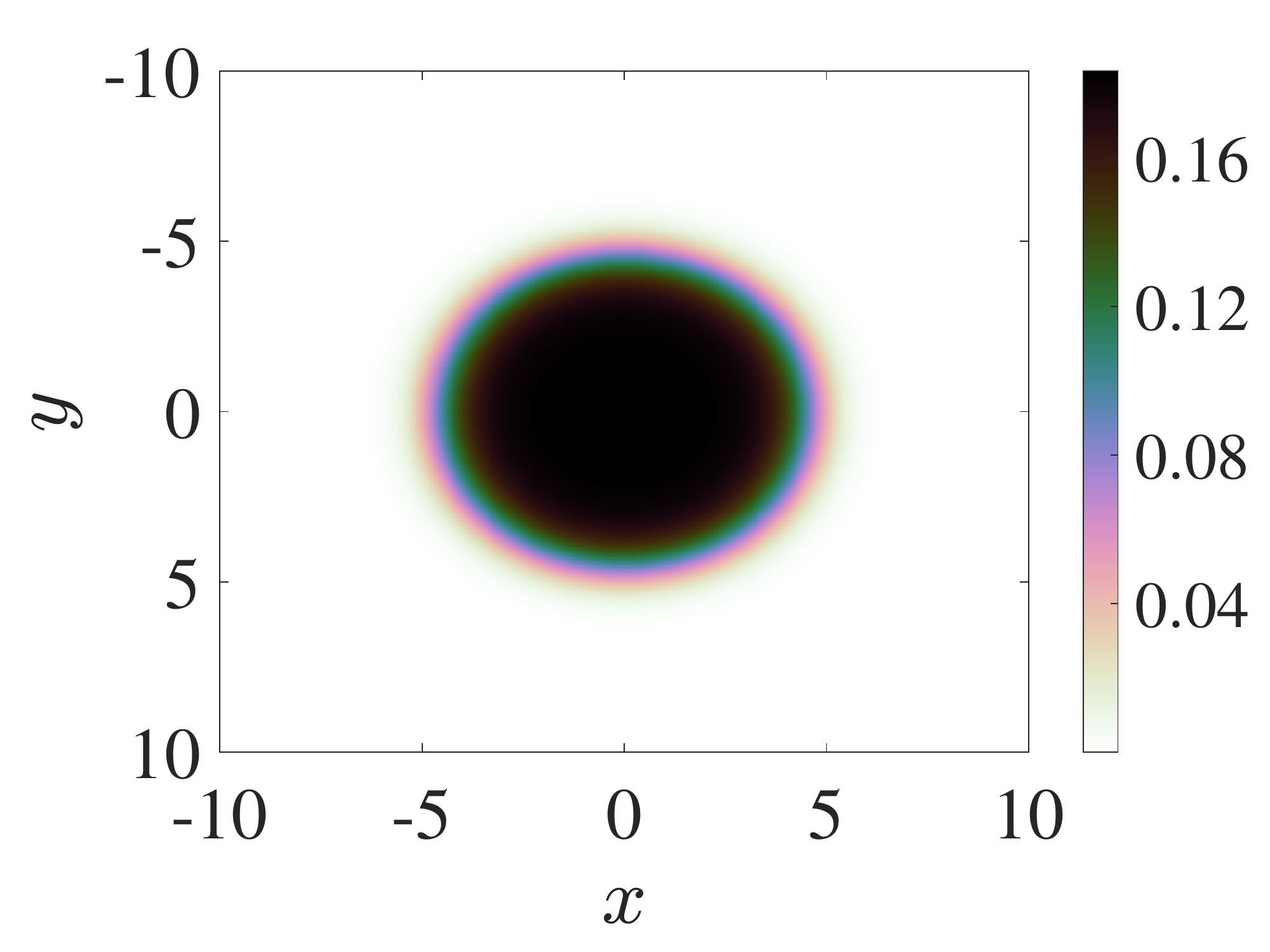}
\put(20,61){$t=0$}
\end{overpic}
\begin{overpic}
[height=.14\textheight, angle =0]{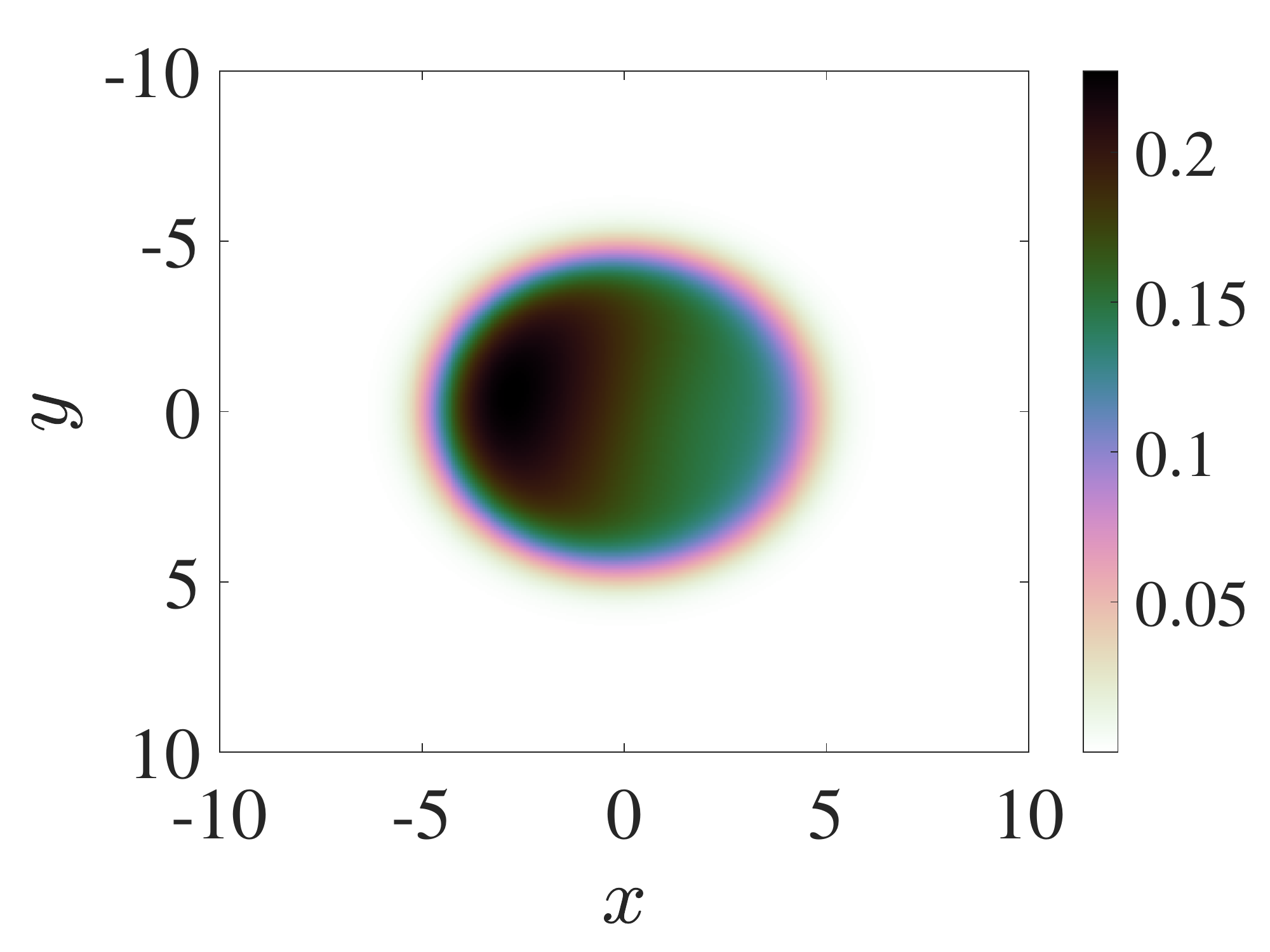}
\put(20,61){$t=230$}
\end{overpic}
\begin{overpic}
[height=.14\textheight, angle =0]{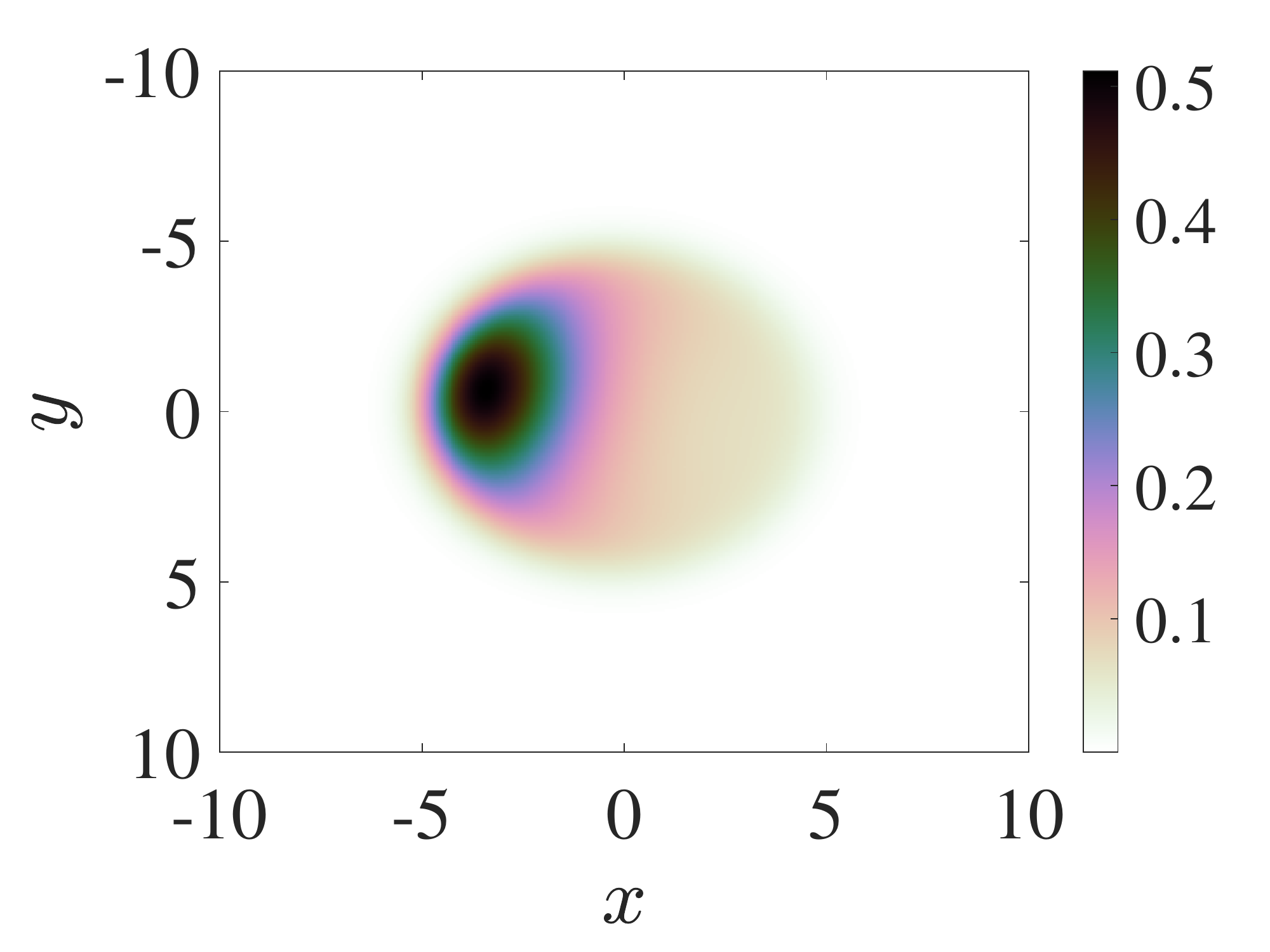}
\put(20,61){$t=240$}
\end{overpic}
\begin{overpic}
[height=.14\textheight, angle =0]{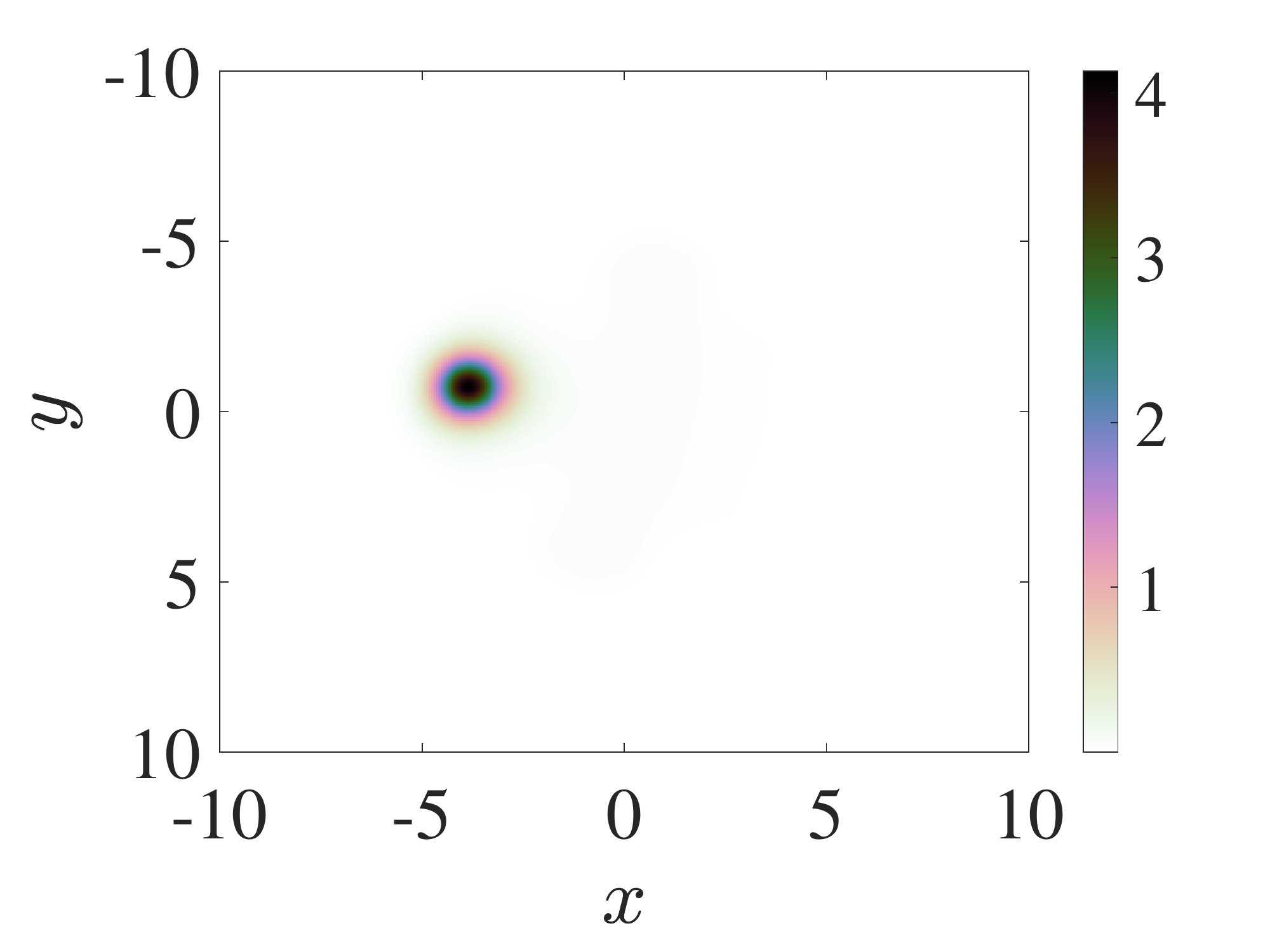}
\put(20,61){$t=250$}
\end{overpic}
\caption{
Snapshots of densities $\rho(x,y)$ of linearly unstable square (top panels) 
and radial (bottom panels) kovaton solutions with $M=12$, and $g=-1$. The rest
of the parameter values are the same as in Fig.~\ref{fig6}.
\label{fig7}
}
\end{center}
\end{figure}

Having discussed the spectral stability analysis results for 2D kovatons, we
now present selective case examples of the dynamics for square and radial kovaton 
solutions in Figs.~\ref{fig6} and~\ref{fig7}. We mention in passing that we
perturbed stationary kovaton solutions by adding a random perturbation with 
amplitude $10^{-4}\times\mathrm{max}(|u^{(0)}|)$ for stable solutions, and by adding 
the eigenvector corresponding to the most unstable eigendirection for unstable
solutions. In Fig.~\ref{fig6}, we check the stable square (top panels) and radial
(bottom panels) kovaton solutions for $g=-1$ and $g=1$ in the left and right columns,
respectively of the figure. In particular, for the case with $g=-1$, the square and 
radial kovaton solutions at $M=5$ are deemed stable (see, Figs.~\ref{fig4}(a) and~\ref{fig5}(a)), 
and we depict the density $\rho(x,y)$ at $t=500$ in the left column of Fig.~\ref{fig6}.
In the right column of the figure, we again showcase the density of perturbed square
and radial kovaton solutions with $g=1$ and $M=20$. Recall that in the repulsive case, 
the pertinent waveforms have been found to be stable (see, Figs.~\ref{fig4}(b) and~\ref{fig5}(b)), 
and this is corroborated in the panels of the right column of Fig.~\ref{fig6} where again
the density $\rho(x,y)$ at $t=500$ is shown therein.

\onecolumngrid\
\twocolumngrid\

We conclude this section on numerical results for the 2D GPE by considering 
Fig.~\ref{fig7} which presents snapshots of densities for the square (top panels) 
and radial (bottom panels) kovaton solutions at different instants of time (see, 
the labels at each panel). These results correspond to $g=-1$ and $M=12$ for both 
cases, i.e., square and radial kovaton solutions. At $t=0$ (see the leftmost panels
in Fig.~\ref{fig7}), we perturb the steady-states therein along the most unstable 
eigendirection, and around $t=285$ and $t=230$ we notice the onset of the instability 
for the square and radial kovaton solutions, respectively. As time progresses, the 
instability manifests itself (see the panels in the third column in the figure), 
driving the dynamics towards an almost stationary solution that is shown in the 
rightmost panels. This transition on the dynamics is strongly reminiscent of the 
one we observed in the 1D case, where the dynamics lead to the stationary bright 
solitary profiles of Fig.~\ref{fig2}. Herein, we observe shifted 2D bright solitary
pulses which should be connected with the pitchfork bifurcations we briefly mentioned
previously. In other words, the ``daughter'' branches emanating from the square and 
radial kovaton solutions at $M\approx 6.5$ and $6.8$, respectively, are expected to 
be stable (i.e., they inherit the stability of the respective ``parent'' branches), 
and they form an attractor upon which an unstable solution (such as the ones shown 
in Fig.~\ref{fig7}) is driven to.

%
\section{\label{s:3D}Three dimensions}

The methodology and stability analysis is similar in three dimensions that we briefly
discuss herein. For constant density in a cube one takes the wave function to be a 
product of 1D kovatons. The simplest 3D kovaton is the product of three 1D kovatons in Cartesian coordinates. In this case we can take $\psi(x,y,z,t) = A(M)\,  u(x,y,z) e^{-\rmi \omega t}$ where 
\begin{align}
   u(x,y,z)
   &=
   [\, \tanh(q-x) + \tanh(q+x) \,] \notag \\
   & \hspace{0em} \times [\, \tanh(q-y) + \tanh(q+y) \,] \notag \\
   & \hspace{0em} \times [\, \tanh(q-z) + \tanh(q+z) \,] 
   \label{e:2Dsquare-u}
\end{align}
with
\begin{equation}\label{e:2Dsquare-Mass}
   M
   =
   \int_{-\infty}^{\infty} \!\! \dd[3]{x} |u(x,y,z)|^2
   =
   64\, A^2(M) \, [\,2 q \coth (2 q)-1\,]^3\>.
\end{equation}

\noindent This leads to a confining potential:
\begin{equation}
V(x,y,z) = - 12+ V_2(x,y,z) + V_3(x,y,z),
\end{equation}
where
\begin{eqnarray}
V_2(x,y,z) &&=2 \left[\frac{(-2 \cosh (2 q) \cosh (2 x)+\cosh (4 x)-3)}{(\cosh (2 q)+\cosh (2
   x))^2}  \right. \nonumber \\
   && \left.
   +\frac{(-2 \cosh (2 q) \cosh (2 y)+\cosh (4 y)-3)}{(\cosh (2 q)+\cosh (2
   y))^2} \right. \nonumber \\
 && \left.   +\frac{(-2 \cosh (2 q) \cosh (2 z)+\cosh (4 z)-3)}{(\cosh (2 q)+\cosh (2
   z))^2}\right]\>, \nonumber \\
\end{eqnarray}
and
{\small
\ba
   &&V_3(x,y,z) =  \nonumber \\
   &&  \frac {g M \sinh ^6(2 q) (2 q \coth (2 q)-1)^{-3 } (\cosh (2 q)+\cosh (2 x))^{-2 }          } %
   { (\cosh (2 q)+\cosh (2 y))^2 (\cosh (2 q)+\cosh (2 z))^2}\>.  \nonumber \\
\ea
}
Similarly, in the radial case we obtain:
\ba \label{e:2DRu_3D}
   u(r)
  && =
   A \, [\, \tanh(q-r) + \tanh(q+r) \, ]  \nonumber \\
  && = A \,\sinh (2 q) \text{sech}(q-r) \text{sech}(q+r)\>,
\ea
\noindent where $r = \sqrt{x^2 + y^2+z^2 }$.
In this case, the density is given by $\rho(r) = | u(r) |^2$ and the mass by
\begin{align}
   M
   &=
   4 \pi \int_{0}^{\infty} \hspace{-0.5em} r^2  \dd{r} \rho(r)
   \label{e:3DRmass} \\
   &=\frac{2}{3} \pi  A^2 \left(-12 q^2+2 \left(4 q^2+\pi ^2\right) q \coth (2 q)-\pi
   ^2\right),
\end{align} 
as well as the potential reads:
{\small
\ba
V(r)&=& 
\text{sech}^2(q-r) \left(A^2 g \sinh ^2(2 q) \text{sech}^2(q+r)-1\right) \nonumber \\
&&+\tanh
   ^2(q-r)   +\tanh ^2(q+r) +\frac{2 \tanh (q-r)}{r}\nonumber \\
 &-&2 \tanh (q-r) \tanh (q+r)-\frac{2 \tanh
   (q+r)}{r} \nonumber \\
  && -\text{sech}^2(q+r)+\omega\>,
\ea
}

\noindent with $\omega=-4$ leading to  $V \rightarrow 0$ as $r \rightarrow \infty$. 
Again one can perform a stability analysis using Derrick's theorem, and reach 
the conclusion that the attractive interaction case becomes unstable as one 
increases the mass $M$ whereas the repulsive interaction case is always stable.

%
\section{\label{s:Conclusions}Conclusions}

In this paper we have shown how to find confining potentials such that the 
exact solution of the NLSE in that potential has constant density in a specified domain. This 
``reverse engineering'' method is entirely general, and one could have chosen 
Gaussian solutions~\cite{Cooper-2022} and multi soliton-like solutions. We 
then investigated the stability properties of these solutions using 
a numerical spectral stability analysis approach. We also tried to understand the stability of
these solutions using energy landscape methods such as Derrick's theorem. 
We found that the ``dark solitons'' were always stable to small perturbations 
and the ``bright solitons" exhibited different critical masses for an instability 
to develop depending on the type of perturbation applied. We corroborated these 
findings by performing numerical simulations as well as numerical stability analysis 
computations. In particular, for self-repulsive interactions, both results from 
Derrick's theorem and Bogoliubov-de Gennes (BdG) analysis predict stability. However for the self-attractive case
the BdG stability analysis results showed that for $g=-1$ (bright solutions), the kovaton 
solutions undergo a symmetry-breaking evolution, i.e., a pitchfork bifurcation where the solution 
itself follows the most unstable eigenvalue direction, and eventually reaches a nearby stable 
solution over the course of time integration of the GPEs.   This instability sets in at a much lower mass than the
usual self-similar blowup instability found in the NLSE without an external potential. In that situation,
the critical mass for this instability to set in is well described by Derrick's theorem. 
  Derrick's theorem considers dilations or contractions only which preserve the $x\rightarrow -x$ symmetry. Thus it cannot shed light on 
potential modes that may exhibit an instability at earlier values of the mass.  Another interesting  property that we find on applying Derrick's theorem is that because the external potential is a function of $M$, it is no longer true that
the second derivative becomes zero for $\beta=1$  at the critical mass. In fact it  always stays positive.  What happens is that
near $\beta=1$ an inflection point develops as we increase $M$.  To partially overcome the parity preserving defect of only considering self-similar perturbations,  we considered how the energy changes when we change the position of one of the components of the kovaton (i.e. the kink).  This deformation breaks the parity symmetry of the problem. We found that indeed the energy minimum as a function of this position parameter starts shifting from the origin at a critical mass which
is more in line with the results of the BdG analysis.  
%

%
\section{Acknowledgments}

EGC, FC, and JFD would like to thank the Santa Fe Institute and the Center for Nonlinear Studies at Los Alamos National Laboratory for their hospitality. 
One of us (AK) is grateful to Indian National Science Academy (INSA) for the award of INSA Senior Scientist position at Savitribai Phule Pune University. 
The work at Los Alamos National Laboratory was carried out under the auspices of the U.S. DOE and NNSA under Contract No. DEAC52-06NA25396. 

%
\appendix
%
\section{\label{s:units}Units}

In ordinary units, the time-dependent GPE is given by
\begin{equation}\label{e:units-GPequ}
   \rmi \hbar \pdv{\psi(\vb{r},t)}{t}
   =
   \Bigl \{\,
      - 
      \frac{\hbar^2}{2 m} \, \nabla^2 
      +
      U_0 \, | \psi(\vb{r},t) |^2 \,
      +
      V(\vb{r}) \,
   \Bigr \} \, \psi(\vb{r},t) \>,
\end{equation}
where at low energy we have that the interaction coefficient is given by:
\begin{equation}\label{e:units-U0}
   U_0
   =
   \frac{4 \pi \hbar^2 a}{m} \>,
\end{equation}
with $a$ (either $a>0$ or $a<0$) the scattering length being on the order of atomic size. 
The wave function for the GPE is normalized so that
\begin{equation}\label{e:unitspsinorm}
   N
   =
   \tint \dd[3]{x}  | \psi(\vb{r},t) |^2 \>,
\end{equation}
where $N$ is the particle number.  
We now need to relate a length scale $b$ to a time (or frequency $\omega_0$) scale. 
We take this to be such that:
\begin{equation}\label{e:bdef}
   \frac{\hbar}{2 m \, \omega_0 \, b^2} = 1 \>,
\end{equation}
so that if we set $\hbar = 1$ and $m = 1/2$, we have simply $\omega_0 = 1/b^2$. 
This way, and upon setting:
\begin{equation}\label{e:setscale}
   \vb*{\xi} = \vb{r}/b
   \qc
   \tau = \omega_0 \, t
   \qc
   \phi(\vb*{\xi},\tau) 
   = 
   \sqrt{\frac{b^3}{N_0}} \, \psi(b \,\vb{r},\omega_0 t) \>,
\end{equation}
the GPE [cf.~Eq.~\eqref{e:units-GPequ}] becomes dimensionless, 
that is
\begin{equation}\label{e:units-GPequ-dimensionless}
   \rmi \, \pdv{\phi(\vb*{\xi},\tau)}{\tau}
   =
   \Bigl \{\,
      - 
      \nabla_{\xi}^2 
      +
      g \, | \phi(\vb*{\xi},\tau) |^2 \,
      +
      W(\vb*{\xi}) \,
   \Bigr \} \, \phi(\vb*{\xi},\tau) \>,
\end{equation}
where
\begin{equation}
   g 
   = 
   \frac{N_0 \, U_0}{\hbar \, \omega_0 \, b^3} 
   = 
   8 \pi \, N_0 \,\Bigl ( \frac{a}{b} \Bigr )\>, \quad %
   W(\vb*{\xi})
   =
   \frac{V(\vb{r}/b)}{\hbar \, \omega_0} \>.
\end{equation}
For our case of ``reverse engineering,'' we set
\begin{equation}\label{e:phitou}
   \phi(\vb*{\xi},\tau)
   =
   u(\vb*{\xi}) \, \rme^{-\rmi \, \omega \tau} \>,
\end{equation}
and found that
\begin{equation}\label{e:Vequation2}
   W(\vb*{\xi})
   =
   \omega 
   +
   [\, \nabla^{2}_{\xi} u(\vb*{\xi}) \,] / u(\vb*{\xi})
   -
   g \, | u(\vb*{\xi}) |^{2}  \>.
\end{equation}
The particle number is now given by
\begin{equation}\label{e:Ndimsionless}
   N/N_0
   =
   \tint \dd[3]{\xi} u^2(\vb*{\xi}) \>.
\end{equation}
It would be natural to take $b = q$, which is the range of the external potential. 
Then in order for $g \sim 1$, we should take:
\begin{equation}\label{e:N0}
   N_0
   \sim
   \frac{1}{8\pi} \, \frac{q}{a} \gg 1 \>,
\end{equation}
so that if we take $N/N_0 \sim 1$, we see that since $N_0$ is a large number, 
this is a reasonable scaling. This means that we can take $q/b=1$ in the scaled 
external potential.   
For ${}^7$Li, the positive $s$-wave scattering length is $\sim 34\, a_0$~\cite{PhysRevA.53.R3713} 
whereas the negative scattering length is on the order of $-15 \, a_0$~\cite{PhysRevLett.72.40}, 
where $a_0 = 53 \times 10^{-12}\,\text{m}$ is the Bohr radius. The mass of ${}^7$Li is 7.016 u where 
$\mathrm{u} = 1.660 \times 10^{-27}$ kg is the atomic mass unit, the reciprocal of Avogadro's number. 
The critical temperature for a BEC to form must be on the order of $T \sim 5\,\mu\text{K}$.

%
\section{\label{s:secondderivative}Curvature of Derrick energy function at minimum}

In this appendix, we compute the second derivative of the Derrick energy 
function $h(\beta,M)$ in 1D for the attractive case ($g=-1$) evaluated at $\beta = 1$.  
The first two derivatives can be determined analytically at $\beta =1$. 
Indeed, upon using the fact that
\begin{widetext}
\bq
\frac{\partial |u|^{2}}{\partial\beta} \Big|_{\beta=1}= \frac{M \sinh ^2(2 q) %
(\cosh (2 q)-4 x \sinh (2 x)+\cosh (2 x))}{(2 q \coth (2 q)-1) %
(\cosh (2 q)+\cosh (2 x))^3},
\eq
and
{\small 
\ba
\frac{\partial^2|\tilde u|^{2}}{\partial\beta^2}\Big|_{\beta=1} =
-\frac{4 M x \sinh ^2(2 q) (2 \cosh (2 q) (\sinh (2 x)+x \cosh (2 x))+4 x+\sinh (4 x)-2
   x \cosh (4 x))}{(2 q \coth (2 q)-1) (\cosh (2 q)+\cosh (2 x))^4},
\ea ~~~
} 

\noindent we indeed find that $\frac{\partial h} {\partial \beta} |_{\beta=1} = 0$. For the
second derivative of $h$ with respect to $\beta$, we get contributions from $h_1$, 
$j_1$ and $j_2$ (see, Eqs.~\eqref{1D:Hbeta}) which 
tell us the answer depends on $g$ as well as $M$. The second derivative is explictly 
given by:
\bq
\frac{\partial^2 h}{\partial \beta^2 } |_{\beta=1} = 2 f_1(q) + g M f_2(q) +  f_3(q),
\eq
where 
\begin{align}
   f_1(q)
   &=   
   \frac{ \text{csch}^3(2 q) (9 \sinh (2 q)
      +\sinh (6 q)-24 q \cosh (2 q))}{6 (2 q \coth (2 q)-1)} \>,
   \notag \\
   f_2( q) 
   &=  
   \int \dd{x} 
   |u_{0}(x)|^{2} \, |u_{\beta \beta}(x,\beta,M)|^2 |_{\beta=1}/M^2 \>,
   \notag \\
   f_3(q)  
   &=   
   \int\dd{x} 
   \Bigl (
      \frac{\cosh(4x) - 2 \cosh(2q) \cosh(2x) - 3}
           { 2\, [\, \cosh^2(q) + \sinh^2(x) \,]^2} \, 
   \Bigr )  \, \frac{|u_{\beta \beta}(x,\beta,M)|^2 |_{\beta=1}}{M}  \>.
\end{align}
\end{widetext}
The functions $f_2(q)$ and $f_3(q)$ are explicitly known in terms of PolyLog functions~\cite{polylog} 
but presenting them would not be very informative. The surprise is that the second derivative 
of $h(\beta,M)$ evaluated at $\beta=1$ is positive for all negative values of $g$. Thus one cannot determine the critical number of atoms for an instability to arise from the second derivative alone. The instability caused  by a perturbation in the width degree of freedom is a result of the 
minimum getting shallower and shallower as we increase $M$. This is seen in our numerical 
evaluation of $h(\beta,M)$.

%
\bibliography{kovaton.bib}

\begin{thebibliography}{28}
\expandafter\ifx\csname natexlab\endcsname\relax\def\natexlab#1{#1}\fi
\expandafter\ifx\csname bibnamefont\endcsname\relax
  \def\bibnamefont#1{#1}\fi
\expandafter\ifx\csname bibfnamefont\endcsname\relax
  \def\bibfnamefont#1{#1}\fi
\expandafter\ifx\csname citenamefont\endcsname\relax
  \def\citenamefont#1{#1}\fi
\expandafter\ifx\csname url\endcsname\relax
  \def\url#1{\texttt{#1}}\fi
\expandafter\ifx\csname urlprefix\endcsname\relax\def\urlprefix{URL }\fi
\providecommand{\bibinfo}[2]{#2}
\providecommand{\eprint}[2][]{\url{#2}}

\bibitem[{\citenamefont{Pitaevskii and Stringari}(2003)}]{stringari}
\bibinfo{author}{\bibfnamefont{L.~P.} \bibnamefont{Pitaevskii}}
  \bibnamefont{and}
  \bibinfo{author}{\bibfnamefont{S.}~\bibnamefont{Stringari}},
  \emph{\bibinfo{title}{Bose-Einstein Condensation}}
  (\bibinfo{publisher}{Oxford University Press}, \bibinfo{address}{Oxford},
  \bibinfo{year}{2003}).

\bibitem[{\citenamefont{Pethick and Smith}(2002)}]{pethick}
\bibinfo{author}{\bibfnamefont{C.}~\bibnamefont{Pethick}} \bibnamefont{and}
  \bibinfo{author}{\bibfnamefont{H.}~\bibnamefont{Smith}},
  \emph{\bibinfo{title}{Bose-Einstein condensation in dilute gases}}
  (\bibinfo{publisher}{Cambridge University Press},
  \bibinfo{address}{Cambridge}, \bibinfo{year}{2002}).

\bibitem[{\citenamefont{Howl et~al.}(2019)\citenamefont{Howl, Penrose, and
  Fuentes}}]{Howl-2019}
\bibinfo{author}{\bibfnamefont{R.}~\bibnamefont{Howl}},
  \bibinfo{author}{\bibfnamefont{R.}~\bibnamefont{Penrose}}, \bibnamefont{and}
  \bibinfo{author}{\bibfnamefont{I.}~\bibnamefont{Fuentes}},
  \bibinfo{journal}{New J. Phys} \textbf{\bibinfo{volume}{21}},
  \bibinfo{pages}{043047} (\bibinfo{year}{2019}).

\bibitem[{\citenamefont{Gaunt et~al.}(2013)\citenamefont{Gaunt, Schmidutz,
  Gotlibovych, Smith, and Hadzibabic}}]{Gaunt-2013}
\bibinfo{author}{\bibfnamefont{A.~L.} \bibnamefont{Gaunt}},
  \bibinfo{author}{\bibfnamefont{T.~F.} \bibnamefont{Schmidutz}},
  \bibinfo{author}{\bibfnamefont{I.}~\bibnamefont{Gotlibovych}},
  \bibinfo{author}{\bibfnamefont{R.~P.} \bibnamefont{Smith}}, \bibnamefont{and}
  \bibinfo{author}{\bibfnamefont{Z.}~\bibnamefont{Hadzibabic}},
  \bibinfo{journal}{Phy. Rev. Lett.} \textbf{\bibinfo{volume}{110}},
  \bibinfo{pages}{200406} (\bibinfo{year}{2013}).

\bibitem[{\citenamefont{Lin et~al.}(2009)\citenamefont{Lin, Compton, Perry,
  Phillips, Porto, and Spielman}}]{Lin-2009}
\bibinfo{author}{\bibfnamefont{Y.-J.} \bibnamefont{Lin}},
  \bibinfo{author}{\bibfnamefont{R.~L.} \bibnamefont{Compton}},
  \bibinfo{author}{\bibfnamefont{A.~R.} \bibnamefont{Perry}},
  \bibinfo{author}{\bibfnamefont{W.~D.} \bibnamefont{Phillips}},
  \bibinfo{author}{\bibfnamefont{J.~V.} \bibnamefont{Porto}}, \bibnamefont{and}
  \bibinfo{author}{\bibfnamefont{I.~B.} \bibnamefont{Spielman}},
  \bibinfo{journal}{Phy. Rev. Lett.} \textbf{\bibinfo{volume}{102}},
  \bibinfo{pages}{130401} (\bibinfo{year}{2009}).

\bibitem[{\citenamefont{Gross}(1961)}]{Gross1961}
\bibinfo{author}{\bibfnamefont{E.~P.} \bibnamefont{Gross}},
  \bibinfo{journal}{Il Nuovo Cimento} \textbf{\bibinfo{volume}{20}},
  \bibinfo{pages}{454} (\bibinfo{year}{1961}).

\bibitem[{\citenamefont{Pitaevskii}(1961)}]{Pitaevskii-1961}
\bibinfo{author}{\bibfnamefont{L.~P.} \bibnamefont{Pitaevskii}},
  \bibinfo{journal}{Soviet Phys. JETP} \textbf{\bibinfo{volume}{20}},
  \bibinfo{pages}{451} (\bibinfo{year}{1961}).

\bibitem[{\citenamefont{Cooper et~al.}(2022)\citenamefont{Cooper, Khare,
  Charalampidis, Dawson, and Saxena}}]{Cooper-2022}
\bibinfo{author}{\bibfnamefont{F.}~\bibnamefont{Cooper}},
  \bibinfo{author}{\bibfnamefont{A.}~\bibnamefont{Khare}},
  \bibinfo{author}{\bibfnamefont{E.~G.} \bibnamefont{Charalampidis}},
  \bibinfo{author}{\bibfnamefont{J.~F.} \bibnamefont{Dawson}},
  \bibnamefont{and} \bibinfo{author}{\bibfnamefont{A.}~\bibnamefont{Saxena}},
  \bibinfo{journal}{Phys. Scr.} \textbf{\bibinfo{volume}{98}},
  \bibinfo{pages}{015011} (\bibinfo{year}{2022}),
  \urlprefix\url{https://dx.doi.org/10.1088/1402-4896/aca227}.

\bibitem[{\citenamefont{He}(1999)}]{He-1999}
\bibinfo{author}{\bibfnamefont{J.}~\bibnamefont{He}},
  \bibinfo{journal}{Computer Methods in Applied Mechanics and Engineering}
  \textbf{\bibinfo{volume}{178}}, \bibinfo{pages}{257} (\bibinfo{year}{1999}).

\bibitem[{\citenamefont{Antar and Pamuk}(2013)}]{Antar-2013}
\bibinfo{author}{\bibfnamefont{N.}~\bibnamefont{Antar}} \bibnamefont{and}
  \bibinfo{author}{\bibfnamefont{N.}~\bibnamefont{Pamuk}},
  \bibinfo{journal}{App. Comp. Math.} \textbf{\bibinfo{volume}{2}},
  \bibinfo{pages}{152} (\bibinfo{year}{2013}),
  \urlprefix\url{https://doi.org/10.11648/j.acm.20130206.18}.

\bibitem[{\citenamefont{Bogolyubov}(1947)}]{Bogolyubov-1947}
\bibinfo{author}{\bibfnamefont{N.~N.} \bibnamefont{Bogolyubov}},
  \bibinfo{journal}{Izv. Akad. Nauk SSSR, Ser. Fiz.}
  \textbf{\bibinfo{volume}{11}}, \bibinfo{pages}{77} (\bibinfo{year}{1947}).

\bibitem[{\citenamefont{Derrick}(1964)}]{Derrick-1964}
\bibinfo{author}{\bibfnamefont{G.~H.} \bibnamefont{Derrick}},
  \bibinfo{journal}{J. Math. Phys.} \textbf{\bibinfo{volume}{5}},
  \bibinfo{pages}{1252} (\bibinfo{year}{1964}),
  \urlprefix\url{https://dx.doi.org/10.1063/1.1704233}.

\bibitem[{\citenamefont{Wadati and Tsurumi}(1998)}]{Wadati-1998}
\bibinfo{author}{\bibfnamefont{M.}~\bibnamefont{Wadati}} \bibnamefont{and}
  \bibinfo{author}{\bibfnamefont{T.}~\bibnamefont{Tsurumi}},
  \bibinfo{journal}{Physics Letters A} \textbf{\bibinfo{volume}{247}},
  \bibinfo{pages}{287} (\bibinfo{year}{1998}).

\bibitem[{\citenamefont{Rosenau and Pikovsky}(2005)}]{Rosenau-2005}
\bibinfo{author}{\bibfnamefont{P.}~\bibnamefont{Rosenau}} \bibnamefont{and}
  \bibinfo{author}{\bibfnamefont{A.}~\bibnamefont{Pikovsky}},
  \bibinfo{journal}{Phy. Rev. Lett.} \textbf{\bibinfo{volume}{94}},
  \bibinfo{pages}{174102} (\bibinfo{year}{2005}).

\bibitem[{\citenamefont{Pikovsky and Rosenau}(2006)}]{Pikovsky-2006}
\bibinfo{author}{\bibfnamefont{A.}~\bibnamefont{Pikovsky}} \bibnamefont{and}
  \bibinfo{author}{\bibfnamefont{P.}~\bibnamefont{Rosenau}},
  \bibinfo{journal}{Physica D} \textbf{\bibinfo{volume}{218}},
  \bibinfo{pages}{56} (\bibinfo{year}{2006}).

\bibitem[{\citenamefont{Popov}(2017)}]{Popov-2017}
\bibinfo{author}{\bibfnamefont{S.~P.} \bibnamefont{Popov}},
  \bibinfo{journal}{Comput. Math. Math. Phys.} \textbf{\bibinfo{volume}{57}},
  \bibinfo{pages}{1560} (\bibinfo{year}{2017}).

\bibitem[{\citenamefont{Rosenau and Hyman}(1993)}]{PhysRevLett.70.564}
\bibinfo{author}{\bibfnamefont{P.}~\bibnamefont{Rosenau}} \bibnamefont{and}
  \bibinfo{author}{\bibfnamefont{J.~M.} \bibnamefont{Hyman}},
  \bibinfo{journal}{Phy. Rev. Lett.} \textbf{\bibinfo{volume}{70}},
  \bibinfo{pages}{564} (\bibinfo{year}{1993}),
  \urlprefix\url{https://link.aps.org/doi/10.1103/PhysRevLett.70.564}.

\bibitem[{\citenamefont{Garralon and Villatoro}(2012)}]{Garralon-2012}
\bibinfo{author}{\bibfnamefont{J.}~\bibnamefont{Garralon}} \bibnamefont{and}
  \bibinfo{author}{\bibfnamefont{F.~R.} \bibnamefont{Villatoro}},
  \bibinfo{journal}{Math. Comput. Model} \textbf{\bibinfo{volume}{55}},
  \bibinfo{pages}{1858} (\bibinfo{year}{2012}).

\bibitem[{\citenamefont{Garralon et~al.}(2013)\citenamefont{Garralon, Rus, and
  Villatoro}}]{Garralon-2013}
\bibinfo{author}{\bibfnamefont{J.}~\bibnamefont{Garralon}},
  \bibinfo{author}{\bibfnamefont{F.}~\bibnamefont{Rus}}, \bibnamefont{and}
  \bibinfo{author}{\bibfnamefont{F.~R.} \bibnamefont{Villatoro}},
  \bibinfo{journal}{Commun. Nonlinear Sci. Numer. Simulat.}
  \textbf{\bibinfo{volume}{18}}, \bibinfo{pages}{1576} (\bibinfo{year}{2013}).

\bibitem[{\citenamefont{Sulem and Sulem}(1999)}]{sulem}
\bibinfo{author}{\bibfnamefont{C.}~\bibnamefont{Sulem}} \bibnamefont{and}
  \bibinfo{author}{\bibfnamefont{P.}~\bibnamefont{Sulem}},
  \emph{\bibinfo{title}{The {N}onlinear {S}chr{\"o}dinger {E}quation}}
  (\bibinfo{publisher}{Springer-Verlag}, \bibinfo{address}{New York},
  \bibinfo{year}{1999}).

\bibitem[{\citenamefont{Dawson et~al.}(2017)\citenamefont{Dawson, Cooper,
  Khare, Mihaila, Ar{\'e}valo, Lan, Comech, and Saxena}}]{Dawson2017}
\bibinfo{author}{\bibfnamefont{J.~F.} \bibnamefont{Dawson}},
  \bibinfo{author}{\bibfnamefont{F.}~\bibnamefont{Cooper}},
  \bibinfo{author}{\bibfnamefont{A.}~\bibnamefont{Khare}},
  \bibinfo{author}{\bibfnamefont{B.}~\bibnamefont{Mihaila}},
  \bibinfo{author}{\bibfnamefont{E.}~\bibnamefont{Ar{\'e}valo}},
  \bibinfo{author}{\bibfnamefont{R.}~\bibnamefont{Lan}},
  \bibinfo{author}{\bibfnamefont{A.}~\bibnamefont{Comech}}, \bibnamefont{and}
  \bibinfo{author}{\bibfnamefont{A.}~\bibnamefont{Saxena}},
  \bibinfo{journal}{Journal of Physics A: Mathematical and Theoretical}
  \textbf{\bibinfo{volume}{50}}, \bibinfo{pages}{505202}
  (\bibinfo{year}{2017}),
  \urlprefix\url{https://dx.doi.org/10.1088/1751-8121/aa9006}.

\bibitem[{pol()}]{polylog}
\eprint{https://reference.wolfram.com/language/ref/PolyLog.html}.

\bibitem[{\citenamefont{Kestyn et~al.}(2016)\citenamefont{Kestyn, Polizzi, and
  Peter~Tang}}]{doi:10.1137/15M1026572}
\bibinfo{author}{\bibfnamefont{J.}~\bibnamefont{Kestyn}},
  \bibinfo{author}{\bibfnamefont{E.}~\bibnamefont{Polizzi}}, \bibnamefont{and}
  \bibinfo{author}{\bibfnamefont{P.~T.} \bibnamefont{Peter~Tang}},
  \bibinfo{journal}{SIAM Journal on Scientific Computing}
  \textbf{\bibinfo{volume}{38}}, \bibinfo{pages}{S772} (\bibinfo{year}{2016}),
  \urlprefix\url{https://doi.org/10.1137/15M1026572}.

\bibitem[{\citenamefont{Charalampidis et~al.}(2020)\citenamefont{Charalampidis,
  Boull\'e, Farrell, and Kevrekidis}}]{CHARALAMPIDIS2020105255}
\bibinfo{author}{\bibfnamefont{E.}~\bibnamefont{Charalampidis}},
  \bibinfo{author}{\bibfnamefont{N.}~\bibnamefont{Boull\'e}},
  \bibinfo{author}{\bibfnamefont{P.}~\bibnamefont{Farrell}}, \bibnamefont{and}
  \bibinfo{author}{\bibfnamefont{P.}~\bibnamefont{Kevrekidis}},
  \bibinfo{journal}{Communications in Nonlinear Science and Numerical
  Simulation} \textbf{\bibinfo{volume}{87}}, \bibinfo{pages}{105255}
  (\bibinfo{year}{2020}),
  \urlprefix\url{https://www.sciencedirect.com/science/article/pii/S1007570420300885}.

\bibitem[{\citenamefont{Mithun et~al.}(2022)\citenamefont{Mithun,
  Carretero-Gonz\'alez, Charalampidis, Hall, and
  Kevrekidis}}]{PhysRevA.105.053303}
\bibinfo{author}{\bibfnamefont{T.}~\bibnamefont{Mithun}},
  \bibinfo{author}{\bibfnamefont{R.}~\bibnamefont{Carretero-Gonz\'alez}},
  \bibinfo{author}{\bibfnamefont{E.~G.} \bibnamefont{Charalampidis}},
  \bibinfo{author}{\bibfnamefont{D.~S.} \bibnamefont{Hall}}, \bibnamefont{and}
  \bibinfo{author}{\bibfnamefont{P.~G.} \bibnamefont{Kevrekidis}},
  \bibinfo{journal}{Phy. Rev. A} \textbf{\bibinfo{volume}{105}},
  \bibinfo{pages}{053303} (\bibinfo{year}{2022}),
  \urlprefix\url{https://link.aps.org/doi/10.1103/PhysRevA.105.053303}.

\bibitem[{\citenamefont{Yang}(2012)}]{jyang_2012}
\bibinfo{author}{\bibfnamefont{J.}~\bibnamefont{Yang}},
  \bibinfo{journal}{Studies in Applied Mathematics}
  \textbf{\bibinfo{volume}{129}}, \bibinfo{pages}{133} (\bibinfo{year}{2012}),
  \urlprefix\url{https://doi.org/10.1111/j.1467-9590.2012.00549.x}.

\bibitem[{\citenamefont{Abraham et~al.}(1996)\citenamefont{Abraham,
  McAlexander, Gerton, Hulet, C\^ot\'e, and Dalgarno}}]{PhysRevA.53.R3713}
\bibinfo{author}{\bibfnamefont{E.~R.~I.} \bibnamefont{Abraham}},
  \bibinfo{author}{\bibfnamefont{W.~I.} \bibnamefont{McAlexander}},
  \bibinfo{author}{\bibfnamefont{J.~M.} \bibnamefont{Gerton}},
  \bibinfo{author}{\bibfnamefont{R.~G.} \bibnamefont{Hulet}},
  \bibinfo{author}{\bibfnamefont{R.}~\bibnamefont{C\^ot\'e}}, \bibnamefont{and}
  \bibinfo{author}{\bibfnamefont{A.}~\bibnamefont{Dalgarno}},
  \bibinfo{journal}{Phy. Rev. A} \textbf{\bibinfo{volume}{53}},
  \bibinfo{pages}{R3713} (\bibinfo{year}{1996}),
  \urlprefix\url{https://link.aps.org/doi/10.1103/PhysRevA.53.R3713}.

\bibitem[{\citenamefont{Moerdijk et~al.}(1994)\citenamefont{Moerdijk, Stwalley,
  Hulet, and Verhaar}}]{PhysRevLett.72.40}
\bibinfo{author}{\bibfnamefont{A.~J.} \bibnamefont{Moerdijk}},
  \bibinfo{author}{\bibfnamefont{W.~C.} \bibnamefont{Stwalley}},
  \bibinfo{author}{\bibfnamefont{R.~G.} \bibnamefont{Hulet}}, \bibnamefont{and}
  \bibinfo{author}{\bibfnamefont{B.~J.} \bibnamefont{Verhaar}},
  \bibinfo{journal}{Phy. Rev. Lett.} \textbf{\bibinfo{volume}{72}},
  \bibinfo{pages}{40} (\bibinfo{year}{1994}),
  \urlprefix\url{https://link.aps.org/doi/10.1103/PhysRevLett.72.40}.

\end{thebibliography}
%
%
\end{document}